\documentclass{lmcs}
\pdfoutput=1

\usepackage{lscape}

\usepackage{bm}
\usepackage{amsmath}
\usepackage{amssymb}
\usepackage[b]{esvect}
\usepackage{stmaryrd}
\usepackage{makecell}
\allowdisplaybreaks
\DeclareMathAlphabet{\mathbbm}{U}{bbm}{m}{n}%

\usepackage{comment}

\makeatletter
\def\@copyrightspace{\relax}
\makeatother

\usepackage{pmboxdraw}
\usepackage{fancyvrb}
\newwrite\myOutput
\makeatletter

\def\my@writeToFile#1{
  \immediate\openout\myOutput=output.txt%
  \immediate\write\myOutput{#1}%
  \immediate\closeout\myOutput%
}

\makeatother

\usepackage[leftmargin=1em,rightmargin=1ex,vskip=1ex]{quoting}
\usepackage{hyperref}
\usepackage{cleveref}

\AtEndPreamble{
  \RequirePackage{hyperref}
  \hypersetup{
    breaklinks = true,
    linktocpage,
    colorlinks = true,
    linkcolor = nordnight,
    citecolor = nordgreen,
    urlcolor = nordblue
  }
}

\newcommand{\co}[1]{\text{\reflectbox{\(#1\)}}}

\crefname{defi}{Definition}{Definitions}
\crefname{thm}{Theorem}{Theorems}
\crefname{lem}{Lemma}{Lemmas}
\crefname{prop}{Proposition}{Propositions}
\crefname{cor}{Corollary}{Corollaries}
\crefname{exa}{Example}{Examples}
\crefname{rem}{Remark}{Remarks}
\newcommand{\cat}[1]{\mathbb{#1}}
\newcommand{\ncat}[1]{\mathbf{#1}}
\newcommand{\fun}[1]{\mathsf{#1}}

\newcommand{\st}{\mid}
\newcommand{\defn}{\mathrel{\coloneqq}}
\newcommand{\codefn}{\mathrel{\text{\reflectbox{\(\defn\)}}}}

\newcommand{\id}[1]{\mathsf{id}_{#1}}
\newcommand{\dcomp}{\mathbin{\fatsemi}} %
\newcommand{\op}{^{\mathsf{op}}} %
\newcommand{\iso}{\cong} %
\newcommand{\obj}[1]{{#1}_{\mathsf{obj}}}

\newcommand{\tensor}{\otimes}
\newcommand{\cp}{\textcopypic}
\newcommand{\discard}{\textdiscardpic}
\newcommand{\swap}{\sigma}
\newcommand{\productmap}[2]{\langle #1,#2 \rangle}
\newcommand{\proj}[1]{\pi_{#1}}

\newcommand{\naturals}{\mathbb{N}}

\newcommand{\kleisli}[1]{\mathbf{Kl}(#1)}
\newcommand{\functorCat}[2]{[#1,#2]} %

\newcommand{\given}{\mid}

\newcommand{\bayesinv}[2]{#1_{#2}^{\dagger}}

\newcommand{\str}{\mathsf{str}}

\usepackage{tikz}
\usepackage{tikz-cd}

\usetikzlibrary{circuits.ee.IEC}
\IfFileExists{figures/externalize.token}{
\usetikzlibrary{external}
\tikzexternalize[prefix=figures/]
}{}

 \usepackage{stringdiagrams}
\usepackage{figuremacros}
\definecolor{med0}{HTML}{1C1B1B} %
\definecolor{med1}{HTML}{261D1D} %
\definecolor{med2}{HTML}{362B2B} %
\definecolor{med3}{HTML}{5E5757} %
\definecolor{med4}{HTML}{FFE983} %
\definecolor{med5}{HTML}{FFF4C2} %
\definecolor{med6}{HTML}{FDF6E3} %
\definecolor{med7}{HTML}{DB7842} %
\definecolor{med8}{HTML}{B32E39} %
\definecolor{med9}{HTML}{821529} %
\definecolor{medA}{HTML}{602E51} %
\definecolor{medB}{HTML}{FFC929} %
\definecolor{medC}{HTML}{60C37E} %
\definecolor{medD}{HTML}{89D7D0} %
\definecolor{medE}{HTML}{3E75DA} %
\definecolor{medF}{HTML}{D0ADE1} %

\colorlet{medBlack}{med0}
\colorlet{medWhite}{med6}
\colorlet{medRed}{med8}
\colorlet{medBlue}{medE}

\definecolor{nordred}{HTML}{bf616a}
\definecolor{nordblue}{HTML}{81a1c1}
\definecolor{norddarkblue}{HTML}{5e81ac}
\definecolor{nordgreen}{HTML}{a3be8c}
\definecolor{nordnight}{HTML}{4c566a}

\definecolor{bordeaux}{HTML}{4b1121}
\definecolor{darkyellow}{HTML}{FFC20A}
\definecolor{nicered}{HTML}{9C0D38}
\definecolor{niceblue}{HTML}{0C7BDC}
\definecolor{penBlue}{HTML}{001E59}

\colorlet{localblack}{black}
\colorlet{localwhite}{white}
\colorlet{localcolor}{penBlue}
\colorlet{localgray}{med3}
\colorlet{localblue}{medE}
\colorlet{localred}{med8}
\colorlet{addcolor}{white}
\colorlet{copycolor}{black}
\usepackage[xcolor,no patch,hyperref,quotation,electronic]{knowledge}
\knowledgeconfigure{notion}
\knowledgestyle{notion}{color=nordnight}
\knowledgestyle{intro notion}{emphasize,color=nordnight}

\knowledge{notion}
| cartesian category
| cartesian categories
| Cartesian category
| Cartesian categories
| cartesian monoidal category
| cartesian monoidal categories
| Cartesian monoidal category
| Cartesian monoidal categories

\knowledge{notion}
| copy
| Copy

\knowledge{notion}
| discard
| Discard

\knowledge{notion}
| deterministic
| Deterministic
| copyable
| Copyable
| determinism
| Determinism

\knowledge{notion}
| total
| Total
| discardable
| Discardable
| totality
| Totality

\knowledge{notion}
| copy-discard category
| copy-discard categories
| Copy-discard category
| Copy-discard categories

\knowledge{notion}
| premonoidal category
| premonoidal categories
| Premonoidal category
| Premonoidal categories

\knowledge{notion}
| symmetric monoidal category
| symmetric monoidal categories
| Symmetric monoidal category
| Symmetric monoidal categories

\knowledge{notion}
| effectful triple
| effectful triples
| Effectful triple
| Effectful triples

\knowledge{notion}
| secure channel
| secure channels
| secure

\knowledge{notion}
| strict effectful triple
| strict effectful triples
| Strict effectful triple
| Strict effectful triples

\knowledge{notion}
| SeedStoch
\newcommand{\SeedStoch}{\kl[SeedStoch]{\ncat{SeedStoch}}}

\knowledge{notion}
| EffStoch
\newcommand{\EffStoch}{\kl[EffStoch]{\ncat{EffStoch}}}

\knowledge{notion}
| effectful machine
| effectful machines
| Effectful machine
| Effectful machines
| effectful Mealy machine
| effectful Mealy machines
| Effectful Mealy machine
| Effectful Mealy machines
| machine
| machines
| Machine
| Machines

\newcommand{\Mealy}[1][]{\kl[effectful machine]{\mathsf{Mealy}}_{#1}}

\knowledge{notion}
| homomorphism of effectful machines
| homomorphisms of effectful machines
| Homomorphism of effectful machines
| Homomorphisms of effectful machines
| machine homomorphism
| machine homomorphisms
| Machine homomorphism
| Machine homomorphisms

\knowledge{notion}
| bisimulation
| Bisimulation

\knowledge{notion}
| bisimilar
| bisimilarity
| Bisimilarity

\knowledge{notion}
| MealyBis

\newcommand{\MealyBis}[1][]{\kl[MealyBis]{\mathsf{Mealy}^{\mathsf{bis}}_{#1}}}

\knowledge{notion}
| uniform feedback
| uniform feedback structure
| Uniform feedback
| Uniform feedback structure
| uniformity
| Uniformity
| feedback category
| feedback categories
| feedback monoidal categories
| feedback

\newcommand{\fbk}{\kl[uniform feedback]{\operatorname{fbk}}}

\knowledge{notion}
| dinaturality
| Dinaturality

\knowledge{notion}
| stream dinaturality
| Stream dinaturality

\newcommand{\dinat}{\mathrel{\kl[dinaturality]{\sim}}}

\knowledge{notion}
| profunctor
| Profunctor

\knowledge{notion}
| strong profunctor
| Strong profunctor

\knowledge{notion}
| whiskering
| whiskering operation

\newcommand{\lwhisk}{\kl[whiskering]{\ltimes}}
\newcommand{\rwhisk}{\kl[whiskering]{\rtimes}}

\knowledge{notion}
| raw effectful stream
| raw effectful streams
| Raw effectful stream
| Raw effectful streams

\newcommand{\rawStream}[1][]{\kl[raw effectful stream]{\mathsf{rawStream}_{#1}}}
\newcommand{\stream}[1]{\mathbf{#1}}

\knowledge{notion}
| stream tensoring
| Stream tensoring

\newcommand{\latercomp}{\mathbin{\kl[stream tensoring]{\cdot}}} %

\knowledge{notion}
| effectful stream
| effectful streams
| Effectful stream
| Effectful streams
\newcommand{\Stream}[1][]{\kl[effectful streams]{\fun{Stream}_{#1}}}
\newcommand{\effectfulStream}{\kl{effectful stream}}

\knowledge{notion}
| effectful functor
| effectful functors
| Effectful functor
| Effectful functors

\knowledge{notion}
| monoidal stream
| monoidal streams
| Monoidal stream
| Monoidal streams

\newcommand{\monStream}{\kl[monoidal stream]{\mathsf{monStream}}}

\knowledge{notion}
| cartesian stream
| cartesian streams
| Cartesian stream
| Cartesian streams

\knowledge{notion}
| parametrized effectful stream
| parametrized effectful streams
| Parametrized effectful stream
| Parametrized effectful streams

\knowledge{notion}
| corepeat

\newcommand{\corepeat}[1]{\kl[corepeat]{\llparenthesis} #1\kl[corepeat]{\rrparenthesis}}

\knowledge{notion}
| effectful trace
| effectful traces
| trace
| traces
| Effectful trace
| Effectful traces
| Trace
| Traces

\knowledge{notion}
| causal trace
| causal traces
| Causal trace
| Causal traces

\newcommand{\traceFun}{\kl[effectful trace]{\fun{Trace}}}

\knowledge{notion}
| trace equivalence
| trace equivalent

\knowledge{notion}
| type-invariant streams
| type-invariant effectful streams
| type-invariant stream
| type-invariant effectful stream
| Type-invariant streams
| Type-invariant effectful streams
| Type-invariant stream
| Type-invariant effectful stream

\newcommand{\StreamInv}[1][]{\kl[type-invariant streams]{\fun{Stream}^{\mathsf{inv}}_{#1}}}

\knowledge{notion}
| isolated dinaturality
| isolated stream dinaturality
| Isolated dinaturality
| Isolated stream dinaturality

\newcommand{\isodinat}{\kl[isolated dinaturality]{\mathrel{\smash{\stackrel{\cdot}{\sim}}}}}

\knowledge{notion}
| isolated effectful stream
| isolated effectful streams
| Isolated effectful stream
| Isolated effectful streams
| isolated stream
| isolated streams
| Isolated stream
| Isolated streams

\newcommand{\isoStream}{\kl[isolated effectful stream]{\mathsf{isoStream}}}

\knowledge{notion}
| projection to isolated streams
| isoproj

\newcommand{\isoproj}[1][\bullet]{\kl[isoproj]{\lbrace\! \left| #1 \right|\!\rbrace}}

\knowledge{notion}
| conditional composition
| Conditional composition

\newcommand{\condcomp}{\mathbin{\kl[conditional composition]{\triangleleft}}}

\knowledge{notion}
| conditional equivalence
| conditionally equivalent
| Conditional equivalence
| Conditionally equivalent

\newcommand{\cequiv}{\mathrel{\kl[conditional equivalence]{\simeq}}}

\knowledge{notion}
| inductive conditional equivalence
| inductive conditionally equivalent

\newcommand{\iequiv}{\mathrel{\kl[inductive conditional equivalence]{\simeq}}}

\knowledge{notion}
|procfun

\newcommand{\procfun}{\kl[procfun]{\fun{proc}}} %

\knowledge{notion}
| streamfun

\newcommand{\streamfun}{\kl[streamfun]{\fun{str}}} %

\knowledge{notion}
| sets

\newcommand{\Set}{\kl[sets]{\ncat{Set}}}

\knowledge{notion}
| Maybe
| Par
| partial functions

\newcommand{\Par}{\kl[Par]{\ncat{Par}}}

\knowledge{notion}
| nonempty powerset
| nonempty powerset monad
| Reltot

\newcommand{\Reltot}{\kl[Reltot]{\ncat{tRel}}}

\knowledge{notion}
| powerset
| powerset monad
| Rel

\newcommand{\parti}{\kl[powerset]{\mathcal{P}}} %
\newcommand{\Rel}{\kl[powerset]{\ncat{Rel}}}

\knowledge{notion}
| subdistributions
| subdistribution
| finitary subdistribution monad
| subStoch

\newcommand{\subdistr}{\kl[subdistributions]{\mathcal{D}_{\leq}}}
\newcommand{\subStoch}{\kl[subdistributions]{\ncat{Stoch}_{\leq}}}

\knowledge{notion}
| distributions
| distribution
| finitary distribution monad
| Stoch

\newcommand{\Stoch}{\kl[Stoch]{\ncat{Stoch}}}
\newcommand{\distr}{\kl[distributions]{\mathcal{D}}}

\knowledge{notion}
| StateStoch

\newcommand{\StateStoch}[1]{\kl[StateStoch]{\ncat{StStoch}}_{#1}}

\knowledge{notion}
| standard Borel spaces
| stdBorel

\newcommand{\stdBorel}{\kl[stdBorel]{\ncat{Borel}}}

\knowledge{notion}
| Giry
| BorelStoch

\newcommand{\Giry}{\kl[Giry]{\mathcal{G}}}
\newcommand{\BorelStoch}{\kl[BorelStoch]{\ncat{BorelStoch}}}

\knowledge{notion}
| subGiry
| BorelsubStoch

\newcommand{\subGiry}{\kl[subGiry]{\mathcal{G}_{\leq}}}
\newcommand{\BorelsubStoch}{\kl[BorelsubStoch]{\ncat{BorelStoch}_{\leq}}}

\knowledge{notion}
| BorelStateStoch

\newcommand{\BorelStateStoch}[1]{\kl[BorelStateStoch]{\ncat{StBorelStoch}}_{#1}}

\knowledge{notion}
| conditional
| conditionals
| Conditional
| Conditionals

\knowledge{notion}
| marginal
| marginals
| Marginal
| Marginals

\knowledge{notion}
| range
| ranges
| Range
| Ranges

\knowledge{notion}
| causal process
| causal processes
| Causal process
| Causal processes

\newcommand{\Proc}[1]{\kl[causal process]{\mathsf{Causal}}({#1})}
\newcommand{\Causal}{\kl[causal process]{\mathsf{Causal}}} %

\knowledge{notion}
| raw conditional sequence
| raw conditional sequences
| Raw conditional sequence
| Raw conditional sequences

\newcommand{\rcSeq}{\kl[raw conditional sequence]{\mathsf{rcSeq}}} %

\knowledge{notion}
| platercomp

\newcommand{\platercomp}{\mathbin{\kl[platercomp]{\cdot}}} %

\knowledge{notion}
| conditional sequence
| conditional sequences
| Conditional sequence
| Conditional sequences
| coinductive conditional sequence
| coinductive conditional sequences
| Coinductive conditional sequence
| Coinductive conditional sequences

\newcommand{\cCausal}{\kl[conditional sequences]{\mathsf{cSeq}}} %
\newcommand{\cSeq}{\kl[conditional sequences]{\mathsf{cSeq}}} %

\knowledge{notion}
| inductive conditional sequence
| inductive conditional sequences
| Inductive conditional sequence
| Inductive conditional sequences

\knowledge{notion}
| iseq
| from coinductive to inductive conditional sequences

\newcommand{\iseq}{\kl[iseq]{\fun{iseq}}} %

\knowledge{notion}
| cseq
| from inductive to coinductive conditional sequences

\newcommand{\cseq}{\kl[cseq]{\fun{cseq}}} %

\knowledge{notion}
| iplatercomp

\newcommand{\iplatercomp}{\mathbin{\kl[iplatercomp]{\cdot}}} %

\knowledge{notion}
| conditional preorder

\newcommand{\condleq}{\mathrel{\kl[conditional preorder]{\leq}}}

\knowledge{notion}
| dirac delta
| Dirac delta
| dirac

\newcommand{\dirac}[1]{\kl[dirac]{\delta}_{#1}}

\knowledge{notion}
| convex sets of distributions
| convex powerset of distributions
| convdistr

\newcommand{\StochRel}{\kl[convdistr]{\ncat{StochRel}}}
\newcommand{\convdistr}{\kl[convdistr]{\mathcal{C}}}

\knowledge{notion}
| stream cipher protocol
| stream cipher protocols
| Stream cipher protocol
| Stream cipher protocols
| cipher protocol
| cipher protocols
| Cipher protocol
| Cipher protocols
| alice
| bob
| cipher category

\newcommand{\Cipher}{\kl[cipher protocol]{\ensuremath{\mathbf{Cipher}}}}
\newcommand{\Alice}{\kl[alice]{\ensuremath{\mathsf{Alice}}}}
\newcommand{\Bob}{\kl[bob]{\ensuremath{\mathsf{Bob}}}}
\newcommand{\CIPHER}{\kl[cipher protocol]{\ensuremath{\mathsf{Cipher}}}}

\knowledge{notion}
| stream bob
| stream alice
| stream cipher

\newcommand{\alice}{\kl[stream alice]{\ensuremath{\mathsf{alice}}}}
\newcommand{\bob}{\kl[stream bob]{\ensuremath{\mathsf{bob}}}}
\newcommand{\cipher}{\kl[stream cipher]{\ensuremath{\mathsf{cipher}}}}

\knowledge{notion}
| totals subcategory

\newcommand{\totals}[1]{\kl[totals subcategory]{\fun{Tot}}(#1)} %

\newcommand{\dinaturality}{\kl{dinaturality}}
\newcommand{\streamDinaturality}{\kl{stream dinaturality}}

\newcommand{\print}{\ensuremath{\mathsf{print}}}

\newcommand{\secure}{\kl[secure channel]{\ensuremath{\mathsf{secure}}}}
\newcommand{\Secure}{\kl[secure channel]{\ensuremath{\mathsf{Secure}}}}

\newcommand{\Seed}{\ensuremath{\mathsf{Sd}}}
\newcommand{\Char}{\ensuremath{\mathsf{Char}}}

\newcommand{\memory}[1]{M_{#1}} %
\newcommand{\now}[1]{{#1}^{\circ}} %
\newcommand{\later}[1]{\ensuremath{{#1}^{+}}} %

\newcommand{\whis}[2]{w_{#2}({#1})}

\newcommand{\trace}{\kl{trace}}

\newcommand{\nf}[1]{#1_{i}}

\usepackage[utf8]{inputenc}
\usepackage{cmll}

\newcommand\scalemath[3]{\scalebox{#1}[#2]{\mbox{\ensuremath{\displaystyle #3}}}}

\newcommand{\leftarrowtip}{\ensuremath{\tikz\draw[line width=0.5pt,->] (10pt,0) -- (0,0);}}
\newcommand{\leftarrowtailnotip}{\ensuremath{\tikz\draw[line width=0.5pt,-<] (0,0) -- (10pt,0);}}

\newcommand{\unicodeStar}{\ensuremath{\star}}
\DeclareUnicodeCharacter{2605}{\unicodeStar}

\DeclareUnicodeCharacter{21D2}{\ensuremath{\Rightarrow}}
\DeclareUnicodeCharacter{2218}{\ensuremath{\circ}}
\DeclareUnicodeCharacter{2022}{\ensuremath{\bullet}}
\DeclareUnicodeCharacter{2219}{\ensuremath{\bullet}}
\DeclareUnicodeCharacter{2026}{\ensuremath{\dots}}
\DeclareUnicodeCharacter{2208}{\ensuremath{\in}}
\DeclareUnicodeCharacter{2192}{\ensuremath{\to}}
\DeclareUnicodeCharacter{2190}{\ensuremath{\leftarrowtip}}
\DeclareUnicodeCharacter{2919}{\ensuremath{\leftarrowtailnotip}}

\DeclareUnicodeCharacter{00D7}{\ensuremath{\times}}
\DeclareUnicodeCharacter{00B7}{\ensuremath{\cdot}}
\DeclareUnicodeCharacter{222B}{\ensuremath{\int}}
\DeclareUnicodeCharacter{22A4}{\ensuremath{\top}}
\DeclareUnicodeCharacter{22A5}{\ensuremath{\bot}}
\DeclareUnicodeCharacter{2264}{\ensuremath{\leq}}

\newcommand{\unicodecolon}{\ensuremath{\colon}}
\DeclareUnicodeCharacter{FE55}{\unicodecolon}
\newcommand{\unicodeleftpar}{\ensuremath{\left(}}
\DeclareUnicodeCharacter{27EE}{\unicodeleftpar}
\newcommand{\unicoderightpar}{\ensuremath{\right)}}
\DeclareUnicodeCharacter{27EF}{\unicoderightpar}
\DeclareUnicodeCharacter{2260}{\neq}
\DeclareUnicodeCharacter{22A9}{\Vdash}
\DeclareUnicodeCharacter{2237}{\proportion}
\DeclareUnicodeCharacter{2124}{\mathbb{Z}}
\DeclareUnicodeCharacter{27E8}{\langle}
\DeclareUnicodeCharacter{27E9}{\rangle}
\DeclareUnicodeCharacter{21A6}{\mapsto}
\DeclareUnicodeCharacter{22A2}{\vdash}
\DeclareUnicodeCharacter{2090}{\ensuremath{{}_a}}
\DeclareUnicodeCharacter{A71B}{{}^\uparrow}
\DeclareUnicodeCharacter{A71C}{{}^\downarrow}
\DeclareUnicodeCharacter{27E6}{\llbracket}
\DeclareUnicodeCharacter{27E7}{\rrbracket}

\newcommand{\unicoderightcircle}{\ensuremath{\RIGHTcircle}}
\DeclareUnicodeCharacter{25D1}{\unicoderightcircle}
\newcommand{\unicodeleftcircle}{\ensuremath{\LEFTcircle}}
\DeclareUnicodeCharacter{25D0}{\unicodeleftcircle}
\DeclareUnicodeCharacter{229B}{\circledast}

\newcommand{\unicodebbA}{\ensuremath{\mathbb{A}}}
\DeclareUnicodeCharacter{1D538}{\unicodebbA}
\newcommand{\unicodebbB}{\ensuremath{\mathbb{B}}}
\DeclareUnicodeCharacter{1D539}{\unicodebbB}
\newcommand{\unicodebbC}{\ensuremath{\mathbb{C}}}
\DeclareUnicodeCharacter{2102}{\unicodebbC}
\DeclareUnicodeCharacter{1D53B}{\ensuremath{\mathbb{D}}}
\DeclareUnicodeCharacter{2115}{\ensuremath{\mathbb{N}}}
\DeclareUnicodeCharacter{211D}{\ensuremath{\mathbb{R}}}
\DeclareUnicodeCharacter{1D543}{\ensuremath{\mathbb{L}}}
\newcommand\UnicodeBlackboardP{\ensuremath{\mathbf{P}}} \DeclareUnicodeCharacter{2119}{\UnicodeBlackboardP}
\DeclareUnicodeCharacter{211A}{\ensuremath{\mathbb{Q}}}
\DeclareUnicodeCharacter{1D544}{\ensuremath{\mathbb{M}}}
\DeclareUnicodeCharacter{1D54C}{\ensuremath{\mathbb{U}}}
\DeclareUnicodeCharacter{1D54D}{\ensuremath{\mathbf{V}}}
\DeclareUnicodeCharacter{1D54E}{\ensuremath{\mathbb{W}}}
\DeclareUnicodeCharacter{1D546}{\ensuremath{\mathbb{O}}}
\DeclareUnicodeCharacter{1D540}{\ensuremath{\mathbb{I}}}
\DeclareUnicodeCharacter{1D54A}{\ensuremath{\mathbb{S}}}
\DeclareUnicodeCharacter{1D53C}{\ensuremath{\mathbb{E}}}

\newcommand{\unicodecalS}{\ensuremath{\mathcal{S}}}
\newcommand{\unicodecalT}{\ensuremath{\mathcal{T}}}
\newcommand{\unicodecalC}{\ensuremath{\mathcal{C}}}
\newcommand{\unicodecalD}{\ensuremath{\mathcal{D}}}
\newcommand{\unicodecalX}{\ensuremath{\mathcal{X}}}
\newcommand{\unicodecalN}{\ensuremath{\mathcal{N}}}
\newcommand{\unicodecalE}{\ensuremath{\mathcal{E}}}
\DeclareUnicodeCharacter{1D4D4}{\unicodecalE}
\DeclareUnicodeCharacter{1D4D2}{\unicodecalC}
\DeclareUnicodeCharacter{1D4D3}{\unicodecalD}
\DeclareUnicodeCharacter{1D4DE}{\mathcal{O}}
\DeclareUnicodeCharacter{1D4E2}{\unicodecalS}
\DeclareUnicodeCharacter{1D4E3}{\unicodecalT}
\DeclareUnicodeCharacter{1D4E7}{\unicodecalX}
\DeclareUnicodeCharacter{1D4D0}{\ensuremath{\mathcal{A}}}
\DeclareUnicodeCharacter{1D4D1}{\ensuremath{\mathcal{B}}}
\DeclareUnicodeCharacter{1D4D6}{\ensuremath{\mathcal{G}}}
\DeclareUnicodeCharacter{1D4D7}{\ensuremath{\mathcal{H}}}
\DeclareUnicodeCharacter{1D4DB}{\ensuremath{\mathcal{L}}}
\DeclareUnicodeCharacter{1D4DC}{\ensuremath{\mathcal{M}}}
\DeclareUnicodeCharacter{1D4DD}{\unicodecalN}
\DeclareUnicodeCharacter{1D4B1}{\ensuremath{\mathcal{V}}}
\DeclareUnicodeCharacter{1D4E5}{\ensuremath{\mathcal{V}}}
\DeclareUnicodeCharacter{1D4E6}{\ensuremath{\mathcal{W}}}
\DeclareUnicodeCharacter{1D4E4}{\ensuremath{\mathcal{U}}}

\DeclareUnicodeCharacter{03B1}{\alpha}
\DeclareUnicodeCharacter{03B2}{\beta}
\DeclareUnicodeCharacter{03BC}{\mu}
\DeclareUnicodeCharacter{03B4}{\delta}
\DeclareUnicodeCharacter{03B5}{\varepsilon}
\DeclareUnicodeCharacter{03B7}{\eta}
\DeclareUnicodeCharacter{03BB}{\lambda}
\DeclareUnicodeCharacter{03C1}{\rho}
\DeclareUnicodeCharacter{03C8}{\psi}
\DeclareUnicodeCharacter{03C4}{\tau}
\DeclareUnicodeCharacter{03A8}{\Psi}
\DeclareUnicodeCharacter{03C3}{\sigma}
\DeclareUnicodeCharacter{03C6}{\varphi}
\DeclareUnicodeCharacter{03A6}{\Phi}
\DeclareUnicodeCharacter{03A3}{\Sigma}
\DeclareUnicodeCharacter{03D5}{\phi}
\DeclareUnicodeCharacter{03B8}{\theta}
\DeclareUnicodeCharacter{03C0}{\ensuremath{\pi}}
\DeclareUnicodeCharacter{0393}{\Gamma}
\DeclareUnicodeCharacter{0394}{\Delta}
\DeclareUnicodeCharacter{03BA}{\kappa}
\DeclareUnicodeCharacter{03BD}{\nu}
\DeclareUnicodeCharacter{25A0}{\blacksquare}
\DeclareUnicodeCharacter{25AA}{\blacksquare}

\newcommand{\hirayo}{\scaleobj{0.9}{\text{\usefont{U}{min}{m}{n}\symbol{'210}}}}
\DeclareUnicodeCharacter{3088}{\hirayo}
\DeclareFontFamily{U}{min}{}
\DeclareFontShape{U}{min}{m}{n}{<-> udmj30}{}

\newcommand\UnicodeWhiteRightPointingSmallTriangle{\triangleright}
\DeclareUnicodeCharacter{25B9}{\mathbin{\UnicodeWhiteRightPointingSmallTriangle}}
\newcommand\UnicodeWhiteDownPointingSmallTriangle{\triangledown}
\DeclareUnicodeCharacter{25BF}{\mathbin{\UnicodeWhiteDownPointingSmallTriangle}}
\newcommand\UnicodeWhiteUpPointingSmallTriangle{\scalemath{1}{-1}{{}^{\triangledown}}}
\DeclareUnicodeCharacter{25B5}{\mathbin{\UnicodeWhiteUpPointingSmallTriangle}}

\DeclareUnicodeCharacter{2080}{\ensuremath{{}_0}}
\DeclareUnicodeCharacter{2081}{\ensuremath{{}_1}}
\DeclareUnicodeCharacter{2082}{\ensuremath{{}_2}}
\DeclareUnicodeCharacter{2083}{\ensuremath{{}_3}}

\DeclareUnicodeCharacter{1D62}{\ensuremath{{}_i}}
\DeclareUnicodeCharacter{2C7C}{\ensuremath{{}_j}}
\DeclareUnicodeCharacter{02B3}{\ensuremath{{}^r}}
\DeclareUnicodeCharacter{02E1}{\ensuremath{{}^\ell}}
\DeclareUnicodeCharacter{1D48}{\ensuremath{{}^d}}
\DeclareUnicodeCharacter{1D50}{\ensuremath{{}^m}}
\DeclareUnicodeCharacter{1D58}{\ensuremath{{}^u}}
\DeclareUnicodeCharacter{209A}{\ensuremath{{}_p}}
\DeclareUnicodeCharacter{2096}{\ensuremath{{}_k}}

\DeclareUnicodeCharacter{2245}{\ensuremath{\cong}}
\DeclareUnicodeCharacter{2286}{\subseteq}

\DeclareUnicodeCharacter{22C5}{\cdot}
\DeclareUnicodeCharacter{25C3}{\ensuremath{\triangleleft}}
\DeclareUnicodeCharacter{25B9}{\ensuremath{\triangleright}}

\newcommand\smallmath[2]{#1{\raisebox{\dimexpr \fontdimen 22 \textfont 2
      - \fontdimen 22 \scriptscriptfont 2 \relax}{$\scriptscriptstyle #2$}}}
\newcommand\smalloplus{\smallmath\mathbin\oplus}

\DeclareUnicodeCharacter{2295}{\smalloplus}
\DeclareUnicodeCharacter{2297}{\otimes}
\DeclareUnicodeCharacter{214B}{\parr}
\DeclareUnicodeCharacter{2298}{\oslash}
\DeclareUnicodeCharacter{25C0}{\mathbin{\blacktriangleleft}}
\DeclareUnicodeCharacter{25C1}{\mathbin{\vartriangleleft}}
\DeclareUnicodeCharacter{22B3}{\mathbin{\triangleright}}
\DeclareUnicodeCharacter{22B2}{\mathbin{\triangleleft}}
\DeclareUnicodeCharacter{FF5C}{\mid}
\DeclareUnicodeCharacter{227A}{\mathbin{\prec}}
\DeclareUnicodeCharacter{227B}{\mathbin{\succ}}
\DeclareUnicodeCharacter{22A3}{\mathbin{\dashv}}
\DeclareUnicodeCharacter{219D}{\ensuremath{\leadsto}}
\DeclareUnicodeCharacter{1361}{\colon}

\DeclareUnicodeCharacter{29D1}{\mathrel{\multimapdotbothB}}
\DeclareUnicodeCharacter{29D2}{\mathrel{\multimapdotbothA}}
\DeclareUnicodeCharacter{22C4}{\mathbin{\diamond}}
\DeclareUnicodeCharacter{226B}{\mathrel{\gg}}
\DeclareUnicodeCharacter{25A1}{\Box}
\DeclareUnicodeCharacter{266F}{\sharp}

\DeclareUnicodeCharacter{2099}{_n}
\DeclareUnicodeCharacter{2098}{_m}
\DeclareUnicodeCharacter{1D5AD}{\ensuremath{\mathsf{N}}}

\DeclareUnicodeCharacter{1D4DF}{\mathcal{P}}

\newcommand\mydots{\makebox[0.6em][c]{.\hfil.\hfil.}}
\DeclareUnicodeCharacter{2026}{\mydots}
\DeclareUnicodeCharacter{226B}{\gg}

\usepackage{stmaryrd}
\newcommand{\unicodeRelationalComposition}{\fatsemi}
\DeclareUnicodeCharacter{2A3E}{\unicodeRelationalComposition}

\DeclareUnicodeCharacter{2014}{\,---\,} %
\usepackage{environ}

\newcommand{\mytikzcdcontext}[2]{
  \begin{tikzpicture}[baseline=(maintikzcdnode.base)]
    \node (maintikzcdnode) [inner sep=0, outer sep=0] {\begin{tikzcd}[#2]
        #1
      \end{tikzcd}};
  \end{tikzpicture}%
}

\NewEnviron{mytikzcd}[1][]{%
  \def\myargs{#1}%
  \edef\mydiagram{%
    \noexpand\mytikzcdcontext{\expandonce\BODY}{\expandonce\myargs}
  }%
  \mydiagram%
}


\begin{document}

\title{Effectful Mealy Machines}
\thanks{Filippo Bonchi was supported by the Italian Ministero dell'Università e
  della Ricerca, grant PRIN 2022, PNRR No. P2022HXNSC - RAP (Resource Awareness
  in Programming). This study was carried out within the National Centre on HPC,
  Big Data and Quantum Computing - SPOKE 10 (Quantum Computing) and received
  funding from the European Union Next-GenerationEU - National Recovery and
  Resilience Plan (NRRP) – MISSION 4 COMPONENT 2, INVESTMENT N. 1.4 – CUP N.
  I53C22000690001. Filippo Bonchi,
  Elena Di Lavore, and Mario Román were partially supported by the Advanced
  Research + Invention Agency (ARIA) Safeguarded AI Programme. For the purpose
  of Open Access the Author has applied a Creative Commons public copyright
  license (CC-BY) to any Author Accepted Manuscript version arising from this
  submission.}

\author[F.~Bonchi]{Filippo Bonchi\lmcsorcid{0000-0002-3433-723X}}[a]
\author[E.~Di~Lavore]{Elena Di Lavore\lmcsorcid{0000-0002-7783-5079}}[b,c]
\author[M.~Román]{Mario Román\lmcsorcid{0000-0003-3158-1226}}[b,c]

\address{Dipartimento di Informatica, Università di Pisa}
\address{Computer Science Department, University of Oxford}
\address{Tarkvarateaduse Instituut, Tallinn University of Technology}

\begin{abstract}
  Effectful Mealy machines, which we introduce, are a generalization of Mealy
  machines with global effects determined by an effectful triple. We provide
  semantics of effectful Mealy machines in terms of both bisimilarity and
  traces: bisimilarity is characterized syntactically, via uniform feedback;
  traces are constructed coinductively in terms of streams. We prove that this
  framework characterizes standard causal processes and existing flavours of
  Mealy machine, bisimilarity, and trace equivalence. In the commutative case,
  we introduce a monoidal generalization of Raney's causal functions: monoidal
  causal processes.
\end{abstract}

\maketitle

\section{Introduction}

Mealy machines—or, \emph{labelled transition systems} with outputs—are state
machines that, given an input, transition to the next state and produce an
output \cite{mealy1955method}. Mealy machines run in discrete-time steps,
mapping sequences of inputs to sequences of outputs. This mapping, together
with, possibly, some effects like nondeterminism, probability, or global state
modifications, constitutes their observable behaviour: their \emph{trace}.

Mealy machines can be abstracted in, at least, two different ways.

One may choose to emphasize their compositionality, the idea that
two machines running in parallel form themselves a machine, and that two machines
running in sequence—with the former outputting inputs to the latter—also form
a machine. From this point of view, Mealy machines are morphisms in a monoidal
category~\cite{katis02feedback}: a machine \(X \to Y\) consists of a
state space \(U\), a transition morphism \(f \colon U \tensor X \to U \tensor
Y\) and an initial state \(i \colon I \to U\).

Instead, one may choose to emphasize their behaviour: behavioural equivalence is
neatly defined by coalgebraic methods as equality in the final coalgebra. From
this point of view, Mealy machines are coalgebras \(f \colon U \to {\fun{T}(U
\times Y)}^{X}\) for some monad \(\fun{T}\) on a cartesian closed category. Two
states of a machine—\(i,j \colon 1 \to U\)—are behaviourally equivalent if they
coincide in the final coalgebra; this serves as a notion of
\kl{bisimilarity}~\cite{DBLP:conf/calco/Staton09}.

We propose to reconcile these two views on Mealy machines, to keep the
compositionality of machines as morphisms but to give them functorial semantics
as coalgebraic traces. At the same time, we expand the possible effects to
premonoidal categories. The resulting notion is that of an \kl{effectful
machine} over an \kl{effectful triple}.

\paragraph{Effectful triples}

\kl{Effectful Mealy machines} do not assume that transitions are pure functions:
instead, transitions are represented by the morphisms of a premonoidal category.
Premonoidal categories are a standard semantic universe for effectful sequential
programs \cite{powerR97:premonoidalnotions}: weaker than monoidal categories,
they allow non-commutative effects like global state alteration while keeping
their same intuitive reading as string diagrams
\cite{jeffrey1997:premonoidal,roman:effectful}.

\kl{Effectful triples} are a refinement of premonoidal
categories~\cite{jeffrey1997:premonoidal}: similar to Freyd
categories~\cite{power1997environments,levy2004}, they keep track of which
morphisms do produce a global effect, which ones only produce local
(commutative) effects, and which ones behave like values—that is, they can be
copied and discarded.

We will study Mealy machines—or simply, \emph{\kl{machines}}—in the generality
of an \kl{effectful triple}. They provide a setting that is general enough to
consider most cases of interest while being well-behaved enough for our
characterization results.

\paragraph{Effectful machines}
In this setting, we define \kl{machines} to be morphisms with an internal
state space, $U ⊗ X → U ⊗ Y$, together with a locally effectful initial state $I
→ U$. This description allows us to endow them with a universal property:
effectful machines are the morphisms of the free \kl{feedback category} over an
\kl{effectful triple}~\cite{katis02feedback}.

Still, we also recover the behavioural view on machines: our classical
machines-as-coalgebras induce machines in this sense; morphisms of
coalgebras—the classical notion of bisimulation—are generalized by \kl{machine
homomorphisms}; \kl{bisimulation} is generalized by connectedness by \kl{machine
homomorphisms}. We recover the classical notion of bisimulation from this
definition, and we prove it entails a notion of \kl{trace equivalence}.

\paragraph{Effectful traces and causal processes}

Executions of \kl{effectful machines} form \kl{effectful traces}, and
\kl{trace equivalence} follows as equality of these \kl{effectful traces}.
\kl{Effectful traces} are coinductively defined to be processes—of an
\kl{effectful triple}—followed by \kl{effectful traces}. Indeed, the trace
of an \kl{effectful machine} is defined to be the initial state and the
transition function followed by the coinductive repetition of the transition
function. \kl{Trace} extends to a functor from \kl{machines} to
a suitable \kl{effectful triple} of \kl{effectful traces}.

We show how \kl{effectful traces} coincide with the existing notions of traces
while generalizing them to the effectful case. To do this, we characterise
\kl{traces} as \kl{causal processes}: in well-behaved categories with only
commuting effects, \kl{effectful traces} coincide with a monoidal generalization
of Raney's notion of \kl{causal process} \cite{raney1958sequential}. As a
corollary, we get the usual notions of trace: e.g., stochastic traces coincide
with stochastic processes \cite{ross1996stochastic}.

\paragraph{Example: the stream cipher protocol}
\kl{Stream cipher protocols} encrypt messages of any given length. They are a
repeated version of the \emph{one-time pad protocol}: a perfectly secure
encryption technique that, however, requires sharing a key ahead of time through
a secure channel. In the one-time pad protocol, a first party (e.g. Alice) wants
to send a private message, $\bm{m}$, to a second party (e.g. Bob), through a
public channel. Alice and Bob already share a key $\bm{k}$—generated randomly
by, say, Alice—through a private channel, and this is the ingredient that
allows secure communication: Alice uses the XOR operation to mix the message and
the key, $\bm{m} ⊕ \bm{k}$, and sends that to Bob; Bob uses now the XOR
operation again to mix the received message with the key, $(\bm{m} ⊕ \bm{k}) ⊕
\bm{k}$. Now, because the XOR operation is a nilpotent algebra, Bob obtains the
decrypted message:
$$(\bm{m} ⊕ \bm{k}) ⊕ \bm{k} = \bm{m} ⊕ (\bm{k} ⊕ \bm{k}) = \bm{m} ⊕ 0 = \bm{m}.$$
An attacker listening to the public channel will only receive the encrypted
message, $\bm{m} ⊕ \bm{k}$, which is perfectly uninformative if they do not know
the value of the key. That is, the one-time pad protocol is secure, but it still
comes with a problem: as soon as the key is used once, it cannot be reused
safely; in order to send a long encrypted message, we need an equally long
pre-shared key. 

\begin{figure}[ht!]
  \centering
  \includegraphics*{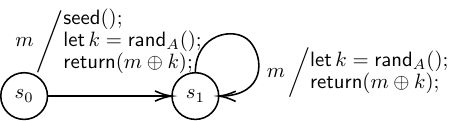} \quad\qquad
  \includegraphics*{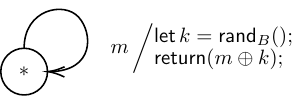}
  \caption{Mealy machines for Alice (left) and Bob (right).}%
  \label{fig:mealy-machines}%
\end{figure}

The \intro{stream cipher protocol} is a solution to this problem. Instead of
using the pre-shared key to encrypt and decrypt, Alice uses it to
$\mathsf{seed}$ two private identical random number generators (called
$\mathsf{rand}_A$ and $\mathsf{rand}_B$): now, Alice and Bob have an
inexhaustible source of shared random numbers that they can use to repeatedly
execute the one-time pad protocol to communicate messages of arbitrary length
(see \Cref{fig:mealy-machines}).  

\begin{exa}[Stream cipher via effectful machines]%
\label{ex:cipher-state-machines}%

Let us first generate an \kl{effectful triple}, $\Cipher{}$, from a signature
(\Cref{eq:fig-cipher-generators}) containing \emph{(i)} a type $C$, representing
messages; \emph{(ii)} effects for initializing a common seed, $s ፡ I ↝ I$,
extracting a random symbol from the first generator, $r_A ፡ I ↝ C$, and from the
second generator, $r_B ፡ I ↝ C$; and \emph{(iii)} pure operators, $(⊕) ፡ C ⊗ C →
C$ and $0 ፡ I → C$, representing the \emph{xor} operation and its unit
(\Cref{eq:fig-cipher-generators}).

Throught the paper, we employ string diagrams for premonoidal categories~\cite{jeffrey1997:premonoidal,roman:effectful}.
Premonoidal string diagrams require an extra wire—here labelled by ``\(R\)'', and additionally coloured in red—that does not represent an object of the category but that prevents effectful morphisms from interchanging. Formally, it ensures that substitution is well-defined \cite[Remark 2.8]{romanS25}; intuitively, it represents the global shared state that only effectful morphisms are allowed to modify.

In this example, the random generators—\(r_{A}\) and \(r_{B}\)—are effectful, as they access the globally shared source of randomness. However, both values and morphisms that only produce local effects cannot access the globally shared state and, thus, are allowed to interchange. In this example, the \emph{xor} operation and its unit do not access the source of randomness.
\begin{equation}
  \begin{split}
    \includegraphics*[scale=0.9]{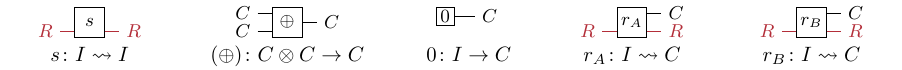}
    \label{eq:fig-cipher-generators}
  \end{split}
\end{equation}
Apart from these, in order to specify two \kl{effectful machines}, we include
\emph{(i)} a type $S$ representing the state space of Alice; \emph{(ii)} two effectful
transition functions, $t_A ፡ S ⊗ C ↝ S ⊗ C$ and $t_B ፡ C ↝ C$; \emph{(iii)} and
two pure states, $s₀ ፡ I → S$ and $s₁ ፡ I → S$. 
\begin{equation*}
  \includegraphics*[scale=0.9]{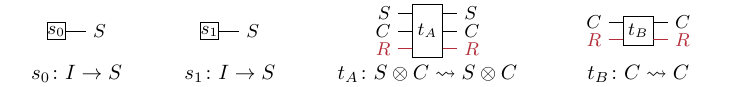}
\end{equation*}

We define $\Alice, \Bob ∈ \Mealy(\Cipher)(C;C)$ as two \kl{effectful machines}:
$\Alice = (S,s_0,t_{A})$ has internal states in $S$, initial state \(s_0\) and
transition morphism \(t_{A}\), while $\Bob = (I, \id{I}, t_{B})$ has no internal
states and transition morphism \(t_{B}\),
\begin{equation*}
    \includegraphics*[scale=0.9]{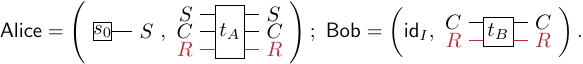}
\end{equation*}  

Transitions (in \Cref{fig:mealy-machines}) are then encoded by equations:
\begin{subequations}
    \begin{align}
    \transitionaliceequationfirstFigLeft & = \transitionaliceequationfirstFigRight\,;\\
    \transitionaliceequationsecondFigLeft & = \transitionaliceequationsecondFigRight\,;\\
    \generatorFig{premorphism=\(t_{B}\)} & = \bobnowFig\,.
    \end{align}%
    \label{eq:fig-cipher-transgenerators}
\end{subequations}

In turn, running $\Alice ∈ \Mealy(\Cipher)(C; C)$ first, copying its output and
then running $\Bob ∈ \Mealy(\Cipher)(C; C)$ over one of the copies produces the
whole-protocol \kl{machine} $\CIPHER ∈ \Mealy(\Cipher)(C; C ⊗ C)$.
\begin{equation*}
  \includegraphics*{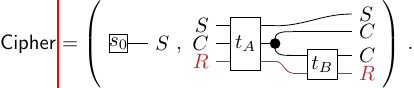}
\end{equation*}

\end{exa}

\subsection{Effectful streams: an effectful trace semantics}

\kl{Traces} record the inputs and outputs of a \kl{machine} along an execution.
They are a semantic universe for dataflow networks: in fact, in the
non-deterministic case, they help extending Kahn's original model
\cite{kahn1974} to a compositional semantics~\cite{jonsson89}. \kl{Traces} could
be an equally fruitful compositional semantics for \kl{machines} and
fully-fledged dataflow programming with effects
\cite{wadge85:lucid,eliotthudak97:reactive}. Yet, categorical semantics for
reactive programming has been initially restricted to the pure cartesian case
\cite{uustalu2008comonadic, krishnaswami12}, and the non-deterministic case
\cite{hildebrandt98}. Only recently, \kl{traces} allowing \emph{commutative}
effects (\emph{monoidal streams}) have been introduced
\cite{carette21,monoidalstreams}. We extend \emph{monoidal streams} beyond
commutative effects in order to introduce a unified notion of \kl{trace}:
\kl{effectful streams}.

\kl{Effectful streams} follow the classical coinductive
\cite{rutten2000universal,kozen17} definition of streams. A stream $s$ is an
element (the \emph{head}, $s^{∘}$) followed by a stream (the \emph{tail},
$s^{+}$). We can define a constant stream, declaring $\mathbf{1}^{∘} = 1$
and $\bm{1}^{+} = \bm{1}$; or instead an alternating stream, declaring
$\bm{alt}^{∘} = 0$, but $\bm{alt}^{+∘} = 1$ and $\bm{alt}^{++} = \bm{alt}$; we
can define the addition of two streams of natural numbers, declaring
$(\bm{x}+\bm{y})^{∘} = \bm{x}^{∘} + \bm{y}^{∘}$ and $(\bm{x}+\bm{y})^{+} =
\bm{x}^{+}+\bm{y}^{+}$; or define the stream of the natural numbers, declaring
$\bm{nats}^{∘} = 0$ and $\bm{nats}^{+} = \bm{1} + \bm{nats}$.

\kl{Effectful streams} follow this coinductive principle, with two important
differences. The first is that, instead of an element of a set, each piece of
the stream will be an \emph{effectful process}: formally, a morphism in an
\kl{effectful triple}. The second is that each piece of the stream—each
process—will not occur in isolation, but will be allowed to communicate with
the next piece via a \emph{memory}. This second principle allows causal
communication: messages can be passed from the past to the future, but not the
other way around. A minimalistic example of an \kl{effectful stream} is a
``counter'' program (\Cref{eq:counter}) printing the natural numbers by counting
in memory. We assume little about the semantics of ``$\print$'': it can be any
morphism $\print ∈ \cat{C}(\naturals; I)$ of any \kl{effectful triple} $(\cat{V},\cat{P},\cat{C})$. This
\kl{effectful stream} does not have inputs nor outputs, $\mathsf{count} ∈
\Stream[\cat{C}](I;I)$, only the memory is returned as an output and received as an
input at each step.
\begin{equation}
  \label{eq:counter}
  \begin{split}
    \includegraphics*[scale=0.9]{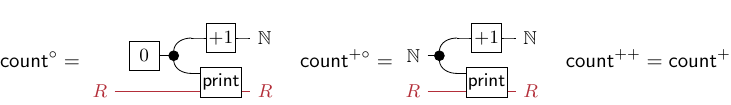}
  \end{split}
\end{equation}

\begin{exa}
\kl{Effectful machines} induce \kl{effectful streams} as \kl{traces}: following
\Cref{eq:fig-cipher-transgenerators}, $\Alice$ and $\Bob$ induce 
streams $\alice, \bob ∈ \Stream(\Cipher)(C;C)$ described by
$$\includegraphics[scale=0.9]{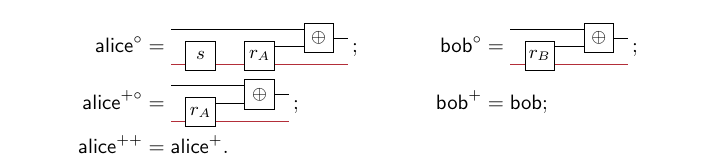}$$
Compositionally, the \kl{trace} of the whole protocol is the \intro[stream cipher]{stream} $\cipher ∈
\Stream(\Cipher)(C ; C ⊗ C)$, in \Cref{fig:cipher}.
\begin{figure}[!h]
  \centering
  \includegraphics*[scale=0.9]{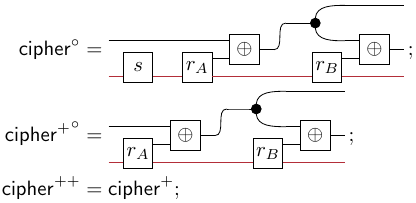}
  \caption{Stream cipher protocol.}
  \label{fig:cipher}
\end{figure}

Security of the \kl{stream cipher} is well-studied, but it allows us to showcase
the framework. In \Cref{sec:cipher-correctness}, we prove that the
\kl{stream cipher protocol} is secure relative to a secure pseudorandom number
generator: we extend the string-diagrammatic correctness for the \emph{one-time
pad} with nilpotent bialgebras due to Broadbent and
Karvonen~\cite{broadbentKarvonen23:composablecryptography}. 
\end{exa}

  In the case of deterministic and total computations, \kl{effectful streams} are
not the first semantics of repeated processes, so we take care of generalizing a
previous one: Raney's \emph{causal functions}~\cite{raney1958sequential}.

\subsection{Causal functions}
\emph{Causal functions} play a key role in the theory of streams since its dawn:
Raney \cite{raney1958sequential} introduces them in 1958, proves that they
compose and that all functions computed by finite \kl{machines} are causal.
Analogous results could be expected for the different flavours of
\kl{machines}, such as non-deterministic or probabilistic, but it is far from
obvious how causality can be defined in effectful settings.

In Raney's definition, a function $f ፡ X^{\omega} → Y^{\omega}$—where $A^\omega$
stands for the set of streams $\{(a_0,a_1,\dots) \mid a_i ∈ A\}$—is
\emph{causal} if the $n$-th output, $y_n$, only depends on the first $n$ inputs
$x_1, \dots, x_n$. We prove that this definition of causality can be generalized
by means of \emph{\kl{conditionals} in \kl{copy-discard categories}}, a notion
from synthetic probability
theory~\cite{jacobs2018probability,fritz2020synthetic}. We define
\emph{\kl{causal processes}} as an $\naturals$-indexed family of morphisms $f_n
፡ X_0 ⊗ \dots ⊗ X_n → Y_0 ⊗ \dots ⊗ Y_n$ such that there exist
\kl{conditionals}: families of morphisms $c_n$ satisfying
\Cref{eq:fig-causality}, for all $n ∈ ℕ$.
\begin{equation}
\begin{split}
    \includegraphics[scale=0.9]{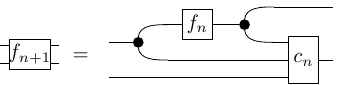}
    \label{eq:fig-causality}
\end{split}
\end{equation}

Our main result (\Cref{th:streams-are-processes}) provides sufficient conditions
guaranteeing that, in the absence of global effects, \kl{effectful streams}
coincide with \kl{causal processes}. The relevance of this result is twofold.
Firstly, while \kl{effectful streams} enjoy multiple relevant properties (neatly
implementable, universally characterized, generalizable to effectful cases) they
are defined modulo an equivalence, called \emph{\kl[stream dinaturality]{dinaturality}}, that could seem
more difficult to handle: our characterization shows that \kl[stream dinaturality]{dinaturality} can be
tamed in most cases. Secondly, \Cref{th:streams-are-processes} is general enough
to include stochastic, partial, non-deterministic, stateful, or deterministic
systems: we do not need to recast \kl{traces} and \kl{bisimilarity} each time we change
the flavour of \kl{machine}. In a sense, \Cref{th:streams-are-processes}
generalizes Raney's results \cite{raney1958sequential} to all these cases
(\Cref{fig:examples-conditionals}).

\subsection{Contributions}
\label{sec:contributions}

We introduce \kl{effectful machines} and show that they form an \kl{effectful
triple} (\Cref{prop:effectful-category-machines}) in the sense of Jeffrey
\cite{jeffrey1997:premonoidal}. 
\kl{Effectful machines} come with a notion of \emph{\kl{bisimilarity}}
(\Cref{def:effectful-bisimulation}) that generalizes the usual coalgebraic one
(\Cref{th:coalgebraic-bisimulation}). Most interestingly, we show that
\emph{uniformity}, first introduced in the context of traced monoidal
categories~\cite{cuazuanescu1994feedback,hasegawa02}, exactly captures
\kl{bisimilarity} once \emph{\kl{trace}} is replaced with \emph{\kl{feedback}}
(\Cref{th:mealyIsFree}).

In order to provide a \kl{trace} semantics to \kl{effectful machines}, we
introduce \emph{\kl{effectful streams}} (\Cref{def:effectfulStreams}): a common
generalization of classical stream transducers, \emph{stateful morphism
sequences} of Sprunger and Katsumata \cite{katsumata19} and the \emph{monoidal
streams} of Di Lavore, de Felice and Román \cite{monoidalstreams}. The
generalization is strict, in the sense that some of its examples could not be
captured by previous approaches: these include partial streams, relational
streams, quantum examples, and partial stochastic streams (apart from streams
with global effects, which we introduce). We prove that our \kl{trace} semantics
is compositional (\Cref{prop:trace-functor}) and that, as expected,
\kl{bisimilarity} implies \kl{trace equivalence}
(\Cref{th:bisimulation-implies-trace}).

We generalize Raney's \emph{causal functions} \cite{raney1958sequential}, to
\emph{\kl{causal processes}} (\Cref{def:causal-processes}) by exploiting
\kl{conditionals} in \kl{copy-discard categories}. Our main result
(\Cref{th:streams-are-processes}) states that, under the additional assumption
of existence of \emph{\kl{ranges}}, \kl{effectful streams} are indeed \kl{causal
processes}. We illustrate \kl{conditionals} and \kl{ranges} for the categories
of relations, partial functions and partial stochastic functions and thus (by
\Cref{th:streams-are-processes}) we obtain characterizations for relational,
partial and partial stochastic streams as \kl{causal processes}
(\Cref{cor:iso,fig:examples-conditionals}). This allows us to conclude that our
semantics based on \kl{effectful streams} generalize existing notions of
\kl{trace}.

\subsection{Related work}

\paragraph{Feedback and machines}

Monoidal machines and free \kl{feedback monoidal categories} have appeared
multiple times in the
literature~\cite{katis1997bicategories,hoshino2014memoryful,DBLP:journals/pacmpl/BonchiHPSZ19},
but their construction seems to be originally due to Katis, Sabadini, and
Walters~\cite{katis02feedback}. This construction was then extended to
\emph{delayed} feedback~\cite{canonicalalgebra} and related to monoidal
streams~\cite{monoidalstreams}. We go one step further: proposing a notion of
\kl{uniform feedback} that follows Hasegawa's \emph{uniform traces} and Căzănescu
and Ştefănescu's feedback~\cite{hasegawa02,cuazuanescu1994feedback}.

\paragraph{Effectful machines}

\kl{Effectful Mealy machines} are a natural extension of the \emph{bicategory of
processes} that Katis, Sabadini and Walters defined as the suspension-loop of a
base symmetric monoidal category~\cite{katis1997bicategories}. Categorical literature on automata is varied, leveraging
functorial semantics~\cite{colcombet20:automata}, coalgebraic techniques~\cite{rutten2000universal,jacobs2006bialgebraic,silva13:determinization,goncharov2020machines}, cartesian categories~\cite{boccali2023completeness}; or expanded Katis, Sabadini and Walters' work~\cite{canonicalalgebra,2021canonicalalgebra}. Finally, note that the term ``monoidal automata'' is also sometimes used to indicate automata recognising monoidal languages~\cite{earnshaw2022regular,earnshaw24contextfree}. Finite state automata can be given a finite axiomatisation~\cite{piedeleu2023finite}, but we are not concerned with this issue in the present manuscript.

Moore machines may also be generalized as tuples of morphisms in a monoidal category, as done by Goguen~\cite{goguen1975discrete}, Arbib and Manes~\cite{arbib1975adjoint}, but also more recently~\cite{adamek2018categorical,urbat2020automata,kocsis2025complete,aristote2026learning}; these approaches usually rely on a closed structure to account for input labels and cannot easily express composition of machines.
In fact, the composition of two monoidal Moore machines is a monoidal \emph{Mealy} machine unless the monoidal product is cartesian.

\paragraph{Effectful streams} 
Kahn~\cite{kahn1974} pioneered stream-based causal dataflow programming. The
particular coalgebraic approach we employ appears briefly in the work of Uustalu
and Vene~\cite{uustalu2008comonadic}, and it was developed in more detail by
Sprunger and Katsumata~\cite{katsumata19}: they first propose cartesian streams
as a categorical model for dataflow programming. This was later refined to
monoidal categories by Di Lavore, de Felice and Román~\cite{monoidalstreams};
however, to define \kl{effectful machines}, we need to significantly differ from
previous design choices: \emph{(i)} by taking the more general, but also more
natural setting of \kl{effectful triples}; \emph{(ii)} by refining feedback
(avoiding its type-theory) which, although sound, is not complete except in the
presence of a coinduction rule \cite{carette21}; and \emph{(iii)} by instead
deriving \kl{bisimilarity} from \emph{uniformity}. Finally, we capture examples
like partial, relational and partial stochastic streams, which were out of reach
for \emph{monoidal streams}, \emph{quantum delayed traces}~\cite{carette21} and
\emph{digital circuits}~\cite{ghica2022compositional}. Causal functions were
introduced by Raney~\cite{raney1958sequential} and later studied in different
aspects~\cite{jacobs:causalfunctions,ghica2022compositional,katsumata19}.

\paragraph{Bisimulation, traces and causal processes}

\kl{Bisimilarity} can be defined for any
coalgebra~\cite{rutten1995calculus,rutten2000universal,sokolova2005coalgebraic,sokolova2011probabilistic}.
When the base category is cartesian closed, transition systems can be expressed as coalgebras for a functor, which induces a notion of bisimulation for them.
The coalgebraic approach to transition systems leverages the universal properties of algebras and coalgebras to uniformly treat bisimilarity and traces~\cite{jacobs2006bialgebraic,silva13:determinization,jacobs2015trace,bonchi2021distribution,bonchi2022traces}.
Probabilistic and metric bisimulation, in particular, have received recent
attention~\cite{prakash1997bisimulation,danos06:bisimulation,bacci17:metric,bonchi2021distribution,bonchi2022traces}.
Different notions of equality for effectful and monoidal programs have been
studied: effectful applicative bisimilarity~\cite{dal2017effectful}, equivalence
for programs with effects and general recursion~\cite{voorneveld2020equality},
and monoidal traces~\cite{jacobs2010coalgebraic}. Our approach is more general
and coincides with the coalgebraic one under reasonable assumptions
(\Cref{th:coalgebraic-bisimulation}).
Traces may also be studied via graded monads~\cite{milius2015generic,dorsch2019graded}.
Generalizations of causal functions have
been studied in coalgebraic terms, see
e.g.~\cite{DBLP:conf/calco/Goncharov13,DBLP:conf/cmcs/PousRT22}. These works
usually require the existence of certain limits and colimits in the underlying
category, while our work relies on a (pre)monoidal structure.

\subsection{Synopsis}
\kl{Effectful triples}—our categorical playground are introduced in
\Cref{sec:effectful-triples}. For any \kl{effectful triple}, we provide a notion
of \kl{effectful machine} in \Cref{sec:machines}, and contextualize it. \kl{Bisimilarity} for
\kl{effectful machines} is introduced in \Cref{sec:feedback}, together with the
universal characterization in terms of \kl{feedback}. \kl{Traces}, as \kl{effectful streams},
are introduced in \Cref{sec:streams}, where we also discuss \kl{trace equivalence}.
\Cref{sec:causal-processes} provides another notion of \kl{traces} for commutative cases:
\kl{causal processes}; \Cref{sec:processes-are-streams} proves their equivalence.

This manuscript is the extended version of ``Effectful Mealy machines: Bisimulation and Trace'' presented at the Fortieth Annual ACM/IEEE Symposium on Logic in Computer Science (LiCS'25)~\cite{2025effectfulmachines}. The present extended version includes detailed proofs and lemmas, and it restates, for clarity, much of the introduction and the material on bisimilarity. It moreover contributes a novel subsection that characterizes causal processes via lax naturality (\Cref{prop:preorder,prop:causal-processes-lax-natural}); this characterization is then employed for a precise comparison with the topos of trees (\Cref{rem:topos-of-trees-enrichment}). %
\section{Effectful triples}%
\label{sec:effectful-triples}

The idea of using a triple of categories for the semantics of values, pure, and
effectful processes is a refinement by Jeffrey~\cite{jeffrey1997:premonoidal} of
the distinction between values and computations in the work of Power and
Thielecke~\cite{power1997environments}, and Levy~\cite{levy2004}.
\kl{Effectful triples} let us deal not only with Kleisli categories of
non-commutative monads, but also those arising from
comonads~\cite{uustalu2008comonadic}, distributive laws~\cite{powerwatanabe99},
or arrows~\cite{hughes00:arrows,hasuoheunenjacobs09:arrows}.

Ultimately, the original notion we develop is that of \emph{premonoidal
categories}~\cite{powerR97:premonoidalnotions,power1997environments}, which
\kl{effectful triples} refine with chosen cartesian and monoidal subcategories.
We assume familiarity with monoidal categories \cite{ml1972maclane,joyal91},
premonoidal categories \cite{powerR97:premonoidalnotions}, and their string
diagrammatic syntax.

\subsection{Cartesian categories and copy-discard categories}

\kl{Cartesian monoidal categories} are \kl{symmetric monoidal categories} whose
tensor constitutes a categorical product. Equivalently, \kl{cartesian monoidal
categories} are categories equipped with a natural and compatible comonoid
structure on each object~\cite{fox1976coalgebras}. 

\begin{defi}[Cartesian category]%
  \label{def:cartesian-category}%
  A \intro{cartesian category} is a symmetric monoidal category $(\cat{V}, \tensor,
  I)$ where each object, $A \in \cat{V}$, has a  \intro{copy}, \((\cp) \colon A
  → A \tensor A\), and \intro{discard}, \((\discard) \colon A → I\), morphisms
  forming a compatible commutative comonoid. All morphisms must be
  \emph{copyable} (or \intro{deterministic}), i.e., $f \dcomp \cp = \cp \dcomp
  (f \tensor f)$, and \emph{discardable} (or \intro{total}), $f \dcomp \discard
  = \discard$.
\end{defi}

While generally useful in pure programming semantics, both axioms of
\kl{cartesian categories} are too strong in the presence of effects. Copyability
fails in the presence of non-determinism: copying the output of a probabilistic
or a non-deterministic program is not the same as executing it twice.
Discardability fails in the presence of partiality: executing a non-halting
program and then discarding its value (which would never terminate) is not the
same as never executing it in the first place.

To allow effects, one needs to weaken copyability or discardability. Weakening
both, we arrive at \kl{copy-discard categories}.\footnote{Also known as
\emph{gs-monoidal categories}~\cite{corradini_gadducci_1999}.} \kl{Copy-discard
categories} do not assume all morphisms to be deterministic and total, but pick
instead a base category of copyable and discardable maps: a \kl{cartesian
monoidal category} $\cat{V}$ that is then included into a symmetric monoidal
category $\cat{P}$ having the same objects but more arrows. The \kl{cartesian
monoidal category} still provides a comonoid structure to every object, but the
symmetric monoidal category now accommodates computations that do not need to be
deterministic nor total.

\begin{defi}[Copy-discard category]%
  \label{def:copy-discard-category}
  A \intro{copy-discard category} is an identity-on-objects symmetric monoidal
  functor $\cat{V} → \cat{P}$ from a \kl{cartesian monoidal category}
  \(\cat{V}\) to a \kl{symmetric monoidal category} \(\cat{P}\).
\end{defi}

\begin{exa}[Kleisli categories are copy-discard categories]%
  \label{ex:copy-discard}
  Kleisli categories of commutative monads on \kl{cartesian categories} are \kl{copy-discard categories}.
  We will consider the category \intro[sets]{\(\Set\)} of sets and functions and the category \intro[stdBorel]{\(\stdBorel\)} of standard Borel spaces and measurable functions.
  In $\Set$, as well as in any of its Kleisli categories, the copy function, $(\cp) \colon X → X \tensor X$, is defined by $(\cp)(x) = (x,x)$, while the discard function, $(\discard) \colon X → I$, is defined by $(\discard)(x) = ()$.
  These functions are measurable, so the copy-discard structure in \(\stdBorel\) and its Kleisli categories is defined in the same way.
  
  We indicate with \intro[Par]{\(\Par\)} the category of sets and partial
  functions; with \intro[Reltot]{\(\Reltot\)} the category of sets and total
  relations; with \intro[Rel]{\(\Rel\)} the category of sets and relations; with
  \intro[Stoch]{\(\Stoch\)} the Kleisli category of the finitely-supported
  distribution monad, \(\distr \colon \Set \to \Set\); with
  \intro[subStoch]{\(\subStoch\)} the Kleisli category of the finitely-supported
  subdistribution monad \(\subdistr = \distr(- + 1) \colon \Set \to \Set\); with \intro[convex powerset of distributions]{\(\StochRel\)} the Kleisli category of the convex powerset of distributions monad \(\convdistr \colon \Set \to \Set\), \(\convdistr(X) = \{S \subseteq \distr(X) \st S \text{ convex}\}\)~\cite{jacobs2008coalgebraic,mio2014expectation,goy2020combining}; with
  \intro[BorelStoch]{\(\BorelStoch\)} the Kleisli category of the Giry monad on
  standard Borel spaces, \(\Giry \colon \stdBorel \to \stdBorel\)~\cite{giry82:categorical}; and with
  \intro[BorelsubStoch]{\(\BorelsubStoch\)} the Kleisli category of the
  Panangaden monad on standard Borel spaces, \(\subGiry = \Giry(- +1) \colon
  \stdBorel \to \stdBorel\)~\cite{panangaden1999}.

  In particular, we will consider the following Kleisli categories: the three
  possibilistic Kleisli categories of partial functions \((\Set, \Par)\), total
  relations \((\Set, \Reltot)\), and relations \((\Set, \Rel)\); the two
  discrete probabilistic Kleisli categories of distributions \((\Set, \Stoch)\),
  and subdistributions \((\Set, \subStoch)\); the Kleisli category of mixed probabilistic and possibilistic choice \((\Set, \StochRel)\); and the two continuous
  probabilistic Kleisli categories of stochastic kernels \((\stdBorel,
  \BorelStoch)\), and substochastic kernels, \((\stdBorel, \BorelsubStoch)\).
\end{exa}

\subsection{Effectful triples}
\label{sec:effectfulCategories}%

Commutative monads give rise to monoidal Kleisli categories. However, when a
strong monad is not commutative, its Kleisli category is only \emph{premonoidal}
\cite{powerR97:premonoidalnotions}. A \kl{premonoidal category} is similar to a
monoidal category but with the difference that the tensor, $(⊗)$, is not
functorial: in other words, \emph{interchange} does not hold: $(f ⊗ \id{}) ⨾
(\id{} ⊗ g)$ is not necessarily equal to $(\id{} ⊗ g) ⨾ (f ⊗ \id{})$. Still, the
tensor of a premonoidal category is separately functorial in each argument: that
is, \emph{left whiskering}, $(X ⊗ -)$, and \emph{right whiskering}, $(- ⊗ X)$,
are functors for each object $X$, even when $(- ⊗ -)$ is not a bifunctor.

Dropping the interchange axiom is important for effects: most monads are not
commutative—the order in which effects are executed does matter—and their
Kleisli categories are premonoidal. However, this does not mean that we need to
relinquish it altogether:  in the same way we separated some \emph{copyable} and
\emph{discardable} morphisms into a different \kl{cartesian monoidal category},
we can now separate some \emph{interchanging} morphisms into a different
\kl{symmetric monoidal category}. This motivates
\kl{effectful triples}~\cite{jeffrey1997:premonoidal}.

\begin{defi}[Effectful triple]%
  \label{def:effectful-triple}
  \AP An \intro{effectful triple}, $(\cat{V}, \cat{P}, \cat{C})$, consists of
  three categories and two identity-on-objects functors, where
  \begin{enumerate}
    \item the first category, $\cat{V}$ is a \kl{cartesian category} of copyable
    and discardable ``values'';
    \item the second category, $\cat{P}$, is a monoidal category representing
    ``pure computations'' or ``local effects'' that can be interchanged without
    altering the result; and
    \item the third category, $\cat{C}$, is a premonoidal category representing
    ``effectful computations'' or ``global effects'' with a fixed order of
    execution.
  \end{enumerate}
  The three categories must have the same objects. The first identity-on-objects functor, $\cat{V} → \cat{P}$, must  preserve monoidal structure strictly; the second identity-on-objects functor, $\cat{P} → \cat{C}$, must preserve monoidal structure strictly: its image must be central.\footnote{A morphism on a premonoidal category is central when it satisfies interchange with any other morphism of the category.} As a result, the tensor of the three categories must coincide.
\end{defi}

For a value, \(v \colon X \to Y\) in \(\cat{V}\), or a pure computation, \(p \colon X \to Y\) in \(\cat{P}\), we will denote their images in \(\cat{C}\) with the same symbols, \(v\) or \(p\), without explicitly writing the functors \(\cat{V} \to \cat{P} \to \cat{C}\).

\begin{rem}[Copy-discard structure of effectful triples]%
  \label{rem:copy-discard-effectful-triples}
  The presence of the \kl{cartesian category} \(\cat{V}\) in an \kl{effectful triple} \((\cat{V}, \cat{P}, \cat{C})\) automatically gives a copy-discard structure via the identity-on-objects functors.
  The construction of \kl{effectful streams} in \Cref{sec:streams} does not depend on the choice of values, \(\cat{V}\); \kl{effectful streams} can be defined for a monoidal Freyd category, \((\cat{P}, \cat{C})\), without the need of a copy-discard structure.
  However, the definitions of \kl{effectful machines} and their bisimilarity does depend on it, so all the constructions are presented as parametric on an \kl{effectful triple}, implicitly assuming a copy-discard structure.
\end{rem}

\begin{defi}[Strict effectful triple]%
  \label{def:strict-effectful-triple}
  \AP An \kl{effectful triple} is \intro[strict effectful triple]{strict}
  whenever its tensor is strictly associative and unital. In other words, an
  \kl{effectful triple} is strict whenever its three categories are.
\end{defi}

A strictification theorem~\cite{ml1972maclane} holds for \kl{effectful triples},
where each \kl{effectful triple} is equivalent to a strict one, $[•] ፡
(\cat{V},\cat{P},\cat{C}) → (\cat{V}_{\str},\cat{P}_{\str},\cat{C}_{\str})$, and
the equivalence preserves the cartesian, monoidal and premonoidal structures. 

\begin{thm}
  \label{thm:strictification}
  Any \kl{effectful triple} is equivalent to a strict one.
\end{thm}
\begin{proof}
  Let $(\cat{V},\cat{P},\cat{C})$ be an \kl{effectful triple} with identity-on-object functors given by $i \colon \cat{V} \to \cat{P}$ and $j \colon \cat{P} \to \cat{C}$. Let $\{\bullet\}_{\cat{V}} \colon \cat{V}_{\str} \to \cat{V}$ %
  be the monoidal equivalence witnessing the strictification of $\cat{V}$~(with the standard construction \cite{ml1972maclane}); and let $\{\bullet\}_{\cat{P}} \colon \cat{P}_{\str} \to \cat{P}$ %
  be the monoidal equivalence witnessing the strictification of $\cat{P}$.

  By construction, $\cat{V}_{\str}$ and $\cat{P}_{\str}$, have the same objects. Now, define $\cat{C}_{\str}$ to have the same objects as $\cat{V}_{\str}$ and $\cat{P}_{\str}$, and to have morphisms given by $\cat{C}_{str}(A;B) = \cat{C}(\{A\}; \{B\})$. There exists thus a premonoidal functor $\{\bullet\} \colon \cat{C}_{\str} \to \cat{C}$ that is the identity on morphisms—thus, full and faithful—and essentially surjective on objects, defining an equivalence.

  Finally, there exist unique identity-on-objects functors $i^{\ast} \colon \cat{V}_{\str} \to \cat{P}_{\str}$ and $j^{\ast} \colon \cat{P}_{\str} \to \cat{C}_{\str}$ commuting with the equivalences. 
\end{proof}

\begin{exa}[Kleisli categories form effectful triples]%
  \label{ex:kleisli-sets-effectful}%
  Any $\Set$-monad $T \colon \Set \to \Set$ induces a Kleisli \kl{effectful
  triple}, $(\Set, \kleisli{\mathcal{Z}(T)}, \kleisli{T})$, consisting of \emph{(i)} the
  category of sets (for values), \emph{(ii)} the Kleisli category of the centre of the
  monad, $\mathcal{Z}(T)$~\cite{lemonnier23:centralSubmonads} (for local effects), and \emph{(iii)} the Kleisli category of the whole monad (for global effects).
  
  More generally, any Freyd category, $(\cat{V}, \cat{C})$~\cite{levy2004},
  induces an \kl{effectful triple}, $(\cat{V}, \mathcal{Z}(\cat{C}), \cat{C})$,
  where $\mathcal{Z}(\cat{C})$ is the monoidal centre of the \kl{premonoidal
  category}.
\end{exa}

\begin{exa}[Beyond the Kleisli construction]
  \label{ex:beyond-kleisli}
  We may also consider \kl{effectful triples} where the monoidal category
  \(\cat{P}\) is smaller than the monoidal centre: e.g., partial functions and
  functions \((\Set, \Set, \Par)\); relations and total relations \((\Set,
  \Reltot, \Rel)\); subdistributions and distributions \((\Set, \Stoch,
  \subStoch)\); convex sets of distributions and functions \((\Set, \Set, \StochRel)\); or partial stochastic channels and stochastic channels \((\stdBorel,
  \BorelStoch, \BorelsubStoch)\).
  
  We may also consider \kl{effectful triples} not coming from a Kleisli
  construction: while \((\Set, \Stoch, \StateStoch{R})\) is the Kleisli category
  of the state monad \({\distr(- \times R)}^{R} \colon \Set \to \Set\), we could
  also consider \intro[BorelStateStoch]{\(\BorelStateStoch{R}\)}, the
  premonoidal category with objects standard Borel spaces and whose morphisms
  \(X \to Y\) are morphisms \(X \times R \to \Giry(Y \times R)\).
  Note, however, that since $\stdBorel$
  is not a closed category, $\BorelStateStoch{R}$ is not the Kleisli category of
  the global state monad but rather an
  \emph{arrow}~\cite{hughes00:arrows,hasuoheunenjacobs09:arrows}.
\end{exa}

\begin{exa}[Stateful stochastic maps]
  We later prove that the \emph{stream cipher protocol} is secure
  \Cref{sec:cipher-correctness}. For this proof, we will work with an
  \kl{effectful triple} whose morphisms are stateful stochastic maps,
  \intro[EffStoch]{}\(\EffStoch = (\Set, \Stoch, \SeedStoch)\). Explicitly,
  $\Seed$ is the set of seeds to a random number generator; and
  \intro[SeedStoch]{}$\SeedStoch = \StateStoch{\Seed \tensor \Seed}$ is the
  Kleisli category of a state monad, whose morphism sets are defined by
  \[\SeedStoch(X;Y) = \Stoch(\Seed ⊗ \Seed ⊗ X ; \Seed ⊗ \Seed ⊗ Y),\] 
  with composition and identity as in $\Stoch$. Intuitively, arrows in
  $\SeedStoch$ are arrows of $\Stoch$ equipped with a \emph{global state},
  $\Seed ⊗ \Seed$: the pair of seeds that Alice and Bob keep.

  It is worth emphasizing that computations with a global state can hardly be
  modelled in monoidal categories—where interchange necessarily holds—but can be
  modelled in \kl{effectful triples}.
\end{exa}
\section{Effectful Machines}%
\label{sec:monoidal-automata}%
\label{sec:machines}%
\label{sec:effMealyMachines}%
 
\kl{Effectful machines} are a unifying notion of Mealy machine that can be
recast in any theory of processes represented by an \kl{effectful triple}.
\kl{Effectful machines} generalize the \emph{monoidal Mealy machines}—or,
\emph{processes}—of Katis, Sabadini and Walters~\cite{katis1997bicategories} for
non-interchanging categorical semantics. Along with \kl{effectful machines}, we
introduce a notion of \kl{bisimilarity}; we later test this definition of
\kl{bisimilarity} against the coalgebraic literature
(\Cref{th:coalgebraic-bisimulation,th:bisimulation-cospans}) and we characterize
it syntactically (\Cref{th:mealyIsFree}).

For the rest of this section, let us fix an \kl{effectful triple} \((\cat{V},
\cat{P}, \cat{C})\). All our constructions will be parametric on this choice.

\subsection{Effectful Machines}

A Mealy machine in a monoidal category $\cat{C}$, from an input space $X \in
\cat{C}$ to an output space $Y \in \cat{C}$, can be formalized as a transition
morphism $U ⊗ X → U ⊗ Y$ together with an initial state $i ፡ I → U$
\cite{katis1997bicategories}. The state space of the Mealy machine, $U \in
\cat{C}$, is part of its data.

\kl{Effectful machines} repeat the same idea in the more general setting of
\kl{effectful triples}: here, transition morphisms are allowed to produce global
effects. Let us first state the definition in the strict case.

\begin{defi}[Effectful machine, in a strict effectful triple]%
  \label{def:effectful-machine-strict}%
  An \intro{effectful machine} $(U,i,f)$ in a \kl{strict effectful triple}
  $(\cat{V},\cat{P},\cat{C})$, taking inputs on $X$ and producing outputs in
  $Y$, consists of a \emph{state object} $U ∈ \cat{C}$, an \emph{initial state}
  $i ∈ \cat{P}(I; U)$, and a \emph{transition morphism}, $f ∈ \cat{C}(U \tensor
  X; U \tensor Y)$.
\end{defi}

Naïvely repeating this definition in the non-strict case presents a problem:
while \kl{effectful machines} can be composed by tensoring their state spaces,
this composition is not associative nor unital in the non-strict case; and one
is then forced to work with bicategories instead \cite{katis1997bicategories}. A
possible remedy would quotient the state space by
isomorphisms~\cite{katis02feedback,canonicalalgebra}, but we would be left to
deal with equivalence classes of states for the rest of the paper. 

Instead, we propose that the state object lives in the strictification: $U ∈
\cat{C}_{\str}$. This makes composition associative and unital \emph{a priori},
and drastically simplifies the rest theory of \kl{effectful machines}, avoding a
bicategorical structure that we will mention later (in
\Cref{def:morphism-effectful-machines}) but not pursue during most of the paper.

\begin{defi}[Effectful machine]%
  \label{def:effectful-machine}%
  An \intro{effectful machine}, taking inputs on $X$ and producing outputs in
  $Y$, is a tuple $(U,i,f)$ consisting of a \emph{state object} $U ∈
  \cat{C}_{\str}$, an \emph{initial state} $i ∈ \cat{P}_{\str}(I; U)$, and a
  \emph{transition morphism}, $f ∈ \cat{C}_{\str}(U \tensor [X]; U \tensor
  [Y])$, where \([X]\) denotes the strictification of the object \(X \in
  \cat{C}\). 
  
  From now on, $\Mealy[\cat{C}](X; Y)$ denotes the set of \kl{effectful
  machines} with inputs in $X$ and outputs in $Y$.
\end{defi}

\begin{rem}%
  \label{rem:relevance-of-values}
  Note that this definition of \kl{effectful machine} does not explicitly use
  the category $\cat{V}$ of values; indeed, it can be repeated for any pair of
  categories $(\cat{P},\cat{C})$. This may be important in cases where a
  copy-discard structure is not available, as in quantum computing semantics.
  However, the category of pure morphisms does become relevant when considering
  \kl{machine homomorphisms} (\Cref{def:morphism-effectful-machines}) and,
  later, \kl{bisimilarity}. From this point of view, the category of values
  only modifies our concept of \kl{bisimilarity}.

  We have chosen to restrict initial states to be local computations, \(i \in \cat{P}(I;U)\), to avoid arbitrary global effects being produced before the machine even starts running.
\end{rem}

Analogously, we may define \emph{monoidal} and \emph{cartesian} machines. A
\emph{monoidal machine}, $(U,i,f) ∈ \Mealy[\cat{P}](X; Y)$, has a monoidal
transition, $f ∈ \cat{P}_{\str}(U \tensor [X]; U \tensor [Y])$
(c.f.~\cite{katis1997bicategories}); a \emph{cartesian machine}, $(U,i,f) ∈
\Mealy[\cat{V}](X; Y)$, has a cartesian transition and initial state, $f ∈
\cat{V}_{\str}(U \tensor [X]; U \tensor [Y])$ and $i ∈ \cat{V}_{\str}(I; U)$.

\begin{prop}[Effectful triple of effectful machines]%
  \label{prop:effectful-category-machines}%
  \kl{Effectful machines}, monoidal machines, and cartesian machines form an
  \kl{effectful triple}, $\Mealy(\cat{V},\cat{P},\cat{C}) $, whose three component
  categories are $\Mealy[\cat{V}]$, $\Mealy[\cat{P}]$ and $\Mealy[\cat{C}]$.
\end{prop}
\begin{proof}
  Composition of two \kl{effectful machines}, $(U,i,f) ∈ \Mealy[\cat{C}](X; Y)$
  and $(V,j,g) ∈ \Mealy[\cat{C}](Y; Z)$, is the \kl{effectful machine} $(U,i,f)
  \dcomp (V,j,g) ∈ \Mealy[\cat{C}](X; Z)$, defined by
  \[(U,i,f) ⨾ (V,j,g) = (U ⊗ V,i ⊗ j, f \bowtie g),\]
  where the transition morphism, $f \bowtie g$, is given by the formula
  \[(σ_{U,V} ⊗ \id{X}) \dcomp (\id{V} \tensor f) \dcomp (σ_{U,V} ⊗ \id{Z}) \dcomp (\id{U} ⊗ g).\]
  \begin{align*}
    \includegraphics[scale=0.9]{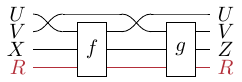}
  \end{align*}
  The identity \kl{effectful machine} is $(I,\id{I},\id{X}) ∈ \Mealy[ℂ](X; X)$.
  Associativity and unitality require strictness of the state spaces; otherwise, the two sides of the following equation would have different state spaces.
  \begin{align*}
    (U ⊗ (V ⊗ W),i ⊗ (j ⊗ k), f \bowtie (g \bowtie h)) =
    ((U ⊗ V) ⊗ W, (i ⊗ j) ⊗ k, (f \bowtie g) \bowtie h).
  \end{align*}
  For transition functions it is straightforward to check unitality, $f \bowtie \id{Y} = \id{X} \bowtie f$, and associativity, $f \bowtie (g \bowtie h) = (f \bowtie g) \bowtie h$ by string diagram manipulation.

  Whiskering of an \kl{effectful machine}, $(U,i,f) ∈ \Mealy[\cat{C}](X; Y)$, by an object $Z ∈ \cat{C}$ is defined by $(U,i,f ⊗ \id{Z}) ∈ \Mealy[\cat{C}](X ⊗ Z; Y ⊗ Z)$, using whiskering on the base premonoidal category $\cat{C}$.
  \begin{align*}
    \includegraphics[scale=0.9]{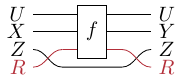}
  \end{align*}
  It is straightforward to check, via string diagrams, that whiskering is functorial and that the monoidal tensor it induces preserves the interchange equation whenever both machines are monoidal.
\end{proof}

\begin{defi}[Machine homomorphism]%
  \label{def:morphism-effectful-machines}%
  A \intro{machine homomorphism} \(\alpha \colon (U,i,f) \Rightarrow (V,j,g)\)
  between two \kl{effectful machines} $(U,i,f), (V,j,g) ∈ \Mealy(X; Y)$, is a
  cartesian morphism $α ∈ \cat{V}_{\str}(U; V)$ that satisfies the
  equations~\ref{eq:fig-mealy-homomorphism}.
\begin{align}
    \begin{tikzpicture}[baseline=-0.1cm]
    \pic(s)at(0,0){state=\(i\)};
    \pic(a)at(0.7,|-s-o){morphismSmall=\(\alpha\)};
    \node[output](o)at(a-o){\(V\)};
  \end{tikzpicture} 
  & =\quad 
  \begin{tikzpicture}[baseline=-0.1cm]
    \pic(t)at(0,0){state=\(j\)};
    \node[output](o)at(t-o){\(V\)};
  \end{tikzpicture}
  &
  \begin{tikzpicture}[baseline=-0.1cm]
    \pic(f)at(0,0){premorphismTwoTwo=\(f\)};
    \pic(a)at(0.8,|-f-o1){morphismSmall=\(\alpha\)};
    \node[output](o1)at(a-o){\(V\)};
    \node[output](o2)at(1.3,|-f-o2){\(Y\)};
    \node[input](i1)at(f-i1){\(U\)};
    \node[input](i2)at(f-i2){\(X\)};
    \node[input,run](i3)at(f-i3){\(R\)};
    \node[output,run](o3)at(1.3,|-f-o3){\(R\)};
    \draw[-](f-o2)to(o2);
    \draw[run](f-o3)to(o3);
  \end{tikzpicture}
  & = 
  \begin{tikzpicture}[baseline=-0.1cm]
    \pic(g)at(0,0){premorphismTwoTwo=\(g\)};
    \pic(a)at(-0.8,|-g-i1){morphismSmall=\(\alpha\)};
    \node[output](o1)at(g-o1){\(V\)};
    \node[output](o2)at(g-o2){\(Y\)};
    \node[input](i1)at(a-i){\(U\)};
    \node[input](i2)at(-1.3,|-g-i2){\(X\)};
    \node[input,run](i3)at(-1.3,|-g-i3){\(R\)};
    \node[output,run](o3)at(g-o3){\(R\)};
    \draw[-](i2)to(g-i2);
    \draw[run](i3)to(g-i3);
  \end{tikzpicture}
  \label{eq:fig-mealy-homomorphism}
  \end{align} \end{defi}

\subsection{Bisimilarity of Effectful Machines}

\kl{Machine homomorphisms} may be seen as witnesses for the simulation of the
first machine into the second: the initial state is translated to the initial
state of the second machine, and transitions are shown to coincide. As such,
\kl{machine homomorphisms} are instrumental in defining \emph{bisimilarity of
effectful machines}: bisimilarity will be the minimal equivalence relation $(\equiv)$
that equates two \kl{effectful machines} if there exists a path of homomorphisms
translating between them. 
In the next section, we prove that this \kl{bisimilarity} of \kl{effectful
machines} $(\equiv)$ coincides with \emph{coalgebraic
bisimilarity}~\cite{rutten2000universal} in well-behaved categories
(\Cref{th:coalgebraic-bisimulation}).

\begin{defi}[Bisimulation]%
  \label{def:effectful-bisimulation}%
  A \intro{bisimulation} between two \kl{effectful machines}, $M$ and $N$, with
  the same input and output types is a zig-zag of homomorphisms, $(α_i ፡ M_i
  \Rightarrow N_i)_{i=0}^{k}$ and $(β_i ፡ M_{i+1} \Rightarrow N_i)_{i=0}^{k-1}$,
  \[ M_0   \overset{α_0}{\Rightarrow} N_0 \overset{β_0}{\Leftarrow} M_1
  \overset{α_1}{\Rightarrow} \dots \overset{β_{k-1}}{\Leftarrow} M_k
  \overset{α_k}{\Rightarrow} N_k, \] where $M_0 = M$ and $N_k = N$. Two
  \kl{effectful machines}, $M$ and $N$, are \intro{bisimilar}, written $M \equiv
  N$, if there is a \kl{bisimulation} between them.
\end{defi}

\begin{rem}[Zig-zag bisimulation]%
  \label{rem:zig-zag-bisimulation}
  Zig-zag of homomorphisms, rather than spans ($M_0 \Leftarrow R \Rightarrow N_n$) or cospans ($M_0 \Rightarrow R \Leftarrow N_n$), are necessary for bisimilarity to be transitive: at this level of generality, we cannot rely on pullbacks or pushouts (and not even on weak pullbacks nor weak pushouts) for composing spans or cospans of homomorphisms.
  Zig-zags can be replaced by sequences of spans or sequences of cospans.
\end{rem}

\begin{prop}[Bisimilarity]
  \kl{Bisimilarity} of \kl{effectful machines} from $X$ to $Y$ is the smallest
  equivalence relation $(\equiv)$ on \(\Mealy(X;Y)\) such that the existence of
  a homomorphism $α ፡ M ⇒ N$ implies $M \equiv N$.
\end{prop}
\begin{proof}
  We check that \kl{bisimilarity} is an equivalence relation. 
  \emph{Reflexivity}. For a machine $M$ with state space $S$, take the zig-zag
  of length $0$ composed of the identity on $S$, $\alpha_{0} = \id{S}$.
  \emph{Symmetry}. For a zig-zag $(\alpha_{i} \colon M_{i} \Rightarrow
  N_{i})_{i=0}^{k}$ and $(\beta_{i} \colon M_{i+1} \Rightarrow
  N_{i})_{i=0}^{k-1}$,
  \[ M_0   \overset{α_0}{\Rightarrow} N_0 \overset{β_0}{\Leftarrow} M_1
  \overset{α_1}{\Rightarrow} \dots \overset{β_{k-1}}{\Leftarrow} M_k
  \overset{α_k}{\Rightarrow} N_k, \] define the zig-zag of length $k+1$ by
  inverting the order of morphisms and adding an identity at the beginning and
  one at the end, $(\id{T_{k}}, \beta_{k-1}, \dots, \beta_{0}, \id{S_{0}})$ and
  $(\alpha_{i})_{i=k}^{0}$,
  \[ 
  N_{k} \overset{\id{T_{k}}}{\Rightarrow} 
  N_{k} \overset{\alpha_{k}}{\Leftarrow} 
  M_{k} \overset{\beta_{k-1}}{\Rightarrow} \dots \overset{\alpha_{0}}{\Leftarrow} 
  M_{0} \overset{\id{S_{0}}}{\Rightarrow} 
  M_{0}. 
  \]
  \emph{Transitivity}. Given two zig-zags, $(\alpha_{i} \colon M_{i} \Rightarrow
  N_{i})_{i=0}^{k}$ and $(\beta_{i} \colon M_{i+1} \Rightarrow
  N_{i})_{i=0}^{k-1}$, and $(\gamma_{i} \colon M'_{i} \Rightarrow
  N'_{i})_{i=0}^{l}$ and $(\delta_{i} \colon M'_{i+1} \Rightarrow
  N'_{i})_{i=0}^{l-1}$ with $N_{k} = M'_{0}$, let us define a zig-zag of length $k+l$
  by composing the last leg of the first with the first leg of the second:
  $(\alpha_{0}, \dots, \alpha_{k-1}, \alpha_{k} \dcomp \gamma_{0}, \gamma_{1},
  \dots, \gamma_{l})$ and $(\beta_{0}, \dots, \beta_{k-1}, \delta_{0}, \dots,
  \delta_{l-1})$.

  Suppose there is an equivalence relation that relates all \kl{effectful
  machines} with a homomorphism between them. By symmetry and transitivity, it
  needs to relate all the \kl{effectful machines} with a zig-zag of
  homomorphisms between them, so it needs to contain \kl{bisimilarity} of
  \kl{effectful machines}.
\end{proof}

\begin{rem}[Fully effectful bisimilarity]%
  \label{rem:effectfulbisimilarity}%
  We choose \kl{machine homomorphisms} to be cartesian: this choice is justified
  because it allows us to recover \kl{bisimilarity}. However, we could also allow
  monoidal and premonoidal homomorphisms, whose zig-zags induce notions of
  \emph{monoidal bisimilarity} and \emph{effectful bisimilarity}; we mention
  them briefly in \Cref{sec:streams}.
\end{rem}

\begin{prop}[Effectful triple of bisimilarity-quotiented machines]%
  \label{prop:mealybisCategory}
  \kl{Effectful machines} quotiented by \kl{bisimilarity} form an \kl{effectful
  triple}, \intro[MealyBis]{}$\MealyBis(\cat{V},\cat{P},\cat{C})$, whose
  component categories are the categories of \kl[effectful machines]{cartesian,
  monoidal and effectful machines} quotiented by \kl{bisimilarity},
  \[\MealyBis[\cat{V}] = \Mealy[\cat{V}]/(\equiv); \quad 
    \MealyBis[\cat{P}] = \Mealy[\cat{P}]/(\equiv); \quad
    \MealyBis[\cat{C}] = \Mealy[\cat{C}]/(\equiv).\]
\end{prop}
\begin{proof}
  We must check that the \kl{bisimilarity} equivalence relation $(\equiv)$ is a
  congruence for composition, tensoring, and whiskering. \kl{Bisimilarity} is
  the smallest equivalence relation generated by homomorphisms; thus, we only
  need to prove that these operations induce homomorphisms.

  Let us prove that composition induces homomorphisms. Consider two pairs of
  \kl{effectful machines} with two \kl{machine homomorphisms} between them,
  \[
    α ፡ (U,i,f) \Rightarrow (U',i',f'), \mbox{ and }
    β ፡ (V,j,g) \Rightarrow (V',j',g').\]
  We claim that
  \[(α ⊗ β) ፡ (U, i, f) ⨾ (V,j,g) \Rightarrow (U', i', f') ⨾ (V', j', g')\]
  is a \kl{machine homomorphism}. Indeed, it suffices to check (\emph{i}) that
  the tensor of two cartesian morphisms is again cartesian; (\emph{ii}) that \(
  (f \bowtie g) ⨾ (α ⊗ β ⊗ \id{Z}) = (α ⊗ β ⊗ \id{Z}) ⨾ (f' \bowtie g')\), which
  is straightforward with string diagrams; and (\emph{iii}) that $(i ⊗ j) ⨾ (α ⊗
  β) = i' ⊗ j'$, which is immediate.

  Finally, let us prove that whiskering induces homomorphisms. Consider an
  object $Z$, and an \kl{effectful machine} with a \kl{machine homomorphism},
  \(α ፡ (U,i,f) \Rightarrow (U',i',f')\). We claim that
  \[ α ፡ (U,i,f) ⊗ Z \Rightarrow (U',i',f') ⊗ Z \]
  is a \kl{machine homomorphism}. Indeed, it suffices to check that (\emph{i})
  the whiskering of a cartesian morphism is again cartesian; (\emph{ii}) that
  $(f ⊗ \id{Z}) ⨾ (α ⊗ \id{Z}) = (α ⊗ \id{Z}) ⨾ (f' ⊗ \id{Z})$, which is
  immediate; (\emph{iii}) and that $α$ still transports the initial state, which
  has not changed.
\end{proof}

\subsection{Case study: \(T\)-machines}%
\label{sec:set-machines}

During this section, we consider $\Set$-monads $T$ and their \kl{effectful
machines} on $(\Set, \mathcal{Z}(\kleisli{T}), \kleisli{T})$; we will prove that
these can be characterized as coalgebras together with an initial state.
While \Cref{th:coalgebraic-bisimulation} and the results leading to it hold for a generic cartesian category \(\cat{V}\), the instances that we intend to generalise are all defined in \(\Set\).

Indeed, \kl{effectful machines} on $(\Set, \mathcal{Z}(\kleisli{T}),
\kleisli{T})$ with inputs in $X$ and outputs in $Y$ are coalgebras for the
endofunctor $\fun{M}_{T} ፡ \cat{V} → \cat{V}$ defined by $\fun{M}_{T} = {(T(- ×
Y))}^{X}$, together with an initial state: indeed, a transition morphism $f ∈
\kleisli{T}(U ⊗ X; U ⊗ Y)$ is, equivalently, a coalgebra \(\hat{f} ፡ U → {(T(U
\times Y))}^{X}\)~\cite{rutten2000universal}. Similarly, \kl{machine
homomorphisms} are \(\fun{M}_{T}\)-coalgebra homomorphisms preserving initial
states.

\begin{prop}
  \label{prop:mealycoalgebra}
  Machine homomorphisms are \(\fun{M}_{T}\)-coalgebra homomorphisms preserving
  the initial states: in other words, in the following equivalence, the leftmost
  diagram commutes in \(\kleisli{T}\) if and only if the rightmost diagram
  commutes in \(\cat{V}\).
  \[
  \begin{mytikzcd}
    {U \tensor X} \arrow{r}{u \tensor \id{X}} \arrow{d}[swap]{f} \& {V \tensor X} \arrow{d}{g}\\
    {U \tensor Y} \arrow{r}[swap]{u \tensor \id{Y}} \& {V \tensor Y}
  \end{mytikzcd}\ \Leftrightarrow\ \begin{mytikzcd}
    {U} \arrow{r}{u} \arrow{d}[swap]{\hat{f}} \& {V} \arrow{d}{\hat{g}}\\
    {T(U \times Y)^X} \arrow{r}[swap]{T(u \times \id{Y})^X} \& {T(V \times Y)^X}
  \end{mytikzcd}
  \]
\end{prop}

In particular, the transition morphisms of deterministic, non-deterministic and
probabilistic machines can be expressed as coalgebras for $\Set$-endofunctors.

\begin{exa}[\(\Set\)-monad machines]%
  \label{ex:set-coalgebras-machines}%
  Deterministic machines use the \emph{identity monad}; partial machines,
  deterministic machines whose transition function may diverge, use the
  \emph{Maybe monad}; non-deterministic machines, as in Baier and
  Katoen~\cite{baier2008principles}, use the \emph{powerset monad} \(\parti\);
  probabilistic machines, or labelled Markov processes, use the \emph{finitary
    distribution monad} \(\distr\); machines with combined probabilistic and nondeterministic choices (also known as Segala systems~\cite{segala1995modeling,segala1995probabilistic}), use the \emph{convex sets of distributions} monad \(\convdistr\); quantum machines use a \emph{quantum branching monad} \(\fun{Q}\)~\cite{hasuo2011quantum}; finally, machines with both local state $U$ and
  fixed shared state $S$ use the \emph{state monad} on $S$. See
  \Cref{fig:monadMealy}.
  \begin{figure}[!h]
  \centering{}
  \begin{tabular}{|l|l|}
    \hline
    Deterministic & $U × X → U × Y$ \\
    Partial & $U × X → U × Y + 1$ \\
    Nondeterministic & $U × X → \parti(U × Y)$ \\
    Probabilistic~\cite{baier2008principles} & $U × X → \distr(U × Y)$ \\
    Nondeterministic and probabilistic & $ U \times X \to \convdistr(U \times Y)$\\
    Quantum~\cite{hasuo2011quantum} & $U × X → \fun{Q}(U × Y)$ \\
    Shared state & $U × X → (U × Y × S)^{S}$ \\
    \hline
  \end{tabular}
  \caption{Machines for various $\Set$-monads.}\label{fig:monadMealy}
  \end{figure}
\end{exa}

For the rest of this section, we characterize our general definition of
bisimilarity in the case of $\Set$-based monads that moreover preserve pullbacks
(\Cref{th:coalgebraic-bisimulation}). Pullback preservation allows us to
substitute zig-zags of homomorphisms by bisimilarity in the coalgebraic sense.

\begin{lem}%
  \label{lemma:times-preserves-pullbacks}%
  The functors \((- \times X) \colon \cat{V} \to \cat{V}\) on a \kl{cartesian
  category} \(\cat{V}\) preserve weak pullbacks.
\end{lem}
\begin{proof}
  Let (1) be a weak pullback square.
  We show that (2) is also a weak pullback square.
  \[
  \begin{mytikzcd}
    \& {U} \arrow[bend right]{ddl}[swap]{s} \arrow[bend left]{ddr}{t} \arrow[dashed]{d}{r}\&\\
    \& {R} \arrow{dl}[swap]{p} \arrow{dr}{q} \&\\
    {S} \arrow{dr}[swap]{u} \& {(1)} \& {T} \arrow{dl}{v} \\
    \& {Q} \&
  \end{mytikzcd}
  \qquad\quad 
  \begin{mytikzcd}
    \& {U} \arrow[bend right]{ddl}[swap]{f} \arrow[bend left]{ddr}{g} \arrow[dashed]{d}{\productmap{r}{x}}\&\\
    \& {R \times X} \arrow{dl}[swap]{p \times \id{}} \arrow{dr}{q \times \id{}} \&\\
    {S \times X} \arrow{dr}[swap]{u \times \id{}} \& {(2)} \& {T \times X} \arrow{dl}{v \times \id{}} \\
    \& {Q \times X} \&
  \end{mytikzcd}
  \] 
  Given \(f ፡ U \to S \times X\) and \(g \colon U \to T \times X\) such that \(f
  \dcomp (u \times \id{X}) = g \dcomp (v \times \id{X})\), we can express them
  as product maps, \(f = \productmap{s}{x}\) and \(g = \productmap{t}{y}\), and
  obtain that \(x = y\) and \(s \dcomp u = t \dcomp v\), by cartesianity of
  \(𝕍\). Then, \(s \colon U \to S\) and \(t \colon U \to T\) form a cone on the
  diagram given by \(u\) and \(v\). This gives a morphism \(r \colon U \to R\)
  to the weak pullback such that \(r \dcomp p = s\) and \(r \dcomp q = t\), and
  a morphism \(h = \productmap{r}{x} \colon U \to R \times X\) such that \(h
  \dcomp (p \times \id{X}) = f\) and \(h \dcomp (q \times \id{X}) = g\), which
  shows that (2) is also a weak pullback.
\end{proof}

\begin{lem}%
  \label{lemma:pullback-preservation}
  Let \(T \colon \cat{V} \to \cat{V}\) be a monad on a cartesian category
  \(\cat{V}\). Then, \(T\) preserves weak pullbacks if and only if the inclusion
  \(\fun{i} \colon \cat{V} \to \kleisli{T}\) preserves them.
\end{lem}
\begin{proof}
  \((\Leftarrow)\) Let \(\fun{U} \colon \kleisli{T} \to \cat{V}\) be the right
  adjoint of \(\fun{i}\) that gives the monad \(T\). The functor \(\fun{U}\) is
  a right adjoint, then it preserves weak limits. The composition of two
  functors that preserve weak pullbacks also preserves weak pullbacks, then \(T
  = \fun{i} \dcomp \fun{U}\) also preserves weak pullbacks.

  \((\Rightarrow)\) Suppose we have a weak pullback square in \(\cat{V}\) (1).
  We need to show that (3) is also a weak pullback square in \(\kleisli{T}\) (we indicate with (\(\rightharpoonup\)) the arrows in the Kleisli category to distinguish them from those in the base category \(\cat{V}\)).
  \[
  \begin{mytikzcd}
    \& {R} \arrow{dl}[swap]{p} \arrow{dr}{q} \&\\
    {X} \arrow{dr}[swap]{u} \& {(1)} \& {Y} \arrow{dl}{v} \\
    \& {Q} \&
  \end{mytikzcd} 
  \qquad
  \begin{mytikzcd}
    \& {S} \arrow[bend right]{ddl}[swap]{x} \arrow[bend left]{ddr}{y} \arrow[dashed]{d}{h}\&\\
    \& {\fun{T}(R)} \arrow{dl}[swap]{\fun{T}(p)} \arrow{dr}{\fun{T}(q)} \&\\
    {\fun{T}(X)} \arrow{dr}[swap]{\fun{T}(u)} \& {(2)} \& {\fun{T}(Y)} \arrow{dl}{\fun{T}(v)} \\
    \& {\fun{T}(Q)} \&
  \end{mytikzcd}
  \qquad
  \begin{mytikzcd}
    \& {S} \arrow[bend right, harpoon]{ddl}[swap]{x} \arrow[bend left, harpoon]{ddr}{y} \arrow[dashed, harpoon]{d}{h}\&\\
    \& {R} \arrow[harpoon]{dl}[swap]{\fun{i}(p)} \arrow[harpoon]{dr}{\fun{i}(q)} \&\\
    {X} \arrow[harpoon]{dr}[swap]{\fun{i}(u)} \& {(3)} \& {Y} \arrow[harpoon]{dl}{\fun{i}(v)} \\
    \& {Q} \&
  \end{mytikzcd}
  \]
  Let \(x \colon S \rightharpoonup X\) and \(y
  \colon S \rightharpoonup Y\) be a cone on the cospan \(\fun{i}(u) \colon X
  \rightharpoonup Q \leftharpoonup Y \co{\colon} \fun{i}(v)\) in
  \(\kleisli{T}\). Then, \(x \colon S \to T(X)\) and \(Y \colon S \to T(Y)\) is
  also a cone on the cospan \(T(u) \colon T(X) \to T(Q) \gets T(Y) \co{\colon}
  T(v)\) in \(\cat{V}\). Since \(T\) preserves weak pullbacks, (2) is also a
  weak pullback square and there is a morphism \(h \colon S \to T(R)\) such that
  \(h \dcomp T(p) = x\) and \(h \dcomp T(q) = y\) in \(\cat{V}\). Then, \(h
  \dcomp \fun{i}(p) = x\) and \(h \dcomp \fun{i}(q) = y\) in \(\kleisli{T}\),
  which shows that (3) is a weak pullback.
\end{proof}

\begin{rem}%
  \label{rem:exp-preserves-pullbacks}
  For a cartesian closed category \(\cat{V}\), the functors \((-)^{X} \colon
  \cat{V} \to \cat{V}\) are right adjoints and preserve weak pullbacks.
\end{rem}

\begin{lem}%
  \label{prop:bisimulation-span}%
  Let $(\cat{V},\cat{P},\cat{C})$ be an \kl{effectful triple} where $\cat{V}$
  has weak pullbacks preserved by the identity-on-objects functor $\cat{V} →
  \cat{C}$. Then, two \kl{effectful machines} are \kl{bisimilar} if and only if
  there is a span of morphisms between them.
\end{lem}
\begin{proof}
  A span determines a zig-zag by attaching an identity at the end, so the `if' direction is trivial.
  For the `only if' direction, we proceed by induction on the length of the zig-zag of morphisms. For the
  base case, \(n = 0\), we have a single morphism, which we can turn into a span
  by adding an identity leg. For the inductive step, we need to show that, given
  two consecutive spans, they can be (weakly) composed into one. Thus, it
  suffices to show that, for a cospan \(u \colon (U,i,f) \to (Q,q,h) \gets
  (V,j,g) \co{\colon} v\), there is a span \(p \colon (R,r,b) \to (U,i,f)\) and
  \(q \colon (R,r,b) \to (V,j,g)\). Suppose we are given such a cospan. Since
  \(\cat{V}\) has weak pullbacks, we can construct the weak pullback (1) of \(u
  \colon U \to Q\) and \(v \colon V \to Q\) and obtain two morphisms \(p \colon
  R \to U\) and \(q \colon R \to V\). The initial states \(i \colon I \to U\)
  and \(j \colon I \to V\) are a cone over the diagram \(u \colon U \to Q \gets
  V \co{\colon} v\). By the property of weak pullbacks, we obtain an initial
  state \(r \colon I \to R\) such that \(r \dcomp p = i\) and \(r \dcomp q =
  j\).
  \[
  \scalebox{0.9}{
    \begin{mytikzcd}
    \& {I} \arrow[bend right]{ddl}[swap]{i} \arrow[bend left]{ddr}{j} \arrow[dashed]{d}{r}\&\\
    \& {R} \arrow{dl}[swap]{p} \arrow{dr}{q} \&\\
    {U} \arrow{dr}[swap]{u} \& {(1)} \& {V} \arrow{dl}{v} \\
    \& {Q} \&
    \end{mytikzcd}
  } \quad\qquad 
  \scalebox{0.9}{
    \begin{mytikzcd}
    \& {R \tensor X} \arrow{dl}[swap]{p \tensor \id{}} \arrow{dr}{q \tensor \id{}} \arrow[bend left, dashed]{ddd}[swap]{b}\&\\
    {U \tensor X} \arrow{dr}[swap]{u \tensor \id{}} \arrow[bend right]{ddd}[swap]{f} \& {(2)} \& {V \tensor X} \arrow{dl}{v \tensor \id{}} \arrow[bend left]{ddd}[swap]{g}\\
    \& {Q \tensor X} \arrow[bend right]{ddd}[swap]{h} \&\\
    \& {R \tensor Y} \arrow{dl}[swap]{p \tensor \id{}} \arrow{dr}{q \tensor \id{}} \&\\
    {U \tensor Y} \arrow{dr}[swap]{u \tensor \id{}} \& {(3)} \& {V \tensor Y} \arrow{dl}{v \tensor \id{}} \\
    \& {Q \tensor Y} \&
  \end{mytikzcd}
  }
  \]
  By \Cref{lemma:times-preserves-pullbacks}, the diagrams (2) and (3) are also
  weak pullbacks. By the definition of morphisms of Mealy machines, the
  morphisms \((p \tensor \id{}) \dcomp f \colon R \tensor X \to S \tensor Y\)
  and \((q \tensor \id{}) \dcomp g \colon R \tensor X \to T \tensor Y\) form a
  cone over \((u \tensor \id{}) \colon S \tensor Y \to Q \tensor Y \gets T
  \tensor Y \co{\colon} (v \tensor \id{})\). By hypothesis, the functor
  \(\cat{V} \to \cat{C}\) also preserves weak pullbacks, which gives a morphism
  \(b \colon R \tensor X \to R \tensor Y\) such that \(b \dcomp (p \tensor
  \id{}) = (p \tensor \id{}) \dcomp f\) and \(b \dcomp (q \tensor \id{}) = (q
  \tensor \id{}) \dcomp g\). Thus, we have constructed a span \(p \colon (R,r,b)
  \to (U,i,f)\) and \(q \colon (R,r,b) \to (V,j,g)\).
\end{proof}

\begin{rem}%
  \label{rem:bisimulation-cospan}
  In the setup of \Cref{prop:bisimulation-span}, with an \kl{effectful triple}
  \((\cat{V},\cat{P},\cat{C})\), a dual statement also holds. If the functor
  \(\cat{V} \to \cat{C}\) preserves weak pushouts and the functors \((- \tensor
  \id{A}) \colon \cat{V} \to \cat{V}\) also preserve them, two \kl{effectful
  machines} are bisimilar if and only if there is a cospan of morphisms between
  them. In this case, we would need the extra condition that \((- \tensor \id{A}) \colon \cat{V} \to \cat{V}\) preserves weak pushouts, which came for free for the case of pullbacks.
\end{rem}

\begin{thm}%
  \label{th:coalgebraic-bisimulation}%
  Let $T$ be a weak-pullback preserving monad on $\Set$. Two \kl{effectful
  machines} on $(\Set, \cat{P}, \kleisli{T})$ are \kl{bisimilar} if and only if
  their associated coalgebras for the endofunctor \(\fun{M}_{T} = {(T(- \times
  Y))}^{X}\) and their initial states are bisimilar in the sense of
  Rutten~\cite{rutten2000universal}.
\end{thm}
\begin{proof}
  Consider two \kl{effectful machines} \((U,i,f), (V,j,g) \colon X \to Y\) in
  the \kl{effectful triple} \((\cat{V}, \cat{P}, \kleisli{T})\). By
  \Cref{lemma:pullback-preservation}, \(T\) preserves weak pullbacks if and only
  if \(\fun{i} \colon \cat{V} \to \kleisli{T}\) does so as well. Then, we can
  apply \Cref{prop:bisimulation-span} and obtain that \((U,i,f)\) and
  \((V,j,g)\) are bisimilar if and only if there is a span of morphisms \(p
  \colon (R,k,b) \to (U,i,f)\) and \(q \colon (R,k,b) \to (V,j,g)\). By
  \Cref{prop:mealycoalgebra}, morphisms of \kl{effectful machines} in
  \((\cat{V}, \cat{P}, \kleisli{T})\) are the same thing as morphisms of
  \(\fun{M}_{T}\)-coalgebras that preserve the initial state. By
  \Cref{lemma:times-preserves-pullbacks} and \Cref{rem:exp-preserves-pullbacks},
  the functor \(\fun{M}_{T}\) is a composition of weak-pullback-preserving
  functors, thus it preserves weak pullbacks. Then, coalgebraic bisimulation
  coincides with a span of \(\fun{M}_{T}\)-coalgebra homomorphisms.
\end{proof}

While it is natural to set the pure computations to be the whole monoidal centre, \(\cat{P} = \mathcal{Z}(\kleisli{T})\), the result above holds for any other choice because the definition of bisimulation only relies on the choice of values.

\begin{cor}%
  \label{cor:set-coalgebras-bisimulation}%
  For deterministic, partial and non-deterministic \kl[effectful
  machines]{machines}, \kl{bisimilarity} coincides with the usual notions of
  bisimilarity as given in, e.g., the monograph by Baier and
  Katoen~\cite{baier2008principles}. For probabilistic machines,
  \kl{bisimilarity} coincides with Larsen and
  Skou's~\cite{LarsenSkou89,prakash1997bisimulation}. For machines with combined
  probabilistic and nondeterministic effects, \kl{bisimilarity} coincides with
  convex bisimilarity~\cite{sokolova2005coalgebraic,sokolova2011probabilistic}.
\end{cor}

We have characterized \kl{bisimilarity} for \kl{effectful machines} whenever the
monad preserves weak pullbacks.  Note, however, that even when the monad does
not preserve weak pullbacks, we still obtain a characterization of
\kl{bisimilarity} in terms of a known definition of coalgebraic bisimilarity:
that of cospans of coalgebra homomorphisms
(e.g~\cite{DBLP:conf/calco/Staton09}). For instance, this result captures
behavioural equivalence for quantum machines; this is analogous to \emph{kernel
bisimilarity} for \emph{quantum labelled transition systems}, as recently
introduced~\cite{ceragioli2024quantum,ceragioli2024coalgebraic}.

\begin{thm}%
  \label{th:bisimulation-cospans}%
  Let \(T\) be a monad on \(\Set\). Two \kl{effectful machines} on \((\Set,
  \cat{P}, \kleisli{T})\) are \kl{bisimilar} if and only if there is a cospan of
  \(\fun{M}_{T}\)-coalgebra homomorphisms between them that preserves the
  initial states.
\end{thm}
\begin{proof}
  The category of coalgebras for a \(\Set\)-endofunctor has all
  colimits~\cite[Theorem~4.2]{rutten2000universal}; as a consequence, zig-zags of coalgebra
  homomorphisms can be composed into a single cospan.
\end{proof}

\begin{cor}
  \kl{Bisimilarity} in the \kl{effectful triple} \((\Set, \cat{P},
  \kleisli{\fun{Q}})\), for the quantum branching monad \(\fun{Q}\)
  (\Cref{ex:set-coalgebras-machines}), gives the analogue of kernel
  bisimilarity~\cite{ceragioli2024quantum} for the quantum branching monad
  \(\fun{Q}\).
\end{cor}

Until here, we have seen how \kl{bisimilarity} in
\Cref{def:effectful-bisimulation} generalizes various existing notions of
bisimilarity. The next section goes one step further: we will show how
\kl{bisimilarity} enjoys a sound and complete string diagrammatic
characterisation by means of \kl{uniform feedback}.
\section{Uniform Feedback and Bisimilarity}%
\label{sec:feedback}

This section gives a universal property to the bisimilarity quotient:
\kl{effectful machines} quotiented by \kl{bisimilarity} are the free uniform feedback structure on an \kl{effectful triple}.

Feedback monoidal categories appear multiple times in the literature as algebras
of transition  
systems~\cite{katis99algebra,katis1997bicategories,katisSW97:spangraph},
where the feedback structure adds a state space to morphisms in the base
category. A feedback structure is an operation that takes a morphism \(f \colon
S \tensor X \to S \tensor Y\) and an initial state \(s \colon I \to S\) to
return a morphism \(\fbk_{s}(f) \colon X \to Y\).
\begin{equation}\label{eq:unif-feedback-diagram}
  \begin{split}
    \fbk_{s}(f) =
    \raisebox{-0.3\height}{\includegraphics[scale =0.9]{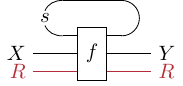}}
  \end{split}
\end{equation}
The morphism \(\fbk_{s}(f)\) represents the process that starts in \(s\), runs \(f\) to transition to the next state, and feeds it back to \(f\) to keep running.
We extend this construction to \kl{effectful triples} and add the uniformity axiom to capture \kl{bisimilarity}.

\begin{defi}%
  \label{def:uniform-feedback}
  A \intro{uniform feedback structure} on $(\cat{V},\cat{P},\cat{C})$ consists of a \kl{premonoidal category} $\cat{F}$ with an identity-on-objects premonoidal functor $\nf{(•)} ፡ \cat{C} → \cat{F}$ and a feedback operator
  \[\fbk ፡ \cat{P}(I;S) × \cat{F}(S ⊗ X; S ⊗ Y) → \cat{F}(X;Y),\]
  denoted by $\fbk_{s}(f)$ for $s ∈ \cat{P}(I;S)$ and $f ∈ \cat{F}(S ⊗ X; S ⊗ Y)$,
  which must satisfy the following axioms (see also
  \Cref{fig:uniform-feedback-string-diagrams}):
\begin{figure}[h!]
  \begin{gather*}
    \begin{aligned}
    \tighteningeffectFigLeft\ &\overset{(1)}{=}\ \tighteningeffectFigRight & \vanishingeffectFig{} &\overset{(3)}{=} \generatorFig{premorphism=\(f\)} \\[6pt]
    \strengtheffectFigLeft &\overset{(2)}{=}  \strengtheffectFigRight & \joiningeffectFigLeft{} &\overset{(3)}{=} \feedbackeffectFig{f}{s \tensor t}
    \end{aligned} \\[7pt]
    \uniformityeffectFigLeftIf = \uniformityeffectFigRightIf  \overset{(4)}{\Rightarrow} \feedbackeffectFig{\nf{c}}{s} = \feedbackeffectFig{\nf{d}}{s \dcomp p}
  \end{gather*}
  \caption{Uniform feedback axioms.}%
  \label{fig:uniform-feedback-string-diagrams}
\end{figure}
  \begin{enumerate}
    \item (\emph{tightening}) for \(u ∈ \cat{F}(X';X)\) and \(v ∈ \cat{F}(Y;Y')\), $\fbk_{s}(((\id{} ⊗ u) \dcomp f \dcomp (\id{} ⊗ v)) ⊗ \id{}) = ((\id{} ⊗ u) \dcomp \fbk_{s}(f) \dcomp (\id{} ⊗ v)) ⊗ \id{}$;
    \item (\emph{strength}) whiskering commutes with feedback, $\whis{\fbk_{s}(f)}{Z} = \fbk_{s}(\whis{f}{Z})$;
    \item (\emph{joining}) multiple applications of feedback can be reduced to a single one, $\fbk_{s}(\fbk_{t}(f)) = \fbk_{s ⊗ t}(f)$ and \(\fbk_{\id{I}}(f) = f\);
    \item (\emph{uniformity}) the existence of a value $p ∈ \cat{V}(S;T)$ and computations, $c ∈ \cat{C}(S ⊗ X; S ⊗ Y)$ and $d ∈ \cat{C}(T ⊗ X; T ⊗ Y)$, such that $c \dcomp (p ⊗ \id{}) =  (p ⊗ \id{}) \dcomp d$ and $\cat{P}(I;p)(s) = t$ implies that $\fbk_{s}(\nf{c}) = \fbk_{t}(\nf{d})$.
  \end{enumerate}
\end{defi}

\begin{rem}[Uniformity, sliding, and traces]%
  \label{rmk:unif}%
  \kl{Uniformity} implies the better known \emph{sliding axiom} of feedback monoidal categories:
  \[\fbk_{s}((p ⊗ \id{X}) \dcomp f)\ =\ \fbk_{(s \dcomp p)}(f \dcomp (p ⊗ \id{Y})).\]
  \begin{equation}
    \begin{split}
      \includegraphics[scale=0.9]{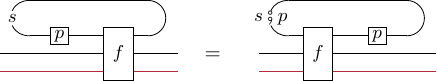}
      \label{eq:fig-sliding}
    \end{split}
  \end{equation}
  This implication is also true in \emph{traced monoidal categories}~\cite{hasegawa02}; in fact, the only difference between the axioms of \kl{uniform feedback} and those of \emph{uniform trace} is the yanking axiom, \(\mathsf{tr}_{X}(\swap_{X,X}) = \id{X}\).
  Yanking distinguishes feedback from trace by imposing that the feedback loop is ``instantaneous''.
  In particular, it distinguishes premonoidal feedback from \emph{premonoidal traces}~\cite{bentonH03}.
\end{rem}

\subsection{Effectful machines: free uniform feedback}%
\label{sec:machines-free-feedback}

The construction of free uniform feedback follows the same idea of the \emph{state construction}~\cite{katis02feedback}, to which we add uniformity.

\begin{lem}%
  \label{lemma:MealyFreeUniform}%
  \kl{Effectful machines} quotiented by \kl{bisimilarity}, $\MealyBis[\cat{C}]$ form a \kl{uniform feedback structure} over $(\cat{V}, \cat{P}, \cat{C})$.
\end{lem}
\begin{proof}
  The identity-on-objects premonoidal functor $J ፡ \cat{C} → \MealyBis[\cat{C}]$ brings any morphism $f ፡ X → Y$ to the \kl[effectful machine]{machine} $J(f) = (I,\id{I},f) ∈ \MealyBis[\cat{C}](X;Y)$.
  Let us construct the feedback operator,
  \[\fbk ፡ \cat{P}(I;T) × \MealyBis[\cat{C}](T ⊗ X; T ⊗ Y) → \MealyBis[\cat{C}](X;Y).\]
  A \kl[effectful machine]{machine} $(S,s,f) ∈ \MealyBis[\cat{C}](T ⊗ X; T ⊗ Y)$ contains a morphism $f ፡ S ⊗ T ⊗ X → S ⊗ T ⊗ Y$ and an initial state $s ∈ S$; together with the initial point $t ∈ \cat{P}(I;T)$, this allows us to define a \kl[effectful machine]{machine} $(S ⊗ T, s ⊗ t, f)$.
  That is to say,
  \[\fbk_t(S,s,f) = (S ⊗ T, s ⊗ t, f).\]
  The only axiom that does not follow by computation is uniformity: the existence of morphisms, $f ∈ \cat{C}(S ⊗ A; S ⊗ B)$ and $g ∈ \cat{C}(T ⊗ A; T ⊗ B)$, satisfying the uniformity equations, $f ⨾ (p ⊗ \id{}) =  (p ⊗ \id{}) ⨾ g$ and $\cat{P}(I;p)(s) = t$, implies that there exists a morphism between their corresponding \kl[effectful machine]{machine}; these must be then \kl{bisimilar}, $\fbk_{s}(\nf{f}) \equiv \fbk_{t}(\nf{g})$, and thus equal in $\MealyBis(\cat{V}, \cat{P}, \cat{C})$.
\end{proof}

\begin{thm}%
  \label{th:mealyIsFree}%
  \kl{Effectful machines} quotiented by \kl{bisimilarity}, $\MealyBis[\cat{C}]$, form the free \kl{uniform feedback structure} over the \kl{effectful triple} $(\cat{V},\cat{P},\cat{C})$.
\end{thm}
\begin{proof}
  The crucial idea of this proof is that each \kl{effectful machine} $(U,i,f)$, for some $f ∈ \cat{C}(U ⊗ X; U ⊗ Y)$ and $i ∈ \cat{P}(I; U)$, arises as a single application of \kl{uniform feedback} over a morphism in the base category:
  $(U,i,f) = \fbk_{i}(f)$.
  Any feedback-preserving functor $H ፡ \Mealy[\cat{C}] → \mathbb{F}$ from the category of \kl{effectful machines} to any other \kl{uniform feedback structure}, $\mathbb{F}$, with an identity-on-objects functor $K ፡ \cat{C} → \mathbb{F}$  is determined by preservation of feedback $H(U,i,f) = H(\fbk_{i}(J(f))) = \fbk_{i}(K(f))$.

  Let us now recall that \kl{effectful machines} form an \kl{effectful triple} themselves (\Cref{prop:effectful-category-machines}), even after quotiented by \kl{bisimilarity} (\Cref{lemma:MealyFreeUniform}).
  Most of the heavy lifting is done by these results: we have shown that we have a \kl{uniform feedback structure} and we built it so that there is a unique possible feedback-preserving mapping to any other \kl{uniform feedback structure}.

  Lastly, we must show that $H$ is indeed a premonoidal functor: checking that
  it preserves composition, for instance, amounts to checking that
  \begin{align*}
    H((U,i,f) ⨾ (V,j,g)) 
    & \overset{(\emph{i})}{=} H(U ⊗ V,i ⊗ j,f \bowtie g)
     \overset{(\emph{ii})}{=} \fbk_{i ⊗ j}(K(f \bowtie g)) \\
    & \overset{(\emph{iii})}{=} \fbk_{i ⊗ j}(K(f) \bowtie K(g))
     \overset{(\emph{iv})}{=} \fbk_{i}(K(f)) ⨾ \fbk_{i}(K(g)),
  \end{align*}
  which follows from (\emph{i}) composition of \kl{effectful machines};
  (\emph{ii}) the only possible construction of $H$; (\emph{iii}) premonoidality
  of $K$; and (\emph{iv}) the axioms of \kl{uniform feedback}. Checking that the
  functor $H$ preserves whiskering is analogous and concludes the proof.
\end{proof}

\begin{rem}[Effectful uniformity]
  As in \Cref{rem:effectfulbisimilarity}, we may also consider \emph{monoidal
  uniformity} and \emph{effectful uniformity}. Effectful uniformity is coarse
  and \kl{effectful machines} quotiented by effectful uniformity do not form a
  category: they can be only composed and tensored with monoidal machines, in a
  structure known as a \emph{strong profunctor}.

  The following \Cref{eq:effectful-uniformity} defines effectful uniformity.
  \begin{equation}
    \begin{split}
      \includegraphics[scale=0.9]{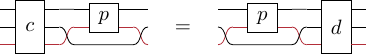}
    \end{split}
    \label{eq:effectful-uniformity}
  \end{equation}
\end{rem}
\section{Effectful Streams}%
\label{sec:streams}

This section introduces \kl{effectful streams}: a semantic universe for the
traces of \kl{effectful machines}. \kl{Effectful streams} are memoryful and
coinductive, but they are also commutative whenever their base category is
commutative: they are an effectful generalization of monoidal streams
\cite{monoidalstreams}. For the following definitions, we fix an
\kl{effectful triple} $(\cat{V}, \cat{P}, \cat{C})$.

Recall that a stream of objects $\stream{X}$ is an object $\now{\stream{X}}$
together with a stream of objects $\later{\stream{X}}$. Similarly,
\kl{effectful streams} are defined coinductively: an \effectfulStream{} is a first
action together with an \effectfulStream{}. However, they have an extra
component—the \emph{memory}—which allows the first action to communicate with
the tail of the stream: two effectful streams are considered equal if there is a
pure transformation between their memories. 
As in \emph{monoidal streams}~\cite{monoidalstreams}, this is formalized by
\emph{\dinaturality{}}, which splits into two different notions in the effectful setting—\kl{stream dinaturality} and \kl{isolated stream dinaturality}.

An \effectfulStream{} consists of \emph{(i)} a first action \(\now{f}\) that
communicates along \emph{(ii)} a memory \(\memory{f}\) with \emph{(iii)} the
tail of the stream \(\later{f}\). 

Let us first define a set of \kl{raw effectful streams}: streams unquotiented by
dinaturality. The set of \emph{effectful streams} will be later defined as the
quotient by dinaturality (\Cref{def:straemDinatural}) of the set of \kl{raw
effectful streams}.

\begin{defi}[Raw effectful stream]%
  \label{def:rawEffectfulStreams}%
  A \intro{raw effectful stream}, $f \in \rawStream[\cat{C}](\stream{X};
  \stream{Y})$, with inputs in $\stream{X} = (X₀,X₁…)$ and outputs in
  $\stream{Y} = (Y₀,Y₁,…)$ is coinductively defined as a tuple consisting of
  \begin{itemize}
    \item $\memory{f} \in \obj{\cat{C}}$, the \emph{memory};
    \item $\now{f} \in \cat{C}(\now{\stream{X}}; \memory{f} \tensor
    \now{\stream{Y}})$, the \emph{head};
    \item $\later{f} \in \rawStream[\cat{C}](\memory{f} \latercomp
    \later{\stream{X}}; \later{\stream{Y}})$, the \emph{tail}.
  \end{itemize}
\end{defi}

Explicitly, a \kl{raw effectful stream}, \(f \colon \stream{X} \to \stream{Y}\), consists of a stream of objects, \(\stream{M} = (M_{0}, M_{1}, M_{2} \dots)\), and a stream of morphisms, \((f_{0} \colon X_{0} \to M_{0} \tensor Y_{0}, f_{1} \colon M_{0} \tensor X_{1} \to M_{1} \tensor Y_{1}, f_{2} \colon M_{1} \tensor X_{2} \to M_{2} \tensor Y_{2}, \dots)\); in the notation of the definition, \(M_{0} = \memory{f}\) and \(f_{0} = \now{f}\).

We depict \kl{raw effectful streams} as in the following diagram, interpreting
that each step in the stream has its own runtime.
\begin{equation*}
  \includegraphics*[scale=0.9]{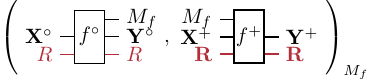}
\end{equation*}
Here, tensoring of an object is coinductively defined by $\now{(M \latercomp
  \stream{X})} = M \tensor \now{\stream{X}}$ and $\later{(M \latercomp
  \stream{X})} = \later{\stream{X}}$. Tensoring extends analogously to morphisms
(\Cref{def:action-streams}).

\begin{defi}[Tensoring]%
  \label{def:action-streams}%
  The \intro[stream tensoring]{tensoring} of a morphism $r \in \cat{P}(M; N)$
  and a raw stream $f \in \rawStream[\cat{C}](N \latercomp \stream{X};
  \stream{Y})$ is the raw stream $r \latercomp f \in \rawStream[\cat{C}](M
  \latercomp \stream{X}; \stream{Y})$ defined coinductively by $\now{(r
  \latercomp f)} = (r \tensor \id{}) ⨾ \now{f}$ and $\later{(r \latercomp f)} =
  \later{f}$.
\end{defi}

\begin{prop}[Tensoring is an action]%
  \label{lemma:monoidal-action-streams}%
  \kl{Stream tensoring} $(\latercomp)$ preserves compositions and identities.
  \[ u \latercomp (v \latercomp s) = (u ⨾ v) \latercomp s \qquad 
  \id{} \latercomp s  = s. \]
\end{prop}

\kl{Stream dinaturality} formalizes the idea that two streams should be equal
whenever their only difference is a pure transformation affecting their memories
at different points in time. It is a particular case of the notion of
dinaturality in category theory.

\begin{defi}[Stream dinaturality]%
  \label{def:straemDinatural}%
  \intro{Stream dinaturality}, $(\dinat)$, is the least equivalence relating two
  streams $(\memory{f},\now{f},\later{f}) \dinat
  (\memory{g},\now{g},\later{g})$, whenever there exists a pure morphism $r \in
  \cat{P}(\memory{g}; \memory{f})$ such that $\now{g} ⨾ (r \tensor \id{}) =
  \now{f}$ and $r \latercomp \later{f} \dinat \later{g}$
  (\Cref{eq:fig-stream-dinaturality}).\footnote{Stream dinaturality is indeed a
  particular case of \kl{dinaturality}. Because of dinaturality, the definition
  of $\Stream[\cat{C}]$ depends on the category $\cat{P}$; we do not write this
  second subscript to avoid confusion.}
  \begin{equation}%
    \label{eq:fig-stream-dinaturality}
    \left( \effectfulstreamnowdinaturalityFig , \effectfulstreamlaterFig \right)_{\memory{f}} \!\!\!
    \dinat \left( \effectfulstreamnowFig{g} ,  \effectfulstreamlaterdinaturalityFig \right)_{\memory{g}}\!\!\!
  \end{equation}
\end{defi}

\Cref{eq:fig-stream-dinaturality} shows that \kl{stream dinaturality}, intuitively, quotients by sliding pure morphisms on the memory.

\begin{defi}%
  \label{def:effectfulStreams}%
  An \intro{effectful stream}, $f \in \Stream[\cat{C}](\stream{X}; \stream{Y})$,
  with inputs in $\stream{X} = (X₀,X₁,…)$ and outputs $\stream{Y} = (Y₀,Y₁,…)$,
  is an equivalence class of \kl{raw effectful streams} under \kl{stream
  dinaturality}. We depict \kl{effectful streams} as in the following diagram,
  interpreting that each step in the stream has its own runtime.
  \begin{equation*}
    \includegraphics*[scale=0.9]{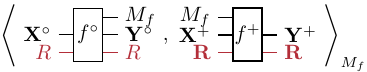}
  \end{equation*}
\end{defi}

Analogously, we define \kl[monoidal streams]{monoidal} and \kl[cartesian streams]{cartesian} streams.
A \intro{monoidal stream}~\cite{monoidalstreams}, $f \in \Stream[\cat{P}](\stream{X}; \stream{Y})$ is a monoidal head morphism $\now{f} \in \cat{P}(\now{\stream{X}};M_f \tensor \now{\stream{Y}})$ followed by a monoidal stream $\later{f} \in \Stream[\cat{P}](M_f \tensor \later{\stream{X}}; \later{\stream{Y}})$, quotiented by
\kl{stream dinaturality}.
A \intro{cartesian stream}, $f \in \Stream[\cat{V}](\stream{X}; \stream{Y})$ is a cartesian head morphism $\now{f} \in \cat{V}(\now{\stream{X}};M_f \tensor \now{\stream{Y}})$ followed by a cartesian stream $\later{f} \in \Stream[\cat{V}](M_f \tensor \later{\stream{X}}; \later{\stream{Y}})$, quotiented by \kl{stream dinaturality} only for cartesian morphisms.

\begin{rem}[Effectful streams are a final coalgebra]%
  \label{rem:finalCoalgebra}
  \Cref{def:effectfulStreams} can be recast in coalgebraic terms.
  The set $\Stream(\stream{X};\stream{Y})$ of effectful streams from $\stream{X}$ to $\stream{Y}$ is the final fixpoint of the functor $ϕ ፡ \functorCat{(\cat{C}^{\omega})\op × \cat{C}^{\omega}}{\Set} → \functorCat{(\cat{C}^{\omega})\op × \cat{C}^{\omega}}{\Set}$ defined by
  \[ ϕ(\fun{Q})(\stream{X},\stream{Y}) =
    ∫^{M \in \cat{P}}\!\!\!
    \cat{C}(\now{\stream{X}}; M \tensor \now{\stream{Y}}) × \fun{Q}(M \latercomp \later{\stream{X}}; \later{\stream{Y}}). \]
  The integral sign denotes a kind of colimit, known as coend, that formalises the \streamDinaturality{} quotient.

  Note that the construction of \kl{effectful streams} relies on the choice of local computations—the morphisms of \(\cat{P}\) are the ones allowed to ``slide'' on the memory—but not on the choice of values, \(\cat{V}\).
\end{rem}

\subsection{The effectful triple of effectful streams}

This section proves that \kl{effectful streams} form an \kl{effectful triple}
(\Cref{th:effectfulStreamsCategory}). We have just shown how the set of
\kl{effectful streams} can be characterized as a final coalgebra: we will use
this characterization to reason coinductively. When reasoning coinductively
about \kl{effectful streams}, we will commonly need a stronger coinductive hypothesis
that parameterizes the first input of the stream: it becomes convenient to explicitly
define \kl{parametrized effectful stream}.

\begin{defi}[Parameterized effectful stream]%
  \label{def:effectfulStreamWithParameters}%
  An \intro{parametrized effectful stream}, from $\stream{X}$ to $\stream{Y}$
  and with parameter $P \in \obj{\cat{C}}$, is an \kl{effectful stream} from $(P
  \latercomp \stream{X}) = (P \tensor X_0, X_1, \dots)$ to $\stream{Y} = (Y_0,
  Y_1, \dots)$.
\end{defi}

The coinductive definitions of sequential composition and whiskering of
\kl{effectful streams} need the stronger coinductive hypothesis given by
\kl{parametrized effectful streams}; the definition for generic streams is,
then, obtained by taking the parameter to be the monoidal unit, \(P = I\).

\begin{defi}[Sequential composition]%
  \label{def:composition-streams}
  The \emph{sequential composition} of two \kl{parametrized effectful streams},
  $f_P \in \Stream[\cat{C}](P \latercomp \stream{X}; \stream{Y})$ and $g_Q \in
  \Stream[\cat{C}](Q \latercomp \stream{Y}; \stream{Z})$, is the
  \kl{parametrized effectful stream} $(f_P ⨾ g_Q) \in \Stream[\cat{C}]((P
  \tensor Q) \latercomp \stream{X}; \stream{Z})$ defined by
  \[%
  \includegraphics{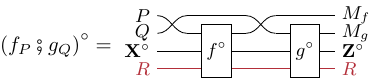}%
  \]%
  and $\later{(f_P ⨾ g_Q)} = \later{f}_{M_f} ⨾ \later{g}_{M_g}$.

  The \emph{sequential composition} of two streams, $f \in \Stream[\cat{C}](\stream{X}; \stream{Y})$ and $g \in \Stream[\cat{C}](\stream{Y}; \stream{Z})$, takes the parameter to be the monoidal unit, $(f ⨾ g) = (f_{I} ⨾ g_{I})$.
\end{defi}

\begin{defi}[Whiskering]%
  \label{def:whiskering-streams}
  The \emph{whiskering} of a \kl{parametrized effectful stream}, $f \in \Stream[\cat{C}](P \latercomp \stream{X}; \stream{Y})$, by a stream of objects \(\stream{U}\) results on the \kl{parametrized effectful stream} $\whis{f_P}{\stream{U}} ፡ P \latercomp \stream{U} \tensor \stream{X} \to \stream{U} \tensor \stream{Y}$.
  \[\includegraphics*[scale=0.9]{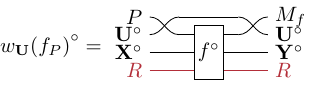}\]
  Whiskering of a stream $f \in \Stream[\cat{C}](\stream{X}; \stream{Y})$ is the stream $\whis{f}{\stream{U}} \in \Stream[\cat{C}](\stream{U} \tensor \stream{X}; \stream{U} \tensor \stream{Y})$ defined by whiskering with the parameter being the monoidal unit, $\whis{f}{\stream{U}} = \whis{f_I}{\stream{U}} \in \Stream[\cat{C}](\stream{U} \tensor \stream{X}; \stream{U} \tensor \stream{Y})$.
\end{defi}

\begin{thm}[Effectful triple of streams]%
  \label{th:effectfulStreamsCategory}%
  \kl{Effectful streams} form an \kl{effectful triple},
  \[ \Stream(\cat{V},\cat{P},\cat{C}) = (\Stream[\cat{V}],\Stream[\cat{P}],\Stream[\cat{C}]).\]
\end{thm}
\begin{proof}
  Let us prove, by coinduction, that composition of
  \kl{parametrized effectful streams} is associative.
  Given three \kl{parametrized effectful streams}, $f ፡ P \latercomp \stream{X} \to \stream{Y}$, $g ፡ Q \latercomp \stream{Y} \to \stream{Z}$, and $h ፡ R \latercomp \stream{Z} \to \stream{W}$, we can see that $\now{(f ⨾ g) ⨾ h}$ and $\now{f ⨾ (g ⨾ h)}$ are equal by string diagrams.

  By the coinductive hypothesis, $\later{((f ⨾ g) ⨾ h)} = ((\later{f} ⨾ \later{g}) ⨾ \later{h}) = (\later{f} ⨾ (\later{g} ⨾ \later{h})) = \later{(f ⨾ (g ⨾ h))}$, where we apply it over three \kl{parametrized effectful streams}:
  $\later{f} ፡ M \latercomp \later{\stream{X}} \to \later{\stream{Y}}$, $\later{g} ፡ N \latercomp \later{\stream{Y}} \to \later{\stream{Z}}$, and $\later{h} ፡ O \latercomp \later{\stream{Z}} \to \later{\stream{W}}$.
  Finally, taking the parameters to be the monoidal units, we prove that the composition of \kl{effectful streams} without parameters is associative.
  The rest of the axioms of an \kl{effectful triple} (unitality, whiskering,...) can be proved similarly.
  This defines a premonoidal category, $\Stream_p(\cat{V},\cat{P},\cat{C})$, which we use to define the \kl{effectful triple} \(\Stream(\cat{V},\cat{P},\cat{C})\).

  The \kl{effectful triple} $\Stream(\cat{V},\cat{P},\cat{C})$ is defined as a triple of categories,
  \[\Stream(\cat{V}) → \Stream(\cat{P}) → \Stream(\cat{C}),\]
  each one of them constructed as a premonoidal category of \kl{effectful streams} over a different base:
  the first one is constructed only used values, $\Stream(\cat{V}) = \Stream_p(\cat{V},\cat{V},\cat{V})$; the second one allows pure computations, $\Stream(\cat{P}) = \Stream_p(\cat{V},\cat{P},\cat{P})$; and the third one allows effectful computations, $\Stream(\cat{C}) = \Stream_p(\cat{V},\cat{P},\cat{C})$.
\end{proof}

\subsection{Effectful bisimulation implies effectful trace equivalence}

Any \kl{effectful machine} induces an \kl{effectful stream} that represents its
execution: its \kl[effectful trace]{trace}. The \kl[effectful trace]{trace} of a
\kl{machine} is a stream that starts with its initial state, and then continues
applying the transition morphism once at each step. Also at each time step, the
current state is passed through the memory to the following time step.

For this construction, explicit notation for infinitely repeating sequences is
convenient. Any object $X$ in $\cat{C}$ can be repeated to form a stream,
\intro[corepeat]{$\corepeat{X}$}, defined by $\now{\corepeat{X}} = X$ and
$\later{\corepeat{X}} = \corepeat{X}$. Analogously, a transition morphism $f \in
\cat{C}(U \tensor X; U \tensor Y)$ can be repeated to an effectful stream
\(\corepeat{f} ፡ U \latercomp \corepeat{X} \to \corepeat{Y}\), defined as
\(\memory{\corepeat{f}} = U\), $\now{\corepeat{f}} = f$ and
$\later{\corepeat{f}} = \corepeat{f}$. The operation $(\latercomp)$ attaches the
initial state to the execution of $f$.

\begin{defi}[Trace]%
  \label{def:effectful-trace}%
  The \intro[effectful trace]{trace} of an \kl{effectful machine} $(U,i,f) \in
  \Mealy(X; Y)$ is the \kl{effectful stream} defined by
  \[\traceFun(U,i,f) = i \latercomp \corepeat{f}.\] 
  Two \kl[effectful machines]{machines} are \intro{trace equivalent} if their
  \kl{traces} coincide.
\end{defi}

\kl{Traces} of \kl{effectful machines} are \emph{type-invariant} \kl{effectful
streams}, meaning that their input types are a constant sequence of the form
$\corepeat{X}$. \intro{Type-invariant effectful streams} can be assembled into a
full subcategory\footnote{\kl{Type-invariant effectful streams} can be given a coinductive definition as well, but it still requires type-variant streams, as the memory objects should not be required to be constant.
Thus, the more natural definition of \kl{type-invariant effectful streams} is the one above, in terms of the category of \kl{effectful streams}.} of \kl[effectful streams]{streams},
\[\StreamInv(\cat{V},\cat{P},\cat{C})(X;Y) = \Stream(\cat{V},\cat{P},\cat{C})(\corepeat{X};\corepeat{Y}).\]

Let us now prove that taking the \kl[effectful trace]{trace} of a \kl{machine}
preserves composition and whiskering: it is a functor between \kl{effectful
triples}.

\begin{lem}[Trace induces a functor]%
  \label{prop:trace-functor}%
  The trace of \kl{effectful machines} defines an effectful functor, 
  $\traceFun ፡ \Mealy[\cat{C}] → \StreamInv[\cat{C}].$
\end{lem}
\begin{proof}
  We know from the literature that $\Mealy[\cat{C}]$ is the free \kl{uniform feedback
  structure} over $(\cat{V},\cat{P},\cat{C})$~\cite[Section
  5.2]{canonicalalgebra}.

  Let us show that $\StreamInv[\cat{C}]$ has a \kl{uniform feedback structure},
  with the identity-on-objects functor being $\corepeat{\bullet} ፡ \cat{C} →
  \StreamInv[\cat{C}]$. Let $(\memory{f},\now{f},\later{f}) ∈
  \StreamInv[\cat{C}](S ⊗ X;S ⊗ Y)$, for some $\now{f} ፡ S ⊗ X → \memory{f} ⊗ S
  ⊗ Y$ and let $s_0 ∈ \cat{P}(I;S)$. We define
  \[\fbk_{s_0}(\memory{M},\now{f},\later{f}) = s₀ \latercomp (\memory{f} ⊗ [S], \now{f}, \later{f});\]
  intuitively, we reinterpret $S$ as being part of the memory. The axioms of
  \kl{uniform feedback} follow immediately; we only highlight \kl{uniformity}
  here: consider $c ∈ \cat{C}(S ⊗ X; S ⊗ Y)$ and $d ∈ \cat{C}(T ⊗ X; T ⊗ Y)$,
  together with $s ∈ \cat{P}(I;S)$ and $p ∈ \cat{P}(S;T)$ satisfying \((p ⊗ \id{X}) ⨾ d = c ⨾ (p ⊗ \id{X})\); we reason that $ p \latercomp \corepeat{d} = \corepeat{c}$ coinductively, because
  \begin{align*}
    p \latercomp \corepeat{d}
      &= p \latercomp (T,d,\corepeat{d})
      = (S,(p ⊗ \id{X}) ⨾ d,\corepeat{d})
      = (S,c ⨾ (p ⊗ \id{X}),\corepeat{d})\\
      &= (S,c, p \latercomp \corepeat{d})
      = (S,c,\corepeat{c})
      = \corepeat{c}.
  \end{align*}
  Then, by definition of feedback, we obtain that
  \[\fbk_{(s ⨾ p)}{\corepeat{d}} = (s ⨾ p) \latercomp \corepeat{d} = s \latercomp (p \latercomp \corepeat{d}) = s \latercomp \corepeat{c} = \fbk_{s}{\corepeat{c}}.\]

  As a consequence of this construction, there exists a unique
  feedback-preserving functor $\traceFun ፡ \Mealy[\cat{C}] →
  \StreamInv[\cat{C}]$, which then must be defined as follows,
  \begin{align*}
    \traceFun(U,i,f) = \traceFun(\fbk_i(f)) 
    = i \latercomp (\memory{f} ⊗ U, \now{\corepeat{f}}, \later{\corepeat{f}}) 
    = i \latercomp (\memory{f} ⊗ U, f, \corepeat{f}) 
    = i \latercomp \corepeat{f}.
  \end{align*}
  This coincides with the definition we gave of $\traceFun$.

  Note that the monoidal case was elaborated in the
  literature~\cite{2021canonicalalgebra,monoidalstreams}; there, \kl{trace}
  preserves the monoidal structure; for the same reasons, in the effectful case,
  \kl{trace} preserves whiskering. Here, we use the universal property of
  \kl{uniform feedback}.
\end{proof}

That \kl{bisimilarity} implies \kl{trace equivalence} is well-known in many
particular settings. For non-deterministic systems, \kl{bisimilarity} entails
trace equivalence. The same happens for coalgebras, when traces are defined as
in both the work of Hasuo, Jacobs, and
Sokolova~\cite{DBLP:journals/lmcs/HasuoJS07}, and in the work of Silva, Bonchi,
Bonsangue, and Rutten~\cite{DBLP:conf/fsttcs/SilvaBBR10}. Our following result
extends this implication to the general case of Mealy machines.

Explicitly, we will show that trace functor preserves \kl{bisimulation}. This means that it
factors through the category of \kl{machines} quotiented by \kl{bisimulation}
(\Cref{th:bisimulation-implies-trace}): whenever two \kl{machines}
are \kl{bisimilar}, they are also \kl{trace equivalent}.

\begin{thm}[Bisimulation implies trace equivalence]%
  \label{th:bisimulation-implies-trace}%
  \kl{Bisimilarity} implies \kl{trace equivalence}: \(\traceFun\) factors
  through the unique feedback preserving functor
  \[\traceFun^{\mathsf{bis}} ፡ \MealyBis[\cat{C}] → \StreamInv[\cat{C}].\]
\end{thm}
\begin{proof}
  We prove that the existence of a morphism $α ፡ (U,i,f) → (V,j,g)$ implies the equality $α \latercomp \corepeat{g} = \corepeat{f}$.
  We proceed by coinduction, noting that
  \begin{align*}
    & \now{(\alpha \latercomp \corepeat{g})} = (\alpha \tensor \id{}) \dcomp g = f \dcomp (\alpha \tensor \id{}) = \now{\corepeat{f}} \dcomp (\alpha \tensor \id{});
  \end{align*}
  and that then, by coinductive hypothesis,
  \(\alpha \latercomp \later{(\alpha \latercomp \corepeat{g})} = \alpha \latercomp \corepeat{g} = \corepeat{f} = \later{\corepeat{f}}\).
  In particular, this implies that $\traceFun(U,i,f) = \traceFun(V,j,g)$ whenever $j = i  \dcomp \alpha$.

  We have shown that the existence of a morphism between two \kl[effectful machines]{machines} implies trace equivalence; we conclude that the existence of a zig-zag of morphisms also implies trace equivalence, by transitivity of equality in \(\Stream{(\cat{V},\cat{P},\cat{C})}\).
\end{proof}

\begin{rem}%
  \label{rem:trace-not-implies-bisimulation}
  The implication of \Cref{th:bisimulation-implies-trace} is strict: \Cref{th:coalgebraic-bisimulation} proves that \kl{effectful bisimilarity} in the case of nondeterministic Mealy machines coincides with their classical notion of bisimilarity.
  \Cref{cor:cortrace} will prove that the traces of nondeterministic Mealy machines given by \kl{effectful streams} also coincide with their classical definition: for a machine \(f \colon U \times X \to \parti(U \times Y)\), a sequence \((y_{0}, \dots, y_{n})\) of outputs is the trace of the sequence \((x_{0}, \dots , x_{n})\) of inputs if there is a sequence of states \((s_{0}, \dots, s_{n+1})\) such that \(s_{0} \in i\) and, for all \(k=0, \dots, n\), \((s_{k+1},y_{k}) \in f(s_{k},x_{k})\).
  It is known that, for nondeterministic Mealy machines, trace equivalence is strictly coarser than bisimilarity~\cite[Remark~7.73]{baier2008principles}.
\end{rem}

\subsection{Isolated-effectful streams}%
\label{sec:isolatedstreams}%

This section discusses a stronger notion of \kl{dinaturality} that can be
imposed to \kl{effectful streams}. This stronger notion—\kl{isolated stream
dinaturality}—represents a realistic assumption that the weaker notion misses:
that, when running a stream, no other process will change the global state.

\kl{Effectful streams}, unlike monoidal streams \cite{monoidalstreams}, can be
quotiented by \dinaturality{} in two different ways: one corresponds to the
assumption that effects are still open (e.g.~other process could use the
runtime), one to the assumption that effects are closed (e.g.~the runtime is
isolated). These assumptions modify the acceptable quotienting: the former gives
rise to coarser equivalence classes.

Why not simply work with \kl{isolated streams}? While a better behaved notion of
equality, it forces us to lose the compositional structure of streams:
\kl{isolated effectful streams} do not form a category—indeed, the
point of isolating them was to prevent their effects from composing—but only a
strong profunctor, meaning that they can be composed only with non-effectful
streams (\Cref{prop:isolatedprofunctor}).

\begin{defi}[Isolated stream dinaturality]%
  \intro{Isolated stream dinaturality}, $(\isodinat)$, is the least equivalence
  relating two streams $\smash(\memory{f},\now{f},\later{f}) \isodinat
  (\memory{g},\now{g},\later{g})$ whenever there exists a premonoidal morphism
  $r ∈ ℂ(M_g;M_f)$ such that $\now{g} ⨾ (r ⊗ \id{}) = \now{f}$ and $r ·
  \later{f} = \later{g}$. 
\end{defi}

We depict isolated dinaturality as in the following
diagram: while the red wire cannot be anymore
formally interpreted in terms of premonoidal string diagrams, we keep it for
intuition.
\begin{equation*}
  \begin{split}
    \includegraphics*[scale=0.9]{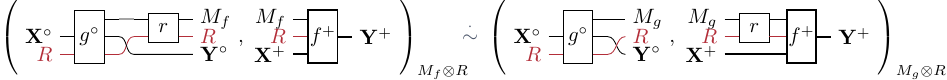}
  \end{split}
  \label{eq:fig-isolated-dinaturality}
\end{equation*}

\begin{defi}%
  \label{def:isolatedEffectfulStreams}%
  An \intro{isolated effectful stream}, $f ∈ \isoStream(\stream{X}; \stream{Y})$
  with inputs $\stream{X} = (X₀,X₁…)$ and outputs $\stream{Y} = (Y₀,Y₁,…)$ is an
  equivalence class of raw effectful streams under \kl{isolated dinaturality}.
\end{defi}

Immediately, we have a \intro[projection to isolated streams]{projection} from
\kl{effectful streams} to \kl{isolated effectful streams}, $\isoproj ፡
\Stream(\stream{X}; \stream{Y}) → \isoStream(\stream{X}; \stream{Y})$. This
projection cannot preserve composition: in fact, \kl{isolated effectful streams}
cannot be composed (for how could we canonically interleave the global effects
of two systems?). While they cannot be composed, isolated effectful streams can
still be composed and tensored with monoidal streams, and that is what we now
state.

\begin{prop}[Isolated effectful streams form a strong profunctor]
  \label{prop:isolatedprofunctor}%
  \kl{Isolated effectful streams} form a strong profunctor over the category of monoidal streams,
  \[\isoStream ፡ \monStream\op × \monStream → \Set\]
\end{prop}

\subsection{Example: the stream cipher is secure}%
\label{sec:cipher-correctness}%

Let us discuss security for the stream cipher protocol
(\Cref{ex:cipher-state-machines}) by giving it
appropriate semantics. For this purpose, we use the Kleisli category of the
finitely-supported distribution monad, $\Stoch$. We employ three fixed sets: a
finite alphabet of characters, $\Char$, a finite set of seeds to our
random number generator, $\Seed$, and a two-element set \(\mathsf{State}
= \{s_{0}, s_{1}\}\) for the internal states of \(\alice\). For both the
character and seed sets, there exist uniform distributions, $u_{\mathsf{c}}
፡ I → \Char$ and $u_{\mathsf{s}} ፡ I → \Seed$.

For the set of characters, there exists moreover a nilpotent and deterministic
``bitwise XOR'' operation $(⊕) ፡ \Char{} × \Char → \Char$,
for which the uniform distribution is a Sweedler integral
\cite{sweedler69:integrals}, meaning that XOR-ing by uniform noise results in
uniform noise,
\[
  \includegraphics{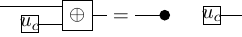}
\] 

Of course, it is impossible to prove that the stream cipher protocol is exactly
equal to a secure channel: it can be easily seen that there exist no perfect
pseudorandom generators in the category of finitely-supported distributions, $\Stoch$.
Instead, we will prove that the protocol is ``approximately equal'' $(\approx)$
to the secure channel.

\begin{asm}[Broadbent and Karvonen,
  {{\cite[\S 7.4]{broadbentKarvonen23:composablecryptography}}}]%
  Let $(\approx)$ be a congruence, preserved by composition and tensoring. An
  \emph{$(\approx)$-pseudorandom number generator} over a finite alphabet is a
  deterministic morphism, $P ፡ \Seed → \Seed ⊗ \Char$, that
  satisfies the following equation.
  \[
    \includegraphics{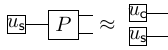}
  \]
\end{asm}

\begin{prop}
  \label{no-pseudorandom}
  There exist no $(=)$-pseudorandom number generators in the category of
  finitary distributions, $\Stoch$.
\end{prop}

The last ingredient we need for this interpretation is to translate the
effectful generators into modifications of a global state, which is a quite
general technique~\cite{mogelbergStaton14:state}. We declare global state to
consist of the pair of seeds that Alice and Bob keep, $\Seed \otimes \Seed$. Our
semantics consists of an \intro{effectful functor} to the \kl{effectful
triple} $\EffStoch$ of stochastic computations with a global state.

\begin{defi}[Interpretation functor]
  The \emph{interpretation functor}, $\llbracket - \rrbracket ፡ \Cipher{} →
  \EffStoch$ is the unique \kl{effectful functor} extending the assignment on
  the generators of the signature generating $\Cipher{}$ we now describe. It
  interprets the object \(C\) as the set of characters $\llbracket C \rrbracket
  = \Char$ and the object \(S\) as the set of states \(\llbracket S \rrbracket =
  \{s₀,s₁\}\). On values, it interprets the XOR symbol as the ``bitwise XOR of
  characters'', $\llbracket\oplus\rrbracket = (\oplus)$. 

  Finally, on effectful generators, it must provide interpretations $\llbracket
  r_a \rrbracket, \llbracket r_b \rrbracket ፡ \Seed \otimes \Seed → \Seed
  \otimes \Seed \otimes \Char$; and $\llbracket s \rrbracket ፡ \Seed \otimes
  \Seed → \Seed \otimes \Seed$; as in the following string diagrams.
  \[
    \includegraphics*[scale=0.9]{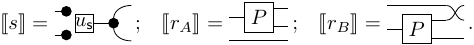}
  \]
\end{defi}

\newcommand{\interpret}[1]{\llbracket #1 \rrbracket}

With this assignment, we can construct \kl{machines} on \(\EffStoch = (\Set,
\Stoch, \SeedStoch)\) corresponding to the syntactic ones described in
\Cref{ex:cipher-state-machines}: by the universal property of \kl{effectful
machines} quotiented by \kl{bisimilarity} (\Cref{th:mealyIsFree}), the functor
\(\interpret{-} ፡ \Cipher{} → \EffStoch\) uniquely lifts to \kl{machines}, defining a feedback-preserving functor
$\interpret{-} ፡ \MealyBis(\Cipher{}) → \MealyBis(\EffStoch)$ that
gives the \kl{trace} semantics  of the protocol, \(\interpret{\CIPHER}\).

Finally, we can state security for the isolated stream cipher: executing it is
approximately equal to executing a secure channel that sends the message
directly from \(\alice\) to \(\bob\) and outputs random noise to an external
attacker.\footnote{Isolated here means that the attacker cannot
access the effectful computation: Alice and Bob's coupled pseudorandom
generators; the attacker still can intercept all messages between them.}

\begin{defi}[Secure channel]
  \AP The \intro{secure channel} is the stateless \kl{machine} in
  \(\Mealy(\Cipher)\) defined by $\Secure = (I, \id{I}, u_c ⊗ \id{C})$. We call
  its \trace{} $\secure = \traceFun\interpret{\Secure}$.
\end{defi}

\begin{thm}%
  \label{th:cipher-secure}%
  The isolated \kl{trace} of the interpretation of the stream cipher is
  approximately equal to that of the \kl{secure channel}, 
  $\isoproj[\traceFun\interpret{\CIPHER}] \approx 
  \isoproj[\traceFun\interpret{\Secure}]$.
\end{thm}
\begin{proof}
  We will prove this statement by coinduction. Let us first compute the
  interpretation of the \kl{stream cipher} from its generators. We have,
  after simplifying the first transition steps,
  \begin{align*}
    \traceFun\llbracket\CIPHER{}\rrbracket 
    & = (\scalebox{0.5}{\initialiseciphersemanticsFig}) · \corepeat{t_{c}} \ \ \text{and}
    & \traceFun\llbracket\Secure{}\rrbracket & = \corepeat{t_{s}},
  \end{align*}
  where the transition functions are specified as
  \begin{align*}
    t_{c} & =\!\!\!\!\!\! \scalebox{0.9}{\transitionciphersemanticsFig{}}; & 
    t_{s} & =\!\!\!\! \scalebox{0.9}{\securenowisolatedFig}.
  \end{align*}
  We start simplifying \((\scalebox{0.5}{\initstateciphersemanticsFig}) \latercomp
  \corepeat{t_{c}}\), using that
  \emph{(i)} the pseudorandom generator \(P\) is
  deterministic, \emph{(ii)} the XOR \(\oplus\) is deterministic, \emph{(iii)}
  the XOR and copy are associative, \emph{(iv)} the XOR is nilpotent,
  \emph{(v, viii)} the XOR and copy are unital, \emph{(vi)} the assumption on
  the pseudorandom generator, and \emph{(vii)} the uniform distribution is a
  Sweedler integral for XOR.
  \begin{align*}
    & \scalebox{1}{\ciphersecureProofFigOne} \overset{(i)}{=} \scalebox{1}{\ciphersecureProofFigTwo}\\
    & \overset{(ii)}{=} \scalebox{1}{\ciphersecureProofFigThree} \overset{(iii)}{=} \scalebox{1}{\ciphersecureProofFigFour}\\
    & \overset{(iv)}{=} \scalebox{1}{\ciphersecureProofFigFive} \overset{(v)}{=} \scalebox{1}{\ciphersecureProofFigSix}\\
    & \overset{(vi)}{\approx} \scalebox{1}{\ciphersecureProofFigSeven} \overset{(vii)}{=} \scalebox{1}{\ciphersecureProofFigEight} \overset{(viii)}{=} \scalebox{1}{\ciphersecureProofFigNine}
  \end{align*}
  
    We now show by coinduction that 
   $\isoproj[\traceFun \interpret{\CIPHER{}}] \approx \isoproj[\traceFun \interpret{\Secure{}}]$ by applying (\emph{i}) the equality just shown, (\emph{ii}) \streamDinaturality{}, and (\emph{iii}) coinduction.
  \begin{align*}
    \isoproj[\traceFun\llbracket\CIPHER{}\rrbracket] & = 
      \isoproj[ (\initialiseciphersemanticsFig) · \corepeat{t_{c}} ]
     = \left\langle \ciphersemanticsoneFig, \later{\corepeat{t_{c}}} \right\rangle_{\Seed \tensor \Seed} \\
    & \overset{(i)}{\approx} 
     \left\langle \ciphersemanticsonelemmaFig, \later{\corepeat{t_{c}}} \right\rangle_{\Seed \tensor \Seed}
      \overset{(ii)}{=} \left\langle \securenowisolatedFig, (\initstateciphersemanticsFig) \latercomp \later{\corepeat{t_{c}}} \right\rangle_{\Seed \tensor \Seed} 
    \\ &      
      \overset{(iii)}{\approx} \left\langle \scalebox{0.5}{\securenowisolatedFig}, \later{\corepeat{t_{s}}} \right\rangle_{\Seed \tensor \Seed}
     = \isoproj[\corepeat{t_{s}}] 
     = \isoproj[\traceFun\llbracket\Secure{}\rrbracket]. 
     \qedhere
  \end{align*}
\end{proof}

\section{Causal Processes}%
\label{sec:causal-processes}

This section introduces \kl{causal processes}: a characterisation of traces with
commutative effects. These \kl[causal traces]{traces} coincide with the usual
trace predicates in the deterministic, nondeterministic and stochastic cases
(\Cref{cor:cortrace}). \Cref{sec:processes-are-streams} will connect \kl[causal
traces]{traces} as \kl{causal processes} with \kl{effectful traces} by proving
that they are isomorphic as \kl{copy-discard categories} (\Cref{def:copy-discard-category}) when the commutative
effects admit \kl{conditionals} and \kl{ranges}
(\Cref{th:streams-are-processes}).

A \kl{causal process} \(f \colon \stream{X} \to \stream{Y}\) is a coherent family of morphisms \(f_{n} \colon X_{0} \tensor \cdots \tensor X_{n} \to Y_{0} \tensor \cdots \tensor Y_{n}\) where each \(f_{n}\) encodes the behaviour of the process until time \(n\).

\begin{defi}%
  \label{def:causal-processes}%
  A \intro{causal process}, $f ፡ \stream{X} → \stream{Y}$, is a sequence of morphisms $\{f_n ፡ X_0 \tensor \cdots \tensor X_n → Y₀ \tensor \cdots \tensor Yₙ\}_{n \in ℕ}$ such that there exist morphisms $\{c_n ፡ Y₀ \tensor \cdots \tensor Y_{n} \tensor X₀ \tensor \cdots \tensor X_{n+1} → Y_{n+1}\}_{n \in ℕ}$ satisfying \Cref{eq:causality-condition}.
  \begin{equation}%
    \label{eq:causality-condition}
    \causalityconditionFig{}
  \end{equation}
\end{defi}

The outputs of the process at different time steps may be correlated and the
causality condition~\eqref{eq:causality-condition} ensures that the output at
time \(n+1\) is coherent with the previous outputs.

A \kl{machine} \((U,i,f) \colon X \to Y\) determines a \kl{causal process}
\(\{f_{n} \colon X^{n+1} \to Y^{n+1}\}_{n \in \naturals}\) that is the execution
of the \kl{machine}.
\[f_{n} = \machineexecutiontraceFig\]
Explicitly, we define \(h_{0} \defn (i \tensor \id{X}) \dcomp f \dcomp \swap_{U,Y}\), \(h_{n+1} \defn (h_{n} \tensor \id{X}) \dcomp f \dcomp \swap_{U,Y}\) and \(f_{n} \defn h_{n} \dcomp (\id{Y^{n+1}} \tensor \discard)\).
If the base category has \kl{conditionals}, the morphisms \(h_{n}\) factor as \(h_{n} = \cp \dcomp ((f_{n} \dcomp \cp) \tensor \id{X^{n+1}}) \dcomp (\id{Y^{n+1}} \tensor d_{n})\), for some \(d_{n} \colon Y^{n+1} \tensor X^{n+1} \to U\).
With these morphisms, we obtain \(c_{n} \defn (d_{n} \tensor \id{X}) \dcomp f \dcomp (\discard \tensor \id{Y})\) that satisfy \Cref{eq:causality-condition}.

The next sections show that \kl{causal processes} form a \kl{copy-discard
category} (\Cref{prop:category-causal-processes}), that the assignment of
processes to machines is functorial and that it coincides with the usual trace
in known examples (\Cref{sec:classical-traces}).

\subsection{The copy-discard category of causal processes}

\kl{Causal processes} may be composed sequentially and in parallel
component-wise, forming a \kl{copy-discard category} (\Cref{def:copy-discard-category}). We need to ensure that
compositions of \kl{causal processes} still satisfy the causality
condition~\eqref{eq:causality-condition}. For this, we ask the base category to
have \emph{\kl{conditionals}}.

\kl{Conditionals} are structure on top of a \kl{copy-discard category} that
originates in categorical
probability~\cite{cho2019disintegration,fritz2020synthetic}. In categories of
probabilistic kernels, \kl{conditionals} allow forming conditional distributions
from joint distributions. Our examples include effects beyond probability, but
all of them still have \kl{conditionals} (\Cref{ex:conditionals}).

\begin{defi}[Conditional]%
  \label{def:conditional}%
  In a \kl{copy-discard category} $(\cat{V},\cat{P})$ a morphism \(f ፡ X → A
  \tensor B\) has a \kl{conditional} when there exists a morphism $c ፡ A \tensor
  X → B$, called \intro{conditional}, such that
  \[\generatorFig{morphismOneTwo=\(f\)} \ = \
  \marginalconditionalsFig{f}{c}\ .\]
  The composition \(f \dcomp \proj{A} \colon X \to A\) is called the \intro{marginal} of \(f\) on \(A\).
  A \kl{copy-discard category} $(\cat{V},\cat{P})$ has conditionals whenever every morphism has a conditional.
\end{defi}

It is convenient to define an \intro[conditional composition]{operation}
\((\condcomp) \colon \cat{P}(X; A) \times \cat{P}(A \tensor X \tensor Y; B) \to
\cat{P}(X \tensor Y; A \tensor B)\); for morphisms $m \in \cat{P}(X; A)$ and $c
\in \cat{P}(A \tensor X \tensor Y; B)$,
\[m \condcomp c \defn \wireconditionalsFig{m}{c}{X}{Y}{A}{B}.\] 
With this notation, the equation for conditionals in \Cref{def:conditional}
becomes \(f = (f \dcomp \proj{A}) \condcomp c\).

\begin{exa}%
  \label{ex:conditionals}
  We recall the expressions of conditionals in some \kl{copy-discard
  categories}. In some of these cases, conditionals can be expressed in string
  diagrams (\Cref{fig:examples-conditionals}).
  \begin{itemize}
    \item \((\Set, \Set)\).
          For a function \(f \colon X \to A \times B\), its conditional on \(A\) is \(\proj{X} \dcomp f \dcomp \proj{B}\).
          More generally, conditionals in \kl{cartesian categories} are of this shape~\cite[Remark~2.4]{fritz2020synthetic}.
    \item \((\Set, \Par)\).
          For a partial function \(f \colon X \to A \times B\), its conditional on \(A\) is \(\proj{X} \dcomp f \dcomp \proj{B}\).
          More generally, conditionals in cartesian restriction categories are of this shape~\cite[Proposition~2.14]{order2025}.
    \item \((\Set, \Rel)\).
          For a relation \(f \colon X \to A \times B\), its conditional on \(A\) is \((\id{A} \times f) \dcomp (\textfrobcappic_{A} \times \id{B})\), where \(\textfrobcappic \colon A \times A \to 1\) denotes the relation \(\{(a,a,\ast) \st a \in A\}\).
          More generally, conditionals in cartesian bicategories of relations are of this shape~\cite[Proposition~2.16]{order2025}.
    \item \((\Set, \Stoch)\).
          For a function \(f \colon X \to \distr(A \times B)\), its conditionals are
          \begin{equation*}
            c(b \mid a,x) =
            \dfrac{f(a,b \mid x)}{\sum_{b' \in B} f(a,b' \mid x)}
          \end{equation*}
          whenever observing \(b\) given \(a\) and \(x\) was possible, \(\sum_{b' \in B} f(a,b' \mid x) \neq 0\), and by $c(b \mid a,x) = \sigma(b)$ for some \(\sigma \in \distr(B)\) otherwise~\cite[Example~11.6]{fritz2020synthetic}.
    \item \((\Set, \subStoch)\).
          For a function \(f \colon X \to \subdistr(A \times B)\), a conditional \(c \colon A \times X \to \subdistr(B)\) is
          \begin{equation*}
            c(b \mid a,x) =
            \dfrac{f(a,b \mid x)}{\sum_{b' \in B} f(a,b' \mid x)} \qquad\text{and}\qquad
            c(\bot \mid a,x) = 0
          \end{equation*}
          whenever defined, \(\sum_{b' \in B} f(a,b' \mid x) \neq 0\), and by $c(b \mid a,x) = 0$ and $c(\bot \mid a,x) = 1$ otherwise~\cite[Proposition~2.13]{2023partialmarkov}.
  \end{itemize}
\end{exa}

\kl{Conditionals} determine a preorder enrichment of \kl{copy-discard categories}, which we use to simplify the definition and proofs about \kl{causal processes}.

\begin{prop}%
  [{\cite[Theorem~3.3]{order2025}}]%
  \label{prop:preorder}
  \kl{Copy-discard categories} with \kl{conditionals} are preorder-enriched with the preorder defined with \kl{conditional composition}: for morphisms \(f,g \colon X \to Y\), we set \intro[conditional preorder]{\(f \condleq g\)} if and only if there is \(s \colon Y \tensor X \to I\) such that \(f = g \condcomp s\).
\end{prop}

\begin{rem}%
  \label{rem:discard-top}
  The discard morphism \(\discard \colon X \to I\) is the top element of the hom-preorder \((\cat{P}(X;I), \condleq)\): for any other \(s \colon X \to I\), unitality of the copy morphism yields \(s = \discard \condcomp s\).
\end{rem}

\begin{exa}%
  [{\cite[Section~3.3]{order2025}}]%
  \label{ex:preorders}
  We recall how the preorder of \Cref{prop:preorder} instantiates in the \kl{copy-discard categories} from \Cref{ex:conditionals}.
  \begin{itemize}
    \item \((\Set, \Set)\).
          The \kl{conditional preorder} in categories where all morphisms are \kl{total} becomes discrete: \(f \condleq g\) if and only if \(f = g\).
    \item \((\Set, \Par)\).
          For two partial functions \(f,g \colon X \to Y\), \(f \condleq g\) if and only if \(f(x) = g(x)\) for all \(x\) where \(f\) is defined, i.e.\ \(f\) is a restriction of \(g\).
    \item \((\Set, \Rel)\).
          For two relations \(f,g \colon X \to Y\), \(f \condleq g\) if and only if \(f \subseteq g\) as subsets of \(X \times Y\).
    \item \((\Set, \Stoch)\).
          As for \((\Set, \Set)\), the \kl{conditional preorder} is discrete because all morphisms are \kl{total}.
    \item \((\Set, \subStoch)\).
          For two functions \(f,g \colon X \to \subdistr(Y)\), \(f \condleq g\) if and only if \(f(y \given x) \leq g(y \given x)\) for all \(x \in X\) and \(y \in Y\).
  \end{itemize}
  These examples justify the intuition that \(f \condleq g\), whenever \(f\) behaves like \(g\) but fails more often than \(g\).
\end{exa}

We characterise \kl{causal processes} as families of morphisms where \(f_{n+1}\) behaves in the same way as \(f_{n}\) on the first \(n\) inputs, but may fail more often.

\begin{prop}%
  \label{prop:causal-processes-lax-natural}
  A sequence of morphisms $\{f_n ፡ X_0 \tensor \cdots \tensor X_n → Y₀ \tensor \cdots \tensor Yₙ\}_{n \in ℕ}$ in a \kl{copy-discard category} with \kl{conditionals} is a \kl{causal process} if and only if \( f_{n+1} \dcomp \proj{n} \condleq \proj{n} \dcomp f_{n} \) for all \(n \in \naturals\).
  \[\causalityinequalityFig\]
\end{prop}
\begin{proof}
  Consider a \kl{causal process} \(f \colon \stream{X} \to \stream{Y}\).
  By discarding the second output in \Cref{eq:causality-condition}, we obtain witnesses \(c_{n} \dcomp \discard\) of the inequalities \(f_{n+1} \dcomp \pi_{n} \condleq \pi_{n} \dcomp f_{n}\).

  For the converse, consider a sequence of morphisms $\{f_n ፡ X_0 \tensor \cdots \tensor X_n → Y₀ \tensor \cdots \tensor Yₙ\}_{n \in ℕ}$.
  By definition of the \kl[conditional preorder]{preorder}, \(f_{n+1} \dcomp \pi_{n+1,n} \condleq \pi_{n+1,n} \dcomp f_{n}\) means that there are morphisms \(s_{n+1} \colon Y_{0} \tensor \cdots \tensor Y_{n} \tensor X_{0} \tensor \cdots \tensor X_{n+1} \to I\) satisfying the equation below.
  \[\laxnaturalityFig{f_{n+1}}{f_{n}}{s_{n+1}}\]
  Using (\emph{i}) conditionals and (\emph{ii}) the inequality, we find morphisms \(c_{n+1}\colon Y_{0} \tensor \cdots \tensor Y_{n} \tensor X_{0} \tensor \cdots \tensor X_{n+1} \to Y_{n+1}\) that satisfy the causality condition (\Cref{eq:causality-condition}).
  \begin{align*}
    \causalityfromnaturalityProofFig{}
  \end{align*}
  Thus, we can take \(c_{n+1} = \cp \dcomp (s_{n+1} \tensor d_{n+1})\).
\end{proof}

\begin{exa}%
  \label{ex:independent-processes}
  A family \(\{f_{n} \colon X_{n} \to Y_{n}\}_{n \in \naturals}\) determines a causal process in which every step is independent of the others, \(\hat{f}_{n} \defn f_{0} \tensor \cdots \tensor f_{n}\).
  It is easy to see that these indeed form a \kl{causal process} thanks to \Cref{prop:causal-processes-lax-natural} and \Cref{rem:discard-top}:
  \[\hat{f}_{n+1} \dcomp \proj{n} = \hat{f}_{n} \tensor (f_{n+1} \dcomp \discard) \condleq \hat{f}_{n} \tensor \discard = \proj{n} \dcomp \hat{f}_{n}.\]
\end{exa}

With \Cref{prop:causal-processes-lax-natural}, we prove that \kl{causal processes} form a \kl{copy-discard category}.

\begin{prop}%
  \label{prop:category-causal-processes}%
  \kl{Causal processes} over a \kl{copy-discard category} $(\cat{V},\cat{P})$ with \kl{conditionals} form a \kl{copy-discard category}, $\Causal(\cat{V},\cat{P})$.
\end{prop}
\begin{proof}
  We show that causal processes are a category where identities and composition are defined component-wise by those in \(\cat{P}\), i.e.\ \((f \dcomp g)_{n} \defn f_{n} \dcomp g_{n}\) and \((\id{\stream{X}})_{n} \defn \id{X_{0} \tensor \cdots \tensor X_{n}}\).
  Since they are defined component wise, they must be associative and unital.
  We proceed to check that they are well-defined, i.e.\ that the identity satisfies the causality condition and that, whenever \(f\) and \(g\) satisfy the causal condition, \(f \dcomp g\) does so too.
  For the identities, we notice that they follow the pattern of \Cref{ex:independent-processes}.
  For compositions, we use the characterisation in \Cref{prop:causal-processes-lax-natural} and that \((\condleq)\) gives a preorder-enrichment of \(\cat{P}\)~\cite{order2025}.
  \begin{equation*}
    (f \dcomp g)_{n+1} \dcomp \proj{n} \defn f_{n+1} \dcomp g_{n+1} \dcomp \proj{n} \condleq f_{n+1} \dcomp \proj{n} \dcomp g_{n} \condleq \proj{n} \dcomp f_{n} \dcomp g_{n} \codefn \proj{n} \dcomp (f \dcomp g)_{n}
  \end{equation*}
  Then, \(\Proc{\cat{V},\cat{P}}\) is a category.
  The monoidal product on objects is defined inductively.
  \begin{align*}
    (\stream{X} \tensor \stream{Y})_{0} & = X_{0} \tensor Y_{0} &
    (\stream{X} \tensor \stream{Y})_{n+1} & = (\stream{X} \tensor \stream{Y})_{n} \tensor X_{n+1} \tensor Y_{n+1}
  \end{align*}
  On morphisms, the definition of monoidal product requires reshufflings, \(\phi\) and \(\phi^{-}\).
  \begin{align*}
    \phi^{n}_{\stream{X}, \stream{Y}} &\colon (\stream{X} \tensor \stream{Y})_{n-1} \tensor X_{n} \tensor Y_{n} \to (\stream{X})_{n-1} \tensor X_{n} \tensor (\stream{Y})_{n-1} \tensor Y_{n}\\
    \phi^{-n}_{\stream{X}, \stream{Y}} & \colon (\stream{X})_{n-1} \tensor X_{n} \tensor (\stream{Y})_{n-1} \tensor Y_{n} \to (\stream{X} \tensor \stream{Y})_{n-1} \tensor X_{n} \tensor Y_{n}
  \end{align*}
  These are also defined by induction, with \(\phi^{0}_{\stream{X},\stream{Y}} \defn \id{X_{0} \tensor Y_{0}}\),
  \begin{align*}
     \phi^{n+1}_{\stream{X},\stream{Y}} & \defn \scalebox{0.9}{\shufflingtensorprocessesFig} &\text{and}&& \phi^{-(n+1)}_{\stream{X},\stream{Y}} & \defn \scalebox{0.9}{\invshufflingtensorprocessesFig}.
  \end{align*}
  It is easy to see, by induction, that they are inverses to each other: \(\phi^{n}_{\stream{X},\stream{Y}} \dcomp \phi^{-n}_{\stream{X},\stream{Y}} = \id{}\) and \(\phi^{-n}_{\stream{X},\stream{Y}} \dcomp \phi^{n}_{\stream{X},\stream{Y}} = \id{}\).
  The monoidal product \(f \tensor g\) is defined on the components by reshuffling the order of the inputs and outputs of the monoidal product \(f_{n} \tensor g_{n}\) in \(\cat{P}\):
  \[(f \tensor g)_{n} \defn \phi^{n}_{\stream{X},\stream{Y}} \dcomp (f_{n} \tensor g_{n}) \dcomp \phi^{-n}_{\stream{X},\stream{Y}} \ .\]
  This operation preserves identities and compositions because it does so in \(\cat{P}\).
  \begin{align*}
    & (\id{\stream{X}} \tensor \id{\stream{Y}})_{n} && ((f \tensor f') \dcomp (g \tensor g'))_{n} \\
    & \defn \phi^{n}_{\stream{X},\stream{Y}} \dcomp ((\id{\stream{X}})_{n} \tensor (\id{\stream{Y}})_{n}) \dcomp \phi^{-n}_{\stream{X},\stream{Y}} && \defn (f \tensor f')_{n} \dcomp (g \tensor g')_{n} \\
    & \defn \phi^{n}_{\stream{X},\stream{Y}} \dcomp (\id{(\stream{X})_{n}} \tensor \id{(\stream{Y})_{n}}) \dcomp \phi^{-n}_{\stream{X},\stream{Y}} && \defn \phi^{n} \dcomp (f_{n} \tensor f'_{n}) \dcomp \phi^{-n} \dcomp \phi^{n} \dcomp (g_{n} \tensor g'_{n}) \dcomp \phi^{-n} \\
    & = \phi^{n}_{\stream{X},\stream{Y}} \dcomp \phi^{-n}_{\stream{X},\stream{Y}} && = \phi^{n} \dcomp (f_{n} \tensor f'_{n}) \dcomp (g_{n} \tensor g'_{n}) \dcomp \phi^{-n} \\
    & = \id{(\stream{X} \tensor \stream{Y})_{n}} && = \phi^{n} \dcomp (f_{n} \dcomp g_{n}) \tensor (f'_{n} \dcomp g'_{n}) \dcomp \phi^{-n} \\
    & \codefn (\id{\stream{X} \tensor \stream{Y}})_{n} && \codefn \phi^{n} \dcomp (f \dcomp g)_{n} \tensor (f' \dcomp g')_{n} \dcomp \phi^{-n}\\
    &&& \codefn ((f \dcomp g) \tensor (f' \dcomp g'))_{n}
  \end{align*}
  We show that the monoidal product using the characterisation in \Cref{prop:causal-processes-lax-natural} and that \((\condleq)\) gives a preorder-enrichment of \(\cat{P}\)~\cite{order2025}.
  \begin{equation*}
    (f \tensor g)_{n+1} \dcomp \proj{n}
    = \causalprocessestensorwelldefProofFigOne
    \condleq \causalprocessestensorwelldefProofFigTwo
    = \proj{n} \dcomp (f \tensor g)_{n}
  \end{equation*}
  Associators and unitors are lifted from \(\cat{P}\) and follow the pattern of \Cref{ex:independent-processes}, so they are well-defined.
  \begin{gather*}
    (\alpha_{\stream{X}, \stream{Y}, \stream{Z}})_{n} \defn \alpha_{X_{0}, Y_{0}, Z_{0}} \tensor \cdots \tensor \alpha_{X_{n}, Y_{n}, Z_{n}} \\
    (\lambda_{\stream{X}})_{n}  \defn \lambda_{X_{0}} \tensor \cdots \tensor \lambda_{X_{n}} \qquad\qquad (\rho_{\stream{X}})_{n} \defn \rho_{X_{0}} \tensor \cdots \tensor \rho_{X_{n}}
  \end{gather*}
  They satisfy the pentagon and triangle equations because they do so component-wise, so \(\Proc{\cat{V},\cat{P}}\) is a monoidal category.
  Finally, the copy-discard structure is also lifted from \((\cat{V},\cat{P})\) following the pattern of \Cref{ex:independent-processes}, so it is well-defined.
  \begin{align*}
    (\discard_{\stream{X}})_{n} & \defn \discard_{X_{0}} \tensor \cdots \tensor \discard_{X_{n}} & (\cp_{\stream{X}})_{n} & \defn \cp_{X_{0}} \tensor \cdots \tensor \cp_{X_{n}}
  \end{align*}
  Coassociativity, counitality, cocommutativity and compatibility of the comonoid structure in \(\Proc{\cat{V},\cat{P}}\) follow from the same properties in \((\cat{V},\cat{P})\) and we obtain a \kl{copy-discard category}.
\end{proof}

\begin{rem}%
  \label{rem:topos-of-trees-enrichment}
  \kl{Causal processes} are enriched over the topos of trees~\cite{birkedal2012topos,birkedal2011topos,kasangian1997topos}.
  Hom-sets can be expressed as functors \(\omega\op \to \Set\), i.e.\ objects in the topos of trees: \(\Causal(\stream{X}, \stream{Y})(n) = \{(p_{0}, \dots, p_{n}) \st (p_{n} \mid n \in \naturals) \in \Causal(\stream{X}, \stream{Y})\}\) with projections \(r_{n} \colon \Causal(\stream{X}, \stream{Y})(n+1) \to \Causal(\stream{X}, \stream{Y})(n)\) defined by \(r_{n}(p_{0}, \dots, p_{n+1}) \defn (p_{0}, \dots, p_{n})\).
  Compositions and monoidal products are defined component-wise, so they are natural transformations \((\dcomp)_{n} \colon \Causal(\stream{X}, \stream{Y})(n) \times \Causal(\stream{Y}, \stream{Z})(n) \to \Causal(\stream{X}, \stream{Z})(n)\) and \((\tensor)_{n} \colon \Causal(\stream{X}, \stream{Y})(n) \times \Causal(\stream{X'}, \stream{Y'})(n) \to \Causal(\stream{X} \tensor \stream{X'}, \stream{Y} \tensor \stream{Y'})(n)\), i.e.\ morphisms in the topos of trees.
\end{rem}

\subsection{Examples of causal processes}
We instantiate causal processes in the \kl{copy-discard categories} from \Cref{ex:conditionals}.
Thanks to the structure in each of them, the causality condition simplifies as shown in the second column in \Cref{fig:examples-conditionals} and we can give simplified descriptions of causal processes in each case.
The last column gives the expression for trace predicates and will be relevant in the next \Cref{sec:classical-traces}.
\begin{figure*}[h!]
  \setlength{\tabcolsep}{3pt}
  \centering
  \begin{tabular}{| l | l l l |}
    \hline
    & Conditionals of \(f\) & Causality condition & Trace predicate \\
    \hline &&&\\[-1.5ex]
    \(\Set\) & \scalebox{0.8}{\conditionalparFig} & \scalebox{0.8}{\causalityconditionsetFig} & \(\begin{aligned} & s_{0} = i \\ & \land \forall k \leq n .\, (s_{k+1}, y_{k}) = f(s_{k},x_{k})\end{aligned}\)\\[1.6em]
    \(\Par\) & \scalebox{0.8}{\conditionalparFig} & \scalebox{0.8}{\causalityconditionparFig} & \(\begin{aligned} &s_{0} = i \\ & \land \forall k \leq n .\, (s_{k+1}, y_{k}) = f(s_{k},x_{k})\end{aligned}\)\\[1.6em]
    \(\Rel\) & \scalebox{0.8}{\conditionalrelFig} & \scalebox{0.8}{\causalityconditionrelationFig} & \(\begin{aligned} & \exists s_{0}, \dots, s_{n+1} \in S .\, s_{0} \in i \\ & \land \forall k \leq n  .\, (s_{k+1}, y_{k}) \in f(s_{k},x_{k})\end{aligned} \) \\[1.6em]
    \(\Stoch\) & \scalebox{0.8}{\qtconditionalsFig} & \scalebox{0.8}{\causalityconditionstochFig} & \(\begin{aligned} & \textstyle\sum_{s_{0}, \dots, s_{n+1} \in S}  i(s_{0}) \\& \cdot \textstyle\prod_{k ≤ n} f(s_{k+1}, y_{k} \mid s_{k},x_{k}) \end{aligned}\)\\[1.6em]
    \(\subStoch\) & \scalebox{0.8}{\qtconditionalsFig} & \scalebox{0.8}{\shortcausalityconditionFig} & \(\begin{aligned} & \textstyle\sum_{s_{0}, \dots, s_{n+1} \in S}  i(s_{0}) \\& \cdot \textstyle\prod_{k ≤ n} f(s_{k+1}, y_{k} \mid s_{k},x_{k}) \end{aligned}\)\\[1em]
    \hline
  \end{tabular}
  \caption{Conditionals, simplified causality condition, and trace
  predicate. The trace predicate determines the behaviour of a machine
  \((U,i,f)\) with inputs \(x_{0}, \dots, x_{n}\) and outputs \(y_{0}, \dots,
  y_{n}\).}\label{fig:examples-conditionals}
\end{figure*}

\begin{rem}
  The cases of $(\Set, \Set)$ and $(\Set, \Stoch)$ were already captured by \emph{monoidal streams}~\cite{monoidalstreams}.
  However, monoidal streams over the category of relations—and any compact closed category—form a posetal category.
  \kl{Effectful streams} prevent this collapse by distinguishing pure and effectful morphisms (here, total relations and arbitrary relations).
  This allows \Cref{th:streams-are-processes} to apply in partial, partial stochastic, and relational settings.
\end{rem}

\paragraph{Cartesian causal processes}
The case of $\Set$ extends to all \kl{cartesian categories}.
Cartesianity ensures that the outputs of a \kl{causal process} are all independent of each other and it simplifies the causality condition~\eqref{eq:causality-condition}.
As a consequence, a cartesian \kl{causal process} \((f_{n} \mid n \in \naturals) ፡ \stream{X} → \stream{Y}\) reduces to a family of morphisms \(p_{n} ፡ X_{0} × \cdots × X_{n} → Y_{n}\)~\cite[Section 6]{monoidalstreams}.
The associated \kl{causal process} is defined by \(f_{0} = p_{0}\) and \(f_{n+1} = (\cp \tensor \id{X_{n+1}}) \dcomp (f_{n} \tensor p_{n+1})\) (see the first row, second column of \Cref{fig:examples-conditionals}).
Cartesian \kl{causal processes} coincide with the classical notion of \emph{causal stream function}~\cite{raney1958sequential,jacobs:causalfunctions,katsumata19} and form the Kleisli category of the \emph{non-empty list comonad}~\cite{uustalu2008comonadic}.
Remarkably, this Kleisli construction works only when the base category is cartesian~\cite[Theorem~6.1]{monoidalstreams}.

\paragraph{Partial deterministic causal processes}
\kl{Causal processes} over the \kl{copy-discard category} \((\Set, \Par)\) are the partial analogue of causal stream functions.
Explicitly, a partial causal process $f \in \Causal(\Set, \Par)(\stream{X}; \stream{Y})$ is a family of functions $f_{n} ፡ X_{0} × \cdots × X_{n} → (Y_{0} × \cdots × Y_{n}) + 1$, indexed by the natural numbers, such that \(f_{n+1} \dcomp \proj{n}\) is a restriction of \(\proj{n} \dcomp f_{n}\) (equation in the second row, second column of \Cref{fig:examples-conditionals}).

\paragraph{Relational causal processes}
Relational \kl{causal processes} are families of functions \(f_{n} ፡ X_0 × \cdots × X_n
→ \parti(Y_{0} × \cdots × Y_{n})\), indexed by the natural numbers, such that \(f_{n+1} \dcomp \proj{n} \subseteq \proj{n} \dcomp f_{n}\) as subsets of \(X_0 × \cdots × X_{n+1} \times Y_{0} \times \cdots \times Y_{n}\) (equation in the third row, second column of \Cref{fig:examples-conditionals}).

\paragraph{Stochastic causal processes}
\kl{Causal processes} over the \kl{copy-discard category} \((\Set, \Stoch)\) coincide with controlled stochastic processes~\cite{fleming1975,ross1996stochastic}.
Explicitly, a stochastic \kl{causal process} is a family of functions \(f_{n} \colon X_{0} \times \cdots \times X_{n} \to \distr(Y_{0} \times \cdots \times Y_{n})\), indexed by the natural numbers, such that the marginal of \(f_{n+1}\) on the first \(n\) outputs is \(\proj{n} \dcomp f_{n}\) (equation in the fourth row, second column of \Cref{fig:examples-conditionals})~\cite[Section 7]{monoidalstreams}.
A causality condition capturing the same intuition has appeared before for strong monads on a finitely distributive category that has list objects~\cite[Definition~17]{DBLP:conf/calco/Goncharov13}; if the monad is, moreover, affine, then the associated causality coincides with the causality condition in \Cref{prop:causal-processes-lax-natural}.

No difficulty arises when considering measurable stochastic kernels: \kl{causal processes} over the \kl{copy-discard category} \((\stdBorel, \BorelStoch)\) may be described as families of measurable functions \(f_{n} \colon X_{0} \times \cdots \times X_{n} \to \Giry(Y_{0} \times \cdots \times Y_{n})\), indexed by the natural numbers, such that the marginal of \(f_{n+1}\) on the first \(n\) outputs is \(\proj{n} \dcomp f_{n}\).
In fact, the same characterisation of causality holds for every \kl{copy-discard category} where all morphisms are \kl{total} (also known as Markov categories~\cite{cho2019disintegration,fritz2020synthetic}).

\paragraph{Partial stochastic causal processes}
\kl{Causal processes} over the \kl{copy-discard category} \((\Set, \subStoch)\) give the partial analogue of controlled stochastic processes.
Explicitly, a partial stochastic \kl{causal process} is a family of functions \(f_{n} \colon X_{0} \times \cdots \times X_{n} \to \subdistr(Y_{0} \times \cdots \times Y_{n})\), indexed by the natural numbers, such that the marginal of \(f_{n+1}\) on the first \(n\) outputs is smaller than \(\proj{n} \dcomp f_{n}\) (equation in the last row, second column of \Cref{fig:examples-conditionals}).
As above, no difficulty arises when considering \kl{causal processes} over the \kl{copy-discard category} \((\stdBorel, \BorelsubStoch)\).

\subsection{Traces: capturing the classical examples}%
\label{sec:classical-traces}%

The execution of an \kl{effectful machine} on a \kl{copy-discard category} forms a \kl{causal process}.
This assignment is functorial.

\begin{prop}%
  \label{prop:trace-process}
  The \intro{causal trace} of an \kl{effectful machine}, \((U,i,f) \colon X \to Y\), in a \kl{copy-discard category}, \((\cat{V}, \cat{P})\), with \kl{conditionals} is the \kl{causal process} \((f_{n} \mid n \in \naturals)\) defined by discarding the state space of the \(n\)-fold composition of the transition morphism \(f\) with itself: \(f_{n} = p_{n} \dcomp (\scalebox{0.6}{\projFig})\), where the morphisms \(p_{n}\) are defined inductively by
  \[p_{0} \defn \traceprocessZeroFig{i}{f} \qquad \mbox{ and }\qquad p_{n+1} \defn \traceprocessIndFig{p_{n}}{f}\,.\]
  This assignment defines a copy-discard functor.
\end{prop}
\begin{proof}
  We check that the assignment is well-defined.
  For an \kl{effectful machine} \((U,i,f)\), consider the morphisms \(f_{n}\) as defined in the statement.
  We use \Cref{prop:causal-processes-lax-natural} and \Cref{rem:discard-top} to obtain
  \[f_{n+1} \dcomp \proj{n} = \traceprocesswelldefProofFigOne \condleq \traceprocesswelldefProofFigTwo = \proj{n} \dcomp f_{n}.\]
  Functoriality will follow from \Cref{th:streams-are-processes}, but it is also easy to check by induction.
\end{proof}

We instantiate \kl{causal traces} in the \kl{copy-discard categories} from \Cref{ex:conditionals} and explicitly write the trace predicates in \Cref{fig:examples-conditionals}.

\begin{cor}\label{cor:cortrace}
  \kl[Causal traces]{Traces} of \kl{machines} over the \kl{copy-discard categories} \((\Set, \Set)\), \((\Set, \Par) \) and \((\Set, \Rel)\) coincide with the standard traces for labelled transition systems~\cite{van2001linear}.
  Traces of \kl{machines} over the \kl{copy-discard categories} \((\Set, \Stoch)\) and \((\Set, \subStoch)\)  coincide with traces of partially observable labelled Markov processes~\cite{castro2009equivalence}.
  The trace predicates in each of these cases are summarised in the last column of \Cref{fig:examples-conditionals}.
\end{cor}
\begin{proof}
  Consider a \kl{machine} \((U,i,f) \colon X \to Y\).
  We describe its trace predicate in each of these cases.
  \begin{itemize}
    \item \((\Set, \Set)\).
          The sequence \((y_{0}, \dots, y_{n})\) of outputs is the trace of the sequence \((x_{0}, \dots , x_{n})\) of inputs if \(s_{0} = i\) and, for all \(k=0, \dots, n\), \((s_{k+1},y_{k}) = f(s_{k},x_{k})\).
    \item \((\Set, \Par)\).
          The sequence \((y_{0}, \dots, y_{n})\) of outputs is the trace of the sequence \((x_{0}, \dots , x_{n})\) of inputs if \(s_{0} = i\) and, for all \(k=0, \dots, n\), \(f(s_{k},x_{k})\) is defined and \((s_{k+1},y_{k}) = f(s_{k},x_{k})\).
    \item \((\Set, \Rel)\).
          The sequence \((y_{0}, \dots, y_{n})\) of outputs is the trace of the sequence \((x_{0}, \dots , x_{n})\) of inputs if there is a sequence of states \((s_{0}, \dots, s_{n+1})\) such that \(s_{0} \in i\) and, for all \(k=0, \dots, n\), \((s_{k+1},y_{k}) \in f(s_{k},x_{k})\).
    \item \((\Set, \Stoch)\).
          The sequence \((y_{0}, \dots, y_{n})\) of outputs is the trace of the sequence \((x_{0}, \dots , x_{n})\) of inputs with probability \(\sum_{(s_{0}, \dots, s_{n+1})} i(s_{0}) \cdot \prod_{k=0}^{n} f(s_{k+1}, y_{k} \given s_{k}, x_{k})\).
    \item \((\Set, \subStoch)\).
          The trace predicate has the same expression as in the \kl{copy-discard category} \((\Set, \Stoch)\), but the total probability mass may be less than \(1\) to account for a probability of failure. \qedhere
  \end{itemize}
\end{proof}

\section{Causal Processes are Streams}%
\label{sec:processes-are-streams}

This section proves the isomorphism between \kl{causal processes} over a \kl{copy-discard category} \((\cat{V}, \cat{P})\) and \kl{effectful streams} over the effectful triple \((\cat{V}, \totals{\cat{P}}, \cat{P})\), where \intro[totals subcategory]{} \(\totals{\cat{P}}\) is the subcategory of \(\cat{P}\) of \kl{total} morphisms,
\[\Causal(\cat{V}, \cat{P}) \iso \Stream(\cat{V}, \totals{\cat{P}}, \cat{P}),\]
when the \kl{copy-discard category} \((\cat{V}, \cat{P})\) has \kl{conditionals} and \kl{ranges}.
The idea behind the isomorphism is that \kl{causal processes} give a normal form for \kl{effectful streams}, where the memory stores all the past inputs and outputs.
\kl{Conditionals} ensure that this normal form exists, while \kl{ranges} ensure that it is unique.

\subsection{Ranges}

The second sufficient condition for the isomorphism between \kl{effectful streams} and \kl{causal processes} (\Cref{th:streams-are-processes}) is the existence of \kl{ranges}.
The conditions for \kl{ranges} enforce that the range of the marginal \(f \dcomp \proj{A}\) of a morphism \(f \colon X \to A \tensor B\)\ \eqref{ax:range-1} does not modify the behaviour of the marginal and\ \eqref{ax:range-2} equalises all conditionals of \(f\).

\begin{defi}[Range]%
  \label{def:ranges}%
  A \intro{range} of a morphism $m \in \cat{P}(X;A)$ is a tuple $(R,r,i)$ where
  $R \in \obj{\cat{P}}$, $r \in \cat{P}(A \tensor X; R)$ is deterministic and $i
  \in \cat{V}(R; A \tensor X)$ such that
  \begin{enumerate}
    \item $m \condcomp \id{A \tensor X} = m \condcomp (r \dcomp i)$;\label{ax:range-1}
    \item for all $c,d \in \cat{P}(A \tensor X \tensor Y; B)$, if  $m \condcomp
    c = m \condcomp d$, then $(i \tensor \id{Y}) \dcomp c = (i \tensor \id{Y})
    \dcomp d$.\label{ax:range-2}
  \end{enumerate}
  A \kl{copy-discard category} \emph{has ranges} if there exists a \kl{range} for
  every morphism.
\end{defi}

The \kl{copy-discard categories} that we considered so far all have \kl{ranges}.

\begin{lem}%
  \label{lem:ranges}
  The \kl{copy-discard categories} from \Cref{ex:conditionals} have \kl{ranges}.
  We write a \kl{range} \((R,r,i)\) of a morphism \(m \colon X \to A\)
  explicitly in \Cref{fig:ranges}, where we always take \(R\) to be a subset of
  \(A \times X\) and \(i \colon R \to A \times X\) to be the inclusion, where
  \(\dirac{x}\) denotes the \intro{Dirac delta}, \(\dirac{x}(x) = 1\) and
  \(\dirac{x}(x') = 0\) for \(x \neq x'\).
  \begin{figure}[h!]
  \begin{tabular}{| l | l l |}
    \hline
    \(\Set\) & \(R = \{(a,x) \in A \times X \st m(x) = a\}\) & \(r(a,x) = (m(x),x)\) \\[0.5em]
    \(\Par\) & \(R = \{(a,x) \in A \times X \st m(x) = a\}\) & \(r(a,x) = \begin{cases} (a,x) & \text{if } a = m(x)\\ \bot & \text{otherwise}\end{cases}\) \\[1.5em]
    \(\Rel\) & \(R = \{(a,x) \in A \times X \st a \in m(x)\}\) & \(r(a,x) = \begin{cases}
        \{(a,x)\} & \text{if } a \in m(x)\\
        \emptyset & \text{otherwise}
      \end{cases}\) \\[1.5em]
    \(\Stoch\) & \(R = \{(a,x) \in A \times X \st m(a \given x) > 0\}\) & \makecell{\(r(a,x) = {\begin{cases} \dirac{(a,x)} & \text{if } m(a \mid x) > 0 \\ \dirac{(a_{x},x)} & \text{otherwise} \end{cases}}\)\\
    for some \(a_{x}\) with \(m(a_{x} \given x) >0\)} \\[1.5em]
    \(\subStoch\) & \(R = \{(a,x) \in A \times X \st m(a \given x) > 0\}\) & \(r(a,x) = \begin{cases} \dirac{(a,x)} & \text{if } m(a \mid x) > 0 \\ \dirac{\bot} & \text{otherwise}\end{cases}\) \\[1.5em]
    \hline
  \end{tabular}
  \caption{Ranges in the example copy-discard categories.}\label{fig:ranges}
  \end{figure}
\end{lem}
\begin{proof}
  We prove the axioms of ranges for those defined in \Cref{fig:ranges}.
  \begin{itemize}
    \item[] (\(\Set\))
          We check the first axiom:
          \(m \condcomp (r \dcomp i) (x) = (m(x), m(x), x) = m \condcomp \id{} (x)\).
          For the second axiom, if \(m \condcomp c = m \condcomp d\), then \(c(m(x),x,y) = d(m(x), x, y)\) for all \(x \in X\) and \(y \in Y\).
          This means that \(c\) and \(d\) coincide on \(R\), i.e.\ that \((i \times \id{Y}) \dcomp c = (i \times \id{Y}) \dcomp d\).
          More in general, one can also construct ranges in any \kl{cartesian category} as in \Cref{lemma:cartesian-ranges}.
    \item[] (\(\Par\))
          The ranges are defined as in \(\Rel\).
          Since \(\Par\) is a monoidal subcategory of \(\Rel\), these ranges also satisfy the same properties, which we now check.
    \item[] (\(\Rel\))
          The relation \(r\) is deterministic and the  relation \(i\) is deterministic and total by definition.
          We check the first condition for ranges as follows.
          \begin{align*}
            m \condcomp (r \dcomp i) (x)
            = \{(a,a',x') \st a \in m(x) \land (x',a') \in r(a,x)\}
            = \{(a,a,x) \st a \in m(x)\}
            = m \condcomp \id{} (x)
          \end{align*}
          Similarly, we check the last condition.
          Suppose that \(m \condcomp c = m \condcomp d\).
          Then, for all \(x \in X\) and \(y \in Y\),
          \begin{equation*}
            \ \ \{(a,b) \in A \times B \st a \in m(x) \land b \in c(a,x,y)\}
            = \{(a,b) \in A \times B \st a \in m(x) \land b \in d(a,x,y)\} ,
          \end{equation*}
          which implies that, for all \(x \in X\),
          \begin{equation*}
            \{(x,a,b) \st (x,a) \in R \land b \in c(a,x,y)\}
            = \{(x,a,b) \st (x,a) \in R \land b \in d(a,x,y)\} .
          \end{equation*}
          This corresponds to saying that \((\id{} \times i) \dcomp c = (\id{} \times i) \dcomp d\).
    \item[] (\(\Stoch\))
          This is essentially~\cite[Proposition~9.9]{coinductivestreams}, but the definition of \kl{ranges} in this text differs slightly, so we rewrite the proof for clarity.
          We check the first axiom of ranges for the pairs \((a',x')\) such that \(m(a' \given x') > 0\) and for those such that \(m(a' \given x') = 0\) separately.
          \begin{align*}
            & \text{if } m(a' \given x') > 0, && \text{if } m(a' \given x') = 0,\\
            & m \condcomp (r \dcomp i) (a,a',x' \mid x) && m \condcomp (r \dcomp i) (a,a',x' \mid x)\\
            & = m(a \mid x) \cdot \dirac{(a,x)}(a',x') && = m(a \mid x) \cdot \dirac{(a_{x},x)}(a',x')\\
            & = m \condcomp \id{} (a,a',x' \mid x) && = m \condcomp \id{} (a,a',x' \mid x)
          \end{align*}
          For the second axiom, suppose that \(m \condcomp c = m \condcomp d\).
          Then, for all \(x \in X\), \(y \in Y\), \(a \in A\) and \(b \in B\),
          \[m(a \mid x) \cdot c(b \mid a,x,y) = m(a \mid x) \cdot d(b \mid a,x,y) ,\]
          which implies that, if \(m(a \mid x) > 0\), then \(c(b \mid a,x,y) = d(b \mid a,x,y)\).
          This means that, if \((a,x) \in R\), then \(c(b \mid a,x,y) = d(b \mid a,x,y)\).
          We can conclude that \((i \times \id{Y}) \dcomp c = (i \times \id{Y}) \dcomp d\).
    \item[] (\(\subStoch\))
          The morphism \(r\) is deterministic and the morphism \(i\) is deterministic and total by definition.
          We check the first condition for ranges in the case in which the computation succeeds and the one in which it fails.
          \begin{align*}
            & m \condcomp (r \dcomp i) (a,a',x' \mid x) && m \condcomp (r \dcomp i) (\bot \mid x)\\
            & = m(a \mid x) \cdot \dirac{(a,x)}(a',x') && = m(\bot \mid x) + \textstyle\sum_{a \in A} m(a \mid x) \cdot r(\bot \given a,x)\\
            & = m \condcomp \id{} (a,a',x' \mid x) && = m \condcomp \id{} (\bot \mid x)
          \end{align*}
          Similarly, we check the last condition.
          Suppose that \(m \condcomp c = m \condcomp d\).
          Then, for all \(x \in X\), \(y \in Y\), \(a \in A\) and \(b \in B\),
          \[m(a \mid x) \cdot c(b \mid a,x,y) = m(a \mid x) \cdot d(b \mid a,x,y) ,\]
          which implies that, if \(m(a \mid x) > 0\), then \(c(b \mid a,x,y) = d(b \mid a,x,y)\).
          This means that, if \((a,x) \in R\), then \(c(b \mid a,x,y) = d(b \mid a,x,y)\).
          We conclude that \((i \times \id{Y}) \dcomp c = (i \times \id{Y}) \dcomp d\). \qedhere
  \end{itemize}
\end{proof}

\begin{lem}%
  \label{lemma:cartesian-ranges}%
    Any \kl{cartesian category} has \kl{ranges}.
\end{lem}
\begin{proof}
  For a morphism \(m \colon X \to A\), a candidate range is simply \(r \defn \proj{X}\) and \(i \defn \cp \dcomp (m \times \id{})\) because \(m\) is both \kl{total} and \kl{deterministic}.
  By \kl{determinism}, we obtain the first axiom of ranges: \(m \condcomp (r \dcomp i) = m \condcomp (\proj{X} \dcomp \cp \dcomp (m \times \id{X})) = \cp \dcomp (\cp \times \id{}) \dcomp (m \times m \times \id{}) = m \condcomp \id{A \times X}\).
  For the second axiom, suppose that \(m \condcomp c = m \condcomp d\);
  by discarding the first output, we obtain that \((i \times \id{}) \dcomp c = (i \times \id{}) \dcomp d\).
\end{proof}

\kl{Conditionals} and \kl{ranges} already give us the first step towards the normal form for \kl{effectful streams}.

\begin{lem}%
  \label{lemma:quasi-total-conditionals-total-on-range}
  Let \(f \colon X \to A \tensor B\) be a morphism in a \kl{copy-discard category} \((\cat{V},\cat{P})\) with \kl{conditionals} and \kl{ranges}.
  Then, all its \kl{conditionals} \(c\) are total on its \kl{range}, i.e.~the composition \(i \dcomp c\) is \kl{total}, for some range \((R,r,i)\) of the marginal \(f \dcomp \proj{A}\).
\end{lem}
\begin{proof}
  For a \kl{conditional} \(c \colon A \tensor X \to B\) of \(f\), we obtain \((f \dcomp \proj{A}) \condcomp \discard_{A \tensor X} = (f \dcomp \proj{A}) \condcomp (c \dcomp \discard_{B})\).
  \begin{gather*}
    \qtconditionalstotalonrangeProofFigFirst
    = \qtconditionalstotalonrangeProofFigSecond
    = \qtconditionalstotalonrangeProofFigThird
  \end{gather*}
  By the property~\eqref{ax:range-2} of ranges and \kl{totality} of \(i\), we obtain the thesis,
  \(i \dcomp c \dcomp \discard = i \dcomp \discard = \discard\).
\end{proof}

Thanks to \kl{ranges}, we can slide \kl{conditionals} on the memory of \kl{effectful streams}.

\begin{cor}%
  \label{cor:sliding-qt-conditionals}
  Consider a morphism \(\now{f} \colon \now{\stream{X}} \to \memory{f} \tensor \now{\stream{Y}}\) in a \kl{copy-discard category} \((\cat{V},\cat{P})\) with \kl{conditionals} and \kl{ranges}.
  A morphism \(\langle \now{f} \mid \later{f} \rangle \colon \stream{X} \to \stream{Y}\) in \(\Stream(\cat{V}, \totals{\cat{P}}, \cat{P})\) with memory \(\memory{f}\) has a representative with memory \(\now{\stream{X}} \tensor \now{\stream{Y}}\):
  \[\langle 
    (\now{f} \dcomp \proj{\now{\stream{Y}}}) \condcomp \id{} \mid m
    \latercomp \later{f} 
  \rangle = 
  \langle
    \now{f} \mid \later{f} 
  \rangle,
  \] 
  for a conditional \(m \colon \now{\stream{X}} \tensor \now{\stream{Y}} \to
  \memory{f}\) of \(\now{f}\).
\end{cor}
\begin{proof}
  Let us compute the representative.
  \begin{align*}
    & \langle (\now{f} \dcomp \proj{\now{\stream{Y}}}) \condcomp \id{} \mid m \latercomp \later{f} \rangle && \\
    & = \langle (\now{f} \dcomp \proj{\now{\stream{Y}}}) \condcomp (r \dcomp i) \mid m \latercomp \later{f} \rangle && \text{(\kl{Ranges})}\\
    & = \langle (\now{f} \dcomp \proj{\now{\stream{Y}}}) \condcomp r \mid i \latercomp (m \latercomp \later{f}) \rangle && \text{(\kl[stream dinaturality]{Dinaturality})}\\
    & = \langle (\now{f} \dcomp \proj{\now{\stream{Y}}}) \condcomp r \mid (i \dcomp m) \latercomp \later{f} \rangle && \text{(\Cref{lemma:monoidal-action-streams})}\\
    & = \langle (\now{f} \dcomp \proj{\now{\stream{Y}}}) \condcomp (r \dcomp i \dcomp m) \mid \later{f} \rangle && \text{(\Cref{lemma:quasi-total-conditionals-total-on-range})}\\
    & = \langle (\now{f} \dcomp \proj{\now{\stream{Y}}}) \condcomp m \mid \later{f} \rangle && \text{(\kl{Ranges})}\\
    & = \langle \now{f} \mid \later{f} \rangle. && \text{(\(m\) conditional of \(\now{f}\))} \qedhere
  \end{align*}
\end{proof}

\begin{rem}%
  \label{rem:range-composition}
  Ranges are idempotent: for two ranges \((R,r,i)\) and \((S,s,j)\) of a morphism \(m\), we have \(i \dcomp s \dcomp j = i\) because, by property~\eqref{ax:range-1} of \kl{ranges} for \((S,s,j)\), \(m \condcomp (s \dcomp j) = m \condcomp \id{}\) and, by property~\eqref{ax:range-2} of \kl{ranges} for \((R,r,i)\), this implies that \(i \dcomp s \dcomp j = i\).
\end{rem}

\subsection{Conditional sequences}

\kl{Conditional sequences} are the normal form of \kl{effectful streams} where the memory stores all the past inputs and outputs.
They are defined coinductively and quotiented by \kl{conditional equivalence}: two \kl{conditional sequences} should be considered equal if their tails are equal up to a \kl{range} of their head.

\begin{defi}%
  \label{def:raw-conditional-sequence}%
  A \intro{raw conditional sequence} $c \in \rcSeq(\stream{X}; \stream{Y})$ in a \kl{copy-discard category} $(\cat{V},\cat{P})$ is given by
  \begin{itemize}
    \item $\now{c} \in \cat{P}(\now{\stream{X}}; \now{\stream{Y}})$, the \emph{head};
    \item $\later{c} \in \rcSeq( (\now{\stream{X}} \tensor \now{\stream{Y}}) · \later{\stream{X}}; \later{\stream{Y}} )$,
   the \emph{tail}.
  \end{itemize}
\end{defi}

\begin{defi}%
  \label{def:action-conditional-seq}%
  \label{def:action-conditional-equiv}%
  The \intro[platercomp]{operation} \((\platercomp)\) and the relation \intro[conditional equivalence]{\((\cequiv)\)} are defined by mutual coinduction.
  For \(u \colon N \to M\) in \(\cat{P}\) and \(c, d ፡ M \platercomp \stream{X} → \stream{Y}\), we define \(u \platercomp c\) and \(c \cequiv d\) as
  \begin{align*}
    \now{(M \platercomp \stream{X})} & \defn M \tensor \now{\stream{X}} & \now{(u \platercomp c)} & \defn (u \tensor \id{}) \dcomp \now{c} & \now{c} &= \now{d}\\
    \later{(M \cdot \stream{X})} & \defn \later{\stream{X}} & \later{(u \platercomp c)} & \defn \latercomplaterdefFig{m} \platercomp \later{c} & i \platercomp \later{c} &\cequiv i \platercomp \later{d}
  \end{align*}
  for some bayesian inverse \(m \colon N \tensor \now{\stream{X}} \tensor \now{\stream{Y}} \to M \tensor \now{\stream{X}}\) of \(\now{c}\) with respect to \(u \tensor \id{}\),
  \[\latercompconditionaldefOneFig{u}{\now{c}} \ = \ \latercompconditionaldefTwoFig{u}{\now{c}}{m} ,\]
  and some \kl{range} $(R,r,i)$ of $\now{c}$.
\end{defi}

\kl{Conditional sequences} are the quotient of \kl{raw conditional sequences} by \((\cequiv)\).
Before giving this definition, we need to show that \((\cequiv)\) is an equivalence relation.
The proof requires mutual coinduction with the proof of properties of the operation \((\platercomp)\).

\begin{lem}%
  \label{lemma:quotient-conditional-sequences}%
  \label{lemma:action-conditionalseq-compositions}%
  \label{lemma:action-conditionalseq-identities}%
  \label{lemma:action-conditionalseq-well-def}%
  The relation \((\cequiv)\) on \kl{conditional sequences} is an equivalence relation.
  Tensoring of \kl{conditional sequences}, \((\platercomp)\), is well-defined on \((\cequiv)\)-equivalence classes and it is an action,
  \[v \platercomp (u \platercomp c) \cequiv (v \dcomp u) \platercomp c \qquad \text{and} \qquad \id{} \platercomp c \cequiv c.\]
\end{lem}
\begin{proof}
  We start by proving that \((\cequiv)\) is symmetric.
  Consider \(c, d \in \rcSeq(M \platercomp \stream{X}, \stream{Y})\) and suppose that \(c \cequiv d\).
  Then,
  \begin{align*}
    \now{c} & = \now{d} & i \platercomp \later{c} & \cequiv i \platercomp \later{d}
  \end{align*}
  By coinductive hypothesis, we obtain that \(i \platercomp \later{d} \cequiv i \platercomp \later{c}\) and, by symmetry of equality, that \(\now{d} = \now{c}\).
  Then, \(d \cequiv c\) by definition.

  We are left to show that
  \begin{enumerate}
    \item\label{p:refl} The relation \((\cequiv)\) is reflexive;
    \item\label{p:trans} The relation \((\cequiv)\) is transitive;
    \item\label{p:action-cond} For \(c \in \rcSeq(M \platercomp \stream{X}, \stream{Y})\), for \(u \colon N \to M\) and for two bayesian inverses \(m,n \colon N \tensor \now{\stream{X}} \tensor \now{\stream{Y}} \to M \tensor \now{\stream{X}}\) of \(\now{c}\) with respect to \(u \tensor \id{\now{\stream{X}}}\), the definition of \(u \platercomp c\) with \(m\) is \kl{conditionally equivalent} to the definition of \(u \platercomp c\) with \(n\);
    \item\label{p:action-equiv} If \(c \cequiv d\), then \(u \platercomp c \cequiv u \platercomp d\);
    \item\label{p:action-comp} For \(c \in \rcSeq(M \platercomp \stream{X}, \stream{Y})\), \(u \colon N \to M\) and \(v \colon P \to N\), \(v \platercomp (u \platercomp c) \cequiv (v \dcomp u) \platercomp c\);
    \item\label{p:action-id} For \(c \in \rcSeq(M \platercomp \stream{X}, \stream{Y})\), \(\id{} \platercomp c \cequiv c\).
  \end{enumerate}
  We prove these by mutual coinduction.
  Consider \(c, c', d, e \in \rcSeq(M \platercomp \stream{X}, \stream{Y})\), \(u \colon N \to M\), \(v \colon P \to N\) and two bayesian inverses \(m,n \colon N \tensor \now{\stream{X}} \tensor \now{\stream{Y}} \to M \tensor \now{\stream{X}}\) of \(\now{c}\) with respect to \(u \tensor \id{\now{\stream{X}}}\).
  Suppose that \(c = c'\), \(c \cequiv d\) and \(d \cequiv e\).
  We start by looking at the heads.
  \begin{enumerate}
    \item Since \(c = c'\), then \(\now{c} = \now{c'}\).
    \item Since \(c \cequiv d\) and \(d \cequiv e\), then \(\now{c} = \now{d} = \now{e}\); by transitivity of equality, \(\now{c} = \now{e}\).
    \item The head of \(u \platercomp c\) does not depend on the choice of bayesian inverse, thus \(\now{(u \platercomp c)}_{m} \defn (u \tensor \id{}) \dcomp \now{c} \codefn \now{(u \platercomp c)}_{n}\).
    \item Since \(c \cequiv d\), then \(\now{c} = \now{d}\) and, by definition of \((\platercomp)\), we obtain that
          \[\now{(u \platercomp c)} = (u \tensor \id{}) \dcomp \now{c} = (u \tensor \id{}) \dcomp \now{d} = \now{(u \platercomp d)}.\]
    \item By definition of \((\platercomp)\), we obtain that
          \[\now{(v \platercomp (u \platercomp c))} = (v \tensor \id{}) \platercomp \now{(u \platercomp c)} = ((v \dcomp u) \tensor \id{}) \platercomp \now{c} = \now{((v \dcomp u) \platercomp c)}.\]
    \item By definition of \((\platercomp)\), we obtain that \(\now{(\id{} \platercomp c)} = (\id{} \tensor \id{}) \dcomp \now{c} = \now{c}\).
  \end{enumerate}
  We now look at the tails.
  \begin{enumerate}
    \item Consider a range \((R,r,i)\) of \(\now{c}\).
          By coinductive hypothesis for~\eqref{p:refl}, \(\later{c} \cequiv \later{c'}\); by coinductive hypothesis for~\eqref{p:action-equiv}, we obtain \(i \platercomp \later{c} \cequiv i \platercomp \later{c'}\).
          We have already shown that the heads coincide, then \(c \cequiv c'\).
    \item Since \(c \cequiv d\) and \(d \cequiv e\), then \(i \platercomp \later{c} \cequiv i \platercomp \later{d}\) and \(j \platercomp \later{d} \cequiv j \platercomp \later{e}\) for some ranges \((R,r,i)\) and \((S,s,j)\) of \(\now{c}\).
          By \Cref{rem:range-composition}, \(i \dcomp s \dcomp j = i\); by coinductive hypothesis for~\eqref{p:action-equiv} we obtain that \((i \dcomp s) \platercomp (j \platercomp \later{d}) \cequiv (i \dcomp s) \platercomp (j \platercomp \later{e})\); and by coinductive hypothesis for~\eqref{p:action-comp},
          \begin{gather*}
            (i \dcomp s) \platercomp (j \platercomp \later{d}) \cequiv (i \dcomp s \dcomp j) \platercomp \later{d} = i \platercomp \later{d}\\
            (i \dcomp s) \platercomp (j \platercomp \later{e}) \cequiv (i \dcomp s \dcomp j) \platercomp \later{e} = i \platercomp \later{e}.
          \end{gather*}
          By coinductive hypothesis for~\eqref{p:trans} and by symmetry of \((\cequiv)\), we obtain \(i \platercomp \later{d} \cequiv i \platercomp \later{e}\) and then \(i \platercomp \later{c} \cequiv i \platercomp \later{e}\).
          Since we have already seen that the heads coincide, we obtain \(c \cequiv e\).
    \item For a \kl{range} \((S,s,j)\) of \((u \tensor \id{}) \dcomp \now{c}\), we have that \(j \dcomp m = j \dcomp n\).
          We indicate with \((u \platercomp c)_{m}\) the \kl{raw conditional sequence} obtained by choosing \(m\) as conditional in the definition of \((\platercomp)\); then we obtain:
          \begin{align*}
            & j \platercomp \later{(u \platercomp c)}_{m} &&\\
            & = j \platercomp (((\id{} \tensor \cp) \dcomp (m \tensor \id{})) \platercomp \later{c})&& \text{(Definition of \((\platercomp)\))} \\
            & \cequiv (j \dcomp (\id{} \tensor \cp) \dcomp (m \tensor \id{})) \platercomp \later{c} && \text{(Coinduction for~\eqref{p:action-comp})}\\
            & = (\cp \dcomp ((j \dcomp m) \tensor (j \dcomp \proj{\now{\stream{Y}}}))) \platercomp \later{c} &&\text{(\kl{Determinism} of \(j\))} \\
            & = (\cp \dcomp ((j \dcomp n) \tensor (j \dcomp \proj{\now{\stream{Y}}}))) \platercomp \later{c} && \text{(\kl{Ranges})}\\
            & = (j \dcomp (\id{} \tensor \cp) \dcomp (n \tensor \id{})) \platercomp \later{c} &&\text{(\kl{Determinism} of \(j\))} \\
            & \cequiv j \platercomp (((\id{} \tensor \cp) \dcomp (m \tensor \id{})) \platercomp \later{c}) && \text{(Coinduction for~\eqref{p:action-comp})}\\
            & = j \platercomp \later{(u \platercomp c)}_{n} && \text{(Definition of \((\platercomp)\))}
          \end{align*}
          and, by coinductive hypothesis for~\eqref{p:trans}, we obtain that \(j \platercomp \later{(u \platercomp c)}_{m} \cequiv j \platercomp \later{(u \platercomp c)}_{n}\).
          We have already seen that the heads coincide, then \((u \platercomp c)_{m} \cequiv (u \platercomp c)_{n}\).
    \item Since \(c \cequiv d\), there is a range \((R,r,i)\) of \(\now{c}\) such that \(i \platercomp \later{c} \cequiv i \platercomp \later{d}\).
          Let \((S,s,j)\) be a range of \((u \tensor \id{}) \dcomp \now{c} =  (u \tensor \id{}) \dcomp \now{d}\), and \(m\) be a conditional of \((u \tensor \id{}) \dcomp \cp \dcomp (\id{} \tensor \now{c}) = (u \tensor \id{}) \dcomp \cp \dcomp (\id{} \tensor \now{d})\).
          Then, \((\id{} \tensor \cp) \dcomp (m \tensor \id{}) \dcomp r \dcomp i\) is also a conditional of \((u \tensor \id{}) \dcomp \cp \dcomp (\id{} \tensor \now{c})\) and it must coincide with \(m\) on the range \(j\). We obtain:
          \begin{align*}
            & j \platercomp \later{(u \platercomp c)}  &&\\
            & = j \platercomp (((\id{} \tensor \cp) \dcomp (m \tensor \id{})) \platercomp \later{c}) && \text{(Definition of \((\platercomp)\))}\\
            & \cequiv (j \dcomp (\id{} \tensor \cp) \dcomp (m \tensor \id{})) \platercomp \later{c} && \text{(Coinduction for~\eqref{p:action-comp})}\\
            & = (j \dcomp (\id{} \tensor \cp) \dcomp (m \tensor \id{}) \dcomp r \dcomp i) \platercomp \later{c} && \text{(\kl{Ranges})}\\
            & \cequiv (j \dcomp (\id{} \tensor \cp) \dcomp (m \tensor \id{}) \dcomp r) \platercomp (i \platercomp \later{c}) && \text{(Coinduction for~\eqref{p:action-comp})}\\
            & \cequiv  (j \dcomp (\id{} \tensor \cp) \dcomp (m \tensor \id{}) \dcomp r) \platercomp (i \platercomp \later{d})  && \text{(Coinduction for~\eqref{p:action-equiv})}\\
            & \cequiv (j \dcomp (\id{} \tensor \cp) \dcomp (m \tensor \id{}) \dcomp r \dcomp i) \platercomp \later{d} && \text{(Coinduction for~\eqref{p:action-comp})}\\
            & =  (j \dcomp (\id{} \tensor \cp) \dcomp (m \tensor \id{})) \platercomp \later{d} && \text{(\kl{Ranges})}\\
            & \cequiv j \platercomp (((\id{} \tensor \cp) \dcomp (m \tensor \id{})) \platercomp \later{d}) && \text{(Coinduction for~\eqref{p:action-comp})}\\
            & = j \platercomp \later{(u \platercomp d)}  && \text{(Definition of \((\platercomp)\))}
          \end{align*}
          and, by coinductive hypothesis for~\eqref{p:trans}, we obtain that \(j \platercomp \later{(u \platercomp c)} = j \platercomp \later{(u \platercomp d)}\).
          Since we have already shown that \(\now{(u \platercomp c)} = \now{(u \platercomp d)}\), we obtain \(u \platercomp c \cequiv u \platercomp d\).
    \item Let \(i ፡ R → L \tensor \now{\stream{X}} \tensor \now{\stream{Y}}\) be a range of \(((v \dcomp u) \tensor \id{}) \dcomp \now{c}\).
          Let \(n \colon M \tensor \now{\stream{X}} \tensor \now{\stream{Y}} \to N \tensor \now{\stream{X}}\), \(m \colon L \tensor \now{\stream{X}} \tensor \now{\stream{Y}} \to M \tensor \now{\stream{X}}\) and \(l \colon L \tensor \now{\stream{X}} \tensor \now{\stream{Y}} \to N \tensor \now{\stream{X}}\) be conditionals of (\ref{eq:compositions-n}), (\ref{eq:compositions-m}) and (\ref{eq:compositions-l}), respectively.
          \begin{align}
            \conditionalscompositionsnFig{}  \label{eq:compositions-n} \\
            \conditionalscompositionsmFig{} \label{eq:compositions-m} \\
            \conditionalscompositionslFig{} \label{eq:compositions-l}
          \end{align}
          Then, \((\id{} \tensor \cp) \dcomp (m \tensor \id{}) \dcomp n\) is also a conditional of (\ref{eq:compositions-l}) and \(i \dcomp l = i \dcomp (\id{} \tensor \cp) \dcomp (m \tensor \id{}) \dcomp n\).
          We obtain:
          \begin{align*}
            & i \platercomp \later{(v \platercomp (u \platercomp c))}\\
            &= i \platercomp (((\id{} \tensor \cp) \dcomp (m \tensor \id{})) \platercomp \later{(u \platercomp c)}) && \text{(Definition of \((\platercomp)\))} \\
            & = i \platercomp (((\id{} \tensor \cp) \dcomp (m \tensor \id{})) \platercomp (((\id{} \tensor \cp) \dcomp (n \tensor \id{})) \platercomp \later{c})) && \text{(Definition of \((\platercomp)\))}\\
            &\cequiv  (i \dcomp (\id{} \tensor \cp) \dcomp (((\id{} \tensor \cp) \dcomp (m \tensor \id{}) \dcomp n) \tensor \id{})) \platercomp \later{c} && \text{(Coinduction for~\eqref{p:action-comp})}\\
            & = (\cp \dcomp ((i \dcomp (\id{} \tensor \cp) \dcomp (m \tensor \id{}) \dcomp n) \tensor (i \dcomp \proj{\now{\stream{Y}}} ))) \platercomp \later{c} && \text{(\kl{Determinism} of \(i\))}\\
            & = (\cp \dcomp ((i \dcomp l) \tensor (i \dcomp \proj{\now{\stream{Y}}} ))) \platercomp \later{c} && \text{(\kl{Ranges})}\\
            & = (i \dcomp (\id{} \tensor \cp) \dcomp (l \tensor \id{})) \platercomp \later{c} && \text{(\kl{Determinism} of \(i\))}\\
            & \cequiv i \platercomp (((\id{} \tensor \cp) \dcomp (l \tensor \id{})) \platercomp \later{c})  && \text{(Coinduction for~\eqref{p:action-comp})}\\
            & =  i \platercomp \later{((v \dcomp u) \platercomp c)} && \text{(Definition of \((\platercomp)\))}
          \end{align*}
          and, by coinductive hypothesis for~\eqref{p:trans}, \(i \platercomp \later{(v \platercomp (u \platercomp c))} \cequiv i \platercomp \later{((v \dcomp u) \platercomp c)}\).
          Since we have already seen that \(\now{(v \platercomp (u \platercomp c))} =  \now{((v \dcomp u) \platercomp c)}\), we obtain that \((v \platercomp (u \platercomp c)) \cequiv ((v \dcomp u) \platercomp c)\).
    \item A conditional of \(\cp \dcomp (\id{} \tensor \now{c})\) is \(\id{M \tensor \now{\stream{X}}} \tensor \discard_{\now{\stream{Y}}}\), then we can choose a \kl{raw conditional sequence} for \(\id{} \platercomp c\) that coincides with \(c\):
          \begin{align*}
            & \later{(\id{} \platercomp c)} &&\\
            & = ((\id{} \tensor \cp) \dcomp (\id{} \tensor \discard \tensor \id{})) \platercomp \later{c} &&\text{(Definition of \((\platercomp)\))}\\
            & = \id{} \platercomp \later{c} && \text{(Counitality)}\\
            & = \later{c} && \text{(Coinduction for~\eqref{p:action-id})}
          \end{align*}
          and, since we have already shown that \(\now{(\id{} \platercomp c)} = \now{c}\), we obtain that \(\id{} \platercomp c = c\); reflexivity of \((\cequiv)\)~\eqref{p:refl}, gives us \(\id{} \platercomp c \cequiv c\).
          \qedhere
  \end{enumerate}
\end{proof}

\Cref{lemma:quotient-conditional-sequences} lets us quotient \kl{raw conditional sequences} by \kl{conditional equivalence}.

\begin{defi}%
  \label{def:conditional-sequences}%
  \intro{Conditional sequences} are equivalence classes of raw conditional sequences under conditional equivalence,
  \[
  \cSeq{}(\stream{X}, \stream{Y}) = \rcSeq(\stream{X}, \stream{Y}) \big/ (\cequiv).
  \]
\end{defi}

The next results prove that \kl{conditional sequences} are isomorphic to \kl{causal processes}.
This is done in two steps: \Cref{lemma:inductive-conditional-sequences} gives an inductive presentation of \kl{conditional sequences};  \Cref{prop:coinductive-conditional-sequences} shows an isomorphism between \kl{causal processes} and \kl{inductive conditional sequences}.

\begin{lem}%
  \label{lemma:inductive-conditional-sequences}%
  \kl{Conditional sequences} have an \intro[inductive conditional sequences]{inductive presentation} as sequences \(\{c_{n} \colon X_{0} \tensor Y_{0} \tensor \cdots \tensor X_{n-1} \tensor Y_{n-1} \tensor X_{n} \to Y_{n}\}\) quotiented by equivalent conditionals: \(\{c_{n}\} \iequiv \{d_{n}\}\) if and only if \(p_{0} = c_{0} = d_{0}\) and \(p_{n+1} = p_{n} \condcomp c_{n+1} = p_{n} \condcomp d_{n+1}\).
\end{lem}
\begin{proof}
  Let $(\cat{V},\cat{P})$ be a \kl{copy-discard category}.
  Given an \kl{inductive conditional sequence} \(\{c_{n}\}\), we define one by coinduction, as \intro[cseq]{\(\now{(\cseq(\{c_{n}\}))} = c_{0}\) and \(\later{(\cseq(\{c_{n}\}))} = \cseq(\{c_{n+1}\})\)}.
  This mapping is well-defined: let \(\{c_{n}\} \iequiv \{d_{n}\}\) be two equivalent conditional sequences, with \(p_{n}\) witnessing the equivalence.
  Then, $c_{0} = d_{0}$.
  By induction, it is easy to see that \(p_{n} = c_{0} \condcomp t_{n} = c_{0} \condcomp s_{n}\), where \(t_{0} = s_{0} = \discard_{X_{0} \tensor Y_{0}}\), \(t_{n+1} = t_{n} \condcomp c_{n+1}\) and \(s_{n+1} = s_{n} \condcomp d_{n+1}\).
  Then, for all \(n \in ℕ\), \((i \tensor \id{}) \dcomp t_{n+1} = (i \tensor \id{}) \dcomp s_{n+1}\) for a range \(i\) of \(c_{0} = d_{0}\).
  If we indicate with \intro[iplatercomp]{\(i \iplatercomp \{e_{n}\}\)} the \kl{inductive conditional sequence} defined as \((i \iplatercomp \{e_{n}\})_{n} \defn (i \tensor \id{}) \dcomp e_{n}\), we obtain that \(i \iplatercomp \{c_{n+1}\} \iequiv i \iplatercomp \{d_{n+1}\}\).
  By coinduction, the corresponding coinductive sequences are also equivalent.
  \begin{align*}
    & i \platercomp \later{(\cseq(\{c_{n}\}))}  = i \platercomp \cseq(\{c_{n+1}\}) \cequiv \cseq(i \iplatercomp \{c_{n+1}\}) \\
    & \cequiv \cseq(i \iplatercomp \{d_{n+1}\}) \cequiv i \platercomp \cseq(\{d_{n+1}\}) = i \platercomp \later{(\cseq(\{d_{n}\}))}
  \end{align*}
  This shows that \(\cseq(\{c_{n}\}) \cequiv \cseq(\{d_{n}\})\).

  Conversely, given a \kl{coinductive conditional sequence} \(c\), we define an inductive one, \intro[iseq]{\(\iseq(c) = \{c_{n}\}\)}, by induction: \(c_{0} = \now{c}\), \(t_{0} = \later{c}\), \(c_{n+1} = \now{t_{n}}\) and \(t_{n+1} = \later{t_{n}}\).
  Suppose there are two equivalent \kl{coinductive conditional sequences}, \(c \cequiv d\).
  Then \(c_{0} = \now{c} = \now{d} = d_{0}\) and \(i \platercomp \later{c} \cequiv i \platercomp \later{d}\) for a range \(i\) of \(\now{c}\).
  By coinduction, we obtain that \(i \iplatercomp \iseq(\later{c}) \iequiv \iseq(i \platercomp \later{c}) \iequiv \iseq(i \platercomp \later{d}) \iequiv i \iplatercomp \iseq(\later{d})\).
  By the properties of ranges, we obtain that \(p_{n+1} = c_{0} \condcomp t_{n+1} \iequiv c_{0} \condcomp s_{n+1} = d_{0} \condcomp s_{n+1} = q_{n+1}\), which shows that \(\iseq(c) \iequiv \iseq(d)\).

  It is easy to see that these mappings are inverses to each other and define isomorphisms.
\end{proof}

Finally, we define an identity-on-objects and isomorphism on hom-sets mapping
between \(\Causal(\cat{V},\cat{P})\) and the inductive presentation of
\(\cCausal(\cat{V},\cat{P})\) given in
\Cref{lemma:inductive-conditional-sequences}. Compositions, identities and
monoidal products in \(\cCausal(\cat{V},\cat{P})\) can be defined to make such
mapping a monoidal functor.

\begin{prop}%
  \label{prop:coinductive-conditional-sequences}%
  \kl{Conditional sequences} over a \kl{copy-discard category} $(\cat{V},\cat{P})$ with \kl{conditionals} and \kl{ranges} form a \kl{copy-discard category} isomorphic to that of \kl{causal processes}
  \[\cCausal(\cat{V},\cat{P}) \iso \Causal(\cat{V},\cat{P}).\]
\end{prop}
\begin{proof}
  We define an isomorphism between causal processes \(\{p_{n}\} \colon \stream{X} \to \stream{Y}\) and inductive conditional sequences \(\{c_{n}\} \colon \stream{X} \to \stream{Y}\) as given in \Cref{lemma:inductive-conditional-sequences}.

  Let \(\{p_{n}\} \colon \stream{X} \to \stream{Y}\) be a causal process.
  By definition of \kl{causal processes}, each component splits in terms of the previous component and a conditional.
  This splitting defines a conditional sequence \(\{c_{n} \colon X_{0} \tensor Y_{0} \tensor \cdots \tensor X_{n-1} \tensor Y_{n-1} \tensor X_{n} \to Y_{n}\}\).
  We check that this mapping is well-defined.
  Let \(\{d_{n}\}\) be another choice of conditionals for \(\{p_{n}\}\).
  These sequences are equivalent, \(\{c_{n}\} \cequiv \{d_{n}\}\), because, by their definition, \(p_{n-1} \condcomp c_{n} = p_{n} = p_{n-1} \condcomp d_{n}\) for all \(n \in \naturals\).

  Let \(\{c_{n}\} \colon \stream{X} \to \stream{Y}\) be an inductive conditional sequence.
  Define by induction \(p_{0} \defn c_{0}\) and \(p_{n+1} \defn p_{n} \condcomp c_{n+1}\).
  Then, \(\{p_{n}\}\) is a causal process.
  This mapping is well-defined because, if \(\{c_{n}\} \cequiv \{d_{n}\}\), they define the same causal process by the definition of equivalence relation on conditional sequences.

  We check that these mappings are isomorphisms.
  Clearly, if we start with a causal process \(\{p_{n}\}\), take its conditional sequence and then its causal process again, we obtain \(\{p_{n}\}\).
  Conversely, if we start with a conditional sequence \(\{c_{n}\}\), take its causal process and then its conditional sequence, we may obtain another conditional sequence \(\{d_{n}\}\).
  However, by definition of causal process, these conditional sequences are equivalent: \(\{c_{n}\} \cequiv \{d_{n}\}\).

  Thus, there is an identity-on-objects and isomorphism on hom-sets mapping between causal processes and inductive conditional sequences.
  By \Cref{lemma:inductive-conditional-sequences}, there is also an identity-on-objects and isomorphism on hom-sets mapping between inductive and coinductive conditional sequences.
  We define compositions, identities and monoidal products of inductive and coinductive conditional sequences to make these isomorphisms monoidal functors.
  Explicitly, for coinductive conditional sequences \(c \colon M \platercomp \stream{X} \to \stream{Y}\) and \(d \colon N \platercomp \stream{Y} \to \stream{Z}\), their composition \(c_{M} \dcomp d_{N} \colon (M \tensor N) \platercomp \stream{X} \to \stream{Z}\) is
  \begin{equation}%
    \label{eq:composition-conditional-sequences}
    \begin{aligned}
    \now{(c_{M} \dcomp d_{N})} & \defn \swap_{M,N} \dcomp (\id{N} \tensor \now{c}) \dcomp \now{d}\\
      \later{(c_{M} \dcomp d_{N})} &\defn ((b \condcomp \id{\now{\stream{X}} \tensor \now{\stream{Z}} \tensor \now{\stream{Y}}}) \dcomp \swap) \platercomp (\later{c}_{\now{\stream{X}} \tensor \now{\stream{Y}}} \dcomp \later{d}_{\now{\stream{Y}} \tensor \now{\stream{Z}}}) ,
    \end{aligned}
  \end{equation}
  where \(b\) is a bayesian inverse of \(\now{d}\) with respect to \(\now{c}\), \(\now{c} \dcomp \cp \dcomp (\now{d} \tensor \id{}) = (\now{c} \dcomp \now{d}) \condcomp b\).
  The identity \(\id{\stream{X}} \colon \stream{X} \to \stream{X}\) is
  \begin{equation}%
    \label{eq:identities-conditional-sequences}
    \now{\id{\stream{X}}} \defn \id{\now{\stream{X}}} \qquad
    \later{\id{\stream{X}}} \defn \discard_{\now{\stream{X}} \tensor \now{\stream{X}}} \platercomp \id{\later{\stream{X}}},
  \end{equation}
  and, for coinductive conditional sequences \(c \colon M \platercomp \stream{X} \to \stream{Y}\) and \(c' \colon M' \platercomp \stream{X'} \to \stream{Y'}\), their monoidal product \(c_{M} \tensor c'_{M'} \colon (M \tensor M') \platercomp \stream{X} \tensor \stream{X'} \to \stream{Y} \tensor \stream{Y'}\) is
  \begin{equation}%
    \label{eq:tensor-conditional-sequences}
    \begin{aligned}
      \now{(c_{M} \tensor c_{M'})} & \defn (\id{} \tensor \swap{} \tensor \id{}) \dcomp (\now{c} \tensor \now{c'}) \\
      \later{(c_{M} \tensor c_{M'})} & \defn (\id{} \tensor \swap{} \tensor \id{}) \dcomp (\later{c}_{\now{\stream{X}} \tensor \now{\stream{Y}}} \tensor \later{c'}_{\now{\stream{X'}} \tensor \now{\stream{Y'}}}).
    \end{aligned}
  \end{equation}
  These monoidal functors also preserve the copy-discard structure because copy and discard morphisms are inherited from \((\cat{V},\cat{P})\) in the same way identities are.
  Then, conditional sequences form a copy-discard category \(\cCausal(\cat{V},\cat{P})\) that is isomorphic to causal processes \(\Causal(\cat{V},\cat{P})\).
\end{proof}

\subsection{A normal form for effectful streams}%
\label{sec:normal-form-streams}

We conclude by proving that \kl{conditional sequences} over the \kl{copy-discard category} \((\cat{V}, \cat{P})\) are isomorphic to \kl{effectful streams} over the \kl{effectful triple} \((\cat{V}, \totals{\cat{P}}, \cat{P})\), where \(\totals{\cat{P}}\) denotes the subcategory of \(\cat{P}\) of \kl{total} morphisms.
We define identity-on-objects mappings between the morphisms of the two categories (\Cref{def:iso-streams-conditional-sequences}), prove that the first is a monoidal functor (\Cref{prop:functor-processes-to-stream}) and the second is its right and left inverse (\Cref{prop:processes-to-streams-faithful,prop:normal-form-streams}).

\begin{defi}%
  \label{def:iso-streams-conditional-sequences}
  For a \kl{conditional sequence} $c ፡ \stream{X} → \stream{Y}$, define the \kl{effectful stream}
  \intro[streamfun]{$\streamfun(c) \in \Stream(\stream{X}, \stream{Y})$} coinductively by
  \begin{align*}
    \memory{\streamfun(c)} &= \now{\stream{X}} \tensor \now{\stream{Y}} &
    \now{\streamfun(c)} &= \now{c} \condcomp \id{\now{\stream{X}} \tensor \now{\stream{Y}}} &
    \later{\streamfun(c)} &= \streamfun(\later{c}).
  \end{align*}
  For a stream \(s \colon \stream{X} \to \stream{Y}\), let us pick a \kl{conditional} \(m \colon \now{\stream{Y}} \tensor \now{\stream{X}} \to \memory{s}\) of \(\now{s}\)
  and define the \kl{conditional sequence} \intro[procfun]{\(\procfun(s) \in \cCausal(\stream{X}, \stream{Y})\)} coinductively by
  \begin{align*}
    \now{(\procfun(s))} &= \now{s} \dcomp \proj{\now{\stream{Y}}} &
    \later{(\procfun(s))} &= \procfun(m \latercomp \later{s}).
  \end{align*}
\end{defi}

\begin{lem}%
  \label{lemma:streamfun-morphism-monoidal-action}
  The mapping \(\streamfun\) from \kl{conditional sequences} over \((\cat{V}, \cat{P})\) to \kl{effectful streams} over \((\cat{V}, \totals{\cat{P}}, \cat{P})\) preserves the action \((\platercomp)\).
  For a \kl{conditional sequence} \(p \colon N \platercomp \stream{X} \to \stream{Y}\) and a morphism \(u \colon M \to N\) in \(\cat{P}\),
  \(u \latercomp \streamfun(p) \dinat \streamfun(u \platercomp p)\).
\end{lem}
\begin{proof}
  Proceed by coinduction.
  Let \(m \colon M \tensor \now{\stream{X}} \tensor \now{\stream{Y}} \to N \tensor \now{\stream{X}}\) be a \kl{conditional} of \((u \tensor \id{}) \dcomp \cp \dcomp (\id{} \tensor \now{c})\).
  \begin{align*}
    & \now{(u \latercomp \streamfun(c))} \\
    & = (u \tensor \id{}) \dcomp \now{\streamfun(c)}  && \text{(Definition of \((\latercomp)\))}\\
    &=  (u \tensor \id{}) \dcomp \cp \dcomp (\id{} \tensor \now{c}) \dcomp (\id{} \tensor \cp) && \text{(Definition of \(\streamfun\))}\\
    & = \cp \dcomp (\id{} \tensor ((u \tensor \id{}) \dcomp (\id{} \tensor \now{c}) \dcomp \cp)) \dcomp (m \tensor \cp) && \text{(\kl{Conditionals})}\\
    &=  \cp \dcomp (\id{} \tensor ((u \tensor \id{}) \dcomp (\id{} \tensor \now{c}) \dcomp \cp)) \dcomp ((\id{} \tensor \cp) \dcomp (m \tensor \id{})) &&  \text{(Coassociativity)}\\
    & = \cp \dcomp (\id{} \tensor (\now{(u \platercomp c)} \dcomp \cp)) \dcomp ((\id{} \tensor \cp) \dcomp (m \tensor \id{})) && \text{(\Cref{def:action-conditional-seq})}\\
    & = \now{\streamfun(u \platercomp c)} \dcomp  ((\id{} \tensor \cp) \dcomp (m \tensor \id{})) && \text{(Definition of \(\streamfun\))}
  \end{align*}
  We now compute the tails of the actions.
  \begin{align*}
    & \later{\streamfun(u \platercomp c)} &&\\
    & = \streamfun(\later{(u \platercomp c)}) && \text{(Definition of \(\streamfun\))}\\
    & = \streamfun(((\id{} \tensor \cp) \dcomp (m \tensor \id{})) \platercomp \later{c}) && \text{(Definition of \(\streamfun\))}\\
    & = ((\id{} \tensor \cp) \dcomp (m \tensor \id{})) \latercomp \streamfun(\later{c}) && \text{(Coinduction)}\\
    & = ((\id{} \tensor \cp) \dcomp (m \tensor \id{})) \latercomp \later{\streamfun(c)} && \text{(Definition of \(\streamfun\))}\\
    & = ((\id{} \tensor \cp) \dcomp (m \tensor \id{})) \latercomp \later{(u \latercomp \streamfun(c))} && \text{(Definition of \((\latercomp)\))}
  \end{align*}
  This shows that \(\streamfun(u \platercomp c) \dinat u \latercomp \streamfun(c)\) via the morphism \((\id{} \tensor \cp) \dcomp (m \tensor \id{})\).
\end{proof}

\begin{prop}[Stream of a conditional sequence]%
  \label{prop:functor-processes-to-stream}%
  For a \kl{copy-discard category} $(\cat{V},\cat{P})$ with \kl{conditionals} and \kl{ranges}, the mapping \(\streamfun\) given in \Cref{def:iso-streams-conditional-sequences} assembles into an identity-on-objects monoidal functor $\streamfun ፡ \cSeq(\cat{V},\cat{P}) → \Stream(\cat{V},\totals{\cat{P}},\cat{P})$.
\end{prop}
\begin{proof}
  The candidate functor \(\streamfun \colon \cCausal(\cat{V},\cat{P}) \to \Stream(\cat{V},\totals{\cat{P}},\cat{P})\) is the identity on objects and, for a \kl{conditional sequence} \(c \colon \stream{X} \to \stream{Y}\), is defined coinductively.
  \begin{align*}
    \memory{\streamfun(c)} & \defn \now{\stream{X}} \tensor \now{\stream{Y}} &
    \now{(\streamfun(c))} & \defn \now{p} \condcomp \id{\now{\stream{X}} \tensor \now{\stream{Y}}} &
    \later{(\streamfun(c))} & \defn \streamfun(\later{c})
  \end{align*}
  We check that this mapping is well defined.
  Suppose there are two equivalent conditional sequences \(c \cequiv d\).
  Then, \(\now{c} = \now{d}\) and \(i \platercomp \later{c} \cequiv i \platercomp \later{d}\), for a range \(r \dcomp i\) of \(\now{c} = \now{d}\).
  \begin{align*}
    & \streamfun(c) && \\
    & = \langle \now{\streamfun(c)} \mid \later{\streamfun(c)} && \\
    & = \langle \now{c} \condcomp \id{} \mid \streamfun(\later{c}) \rangle && \text{(Definition of \(\streamfun\))}\\
    & = \langle \now{c} \condcomp (r \dcomp i) \mid \streamfun(\later{c}) \rangle && \text{(\kl{Ranges})}\\
    & \dinat \langle \now{c} \condcomp r \mid i \latercomp \streamfun(\later{c}) \rangle && \text{(\kl[stream dinaturality]{Dinaturality})}\\
    & \dinat \langle \now{c} \condcomp r \mid \streamfun(i \latercomp \later{c}) \rangle && \text{(\Cref{lemma:streamfun-morphism-monoidal-action})}\\
    & \dinat \langle \now{d} \condcomp r \mid \streamfun(i \latercomp \later{d}) \rangle && \text{(Coinduction)}\\
    & \dinat \langle \now{d} \condcomp r \mid i \latercomp \streamfun(\later{d}) \rangle && \text{(\Cref{lemma:streamfun-morphism-monoidal-action})}\\
    & \dinat \langle \now{d} \condcomp (r \dcomp i) \mid \streamfun(\later{d}) \rangle && \text{(\kl[stream dinaturality]{Dinaturality})}\\
    & = \langle \now{d} \condcomp \id{} \mid \streamfun(\later{d}) \rangle && \text{(\kl{Ranges})}\\
    & = \langle \now{\streamfun(d)} \mid \streamfun(\later{d}) \rangle && \text{(Definition of \(\streamfun\))}\\
    & = \streamfun(d)
  \end{align*}
  We now check that \(\streamfun\) preserves compositions.
  Applying \(\streamfun\) to the composition of two conditional sequences \(p \colon \stream{X} \to \stream{Y}\) and \(q \colon \stream{Y} \to \stream{Z}\), we obtain
  \begin{align*}
    \memory{\streamfun(p \dcomp q)} & = \now{\stream{X}} \tensor \now{\stream{Z}} \\
    \now{\streamfun(p \dcomp q)} & = (\now{p} \dcomp \now{q}) \condcomp \id{\now{\stream{X}} \tensor \now{\stream{Z}}} \\
    \later{\streamfun(p \dcomp q)} & = \streamfun(((b \condcomp \id{}) \dcomp \swap{}) \platercomp (\later{p}_{\now{\stream{X}} \tensor \now{\stream{Y}}} \dcomp \later{q}_{\now{\stream{Y}} \tensor \now{\stream{Z}}})) \ ,
  \end{align*}
  while, applying \(\streamfun\) to \(p\) and \(q\) separately, we obtain
  \begin{align*}
    \memory{\streamfun(p) \dcomp \streamfun(q)} & = \now{\stream{X}} \tensor \now{\stream{Y}} \tensor \now{\stream{Y}} \tensor \now{\stream{Z}}\\
    \now{(\streamfun(p) \dcomp \streamfun(q))} & = (\now{p} \condcomp \id{}) \dcomp (\id{} \tensor (\now{q} \condcomp \id{}))\\
    \later{(\streamfun(p) \dcomp \streamfun(q))} & = \streamfun(\later{p})_{\now{\stream{X}} \tensor \now{\stream{Y}}} \dcomp \streamfun(\later{q})_{\now{\stream{Y}} \tensor \now{\stream{Z}}} \ .
  \end{align*}
  We show by coinduction that \(\streamfun(p \dcomp q) \dinat \streamfun(p) \dcomp \streamfun(q)\).
  \begin{align*}
    & \langle (\now{p} \dcomp \now{q}) \condcomp \id{} \mid \streamfun(((b \condcomp \id{}) \dcomp \swap{}) \platercomp (\later{p}_{\now{\stream{X}} \tensor \now{\stream{Y}}} \dcomp \later{q}_{\now{\stream{Y}} \tensor \now{\stream{Z}}})) \rangle \\
    &= \langle (\now{p} \dcomp \now{q}) \condcomp \id{} \mid ((b \condcomp \id{}) \dcomp \swap{}) \latercomp \streamfun(\later{p}_{\now{\stream{X}} \tensor \now{\stream{Y}}} \dcomp \later{q}_{\now{\stream{Y}} \tensor \now{\stream{Z}}}) \rangle &&  \text{(\Cref{lemma:streamfun-morphism-monoidal-action})} \\
    &= \langle (\now{p} \condcomp \id{}) \dcomp (\id{} \tensor (\now{q} \condcomp \id{})) \mid \streamfun(\later{p}_{\now{\stream{X}} \tensor \now{\stream{Y}}} \dcomp \later{q}_{\now{\stream{Y}} \tensor \now{\stream{Z}}}) \rangle &&  \text{(\Cref{cor:sliding-qt-conditionals})} \\
    &=  \langle (\now{p} \condcomp \id{}) \dcomp (\id{} \tensor (\now{q} \condcomp \id{})) \mid \streamfun(\later{p})_{\now{\stream{X}} \tensor \now{\stream{Y}}} \dcomp \streamfun(\later{q})_{\now{\stream{Y}} \tensor \now{\stream{Z}}} \rangle && \text{(Coinduction)}
  \end{align*}
  Similarly, we show by coinduction that \(\streamfun(\id{\stream{X}}) \dinat \id{\stream{X}}\).
  \begin{align*}
    & \streamfun(\id{\stream{X}}) &&\\
    & = \langle \id{\now{X}} \condcomp \id{} \mid \streamfun(\discard \platercomp \id{\later{\stream{X}}}) \rangle && \text{(\Cref{eq:identities-conditional-sequences})}\\
    & = \langle \id{\now{X}} \condcomp \id{} \mid \discard \latercomp \streamfun(\id{\later{\stream{X}}}) \rangle && \text{(\Cref{lemma:streamfun-morphism-monoidal-action})}\\
    & = \langle \id{\now{X}} \condcomp \discard \mid \streamfun(\id{\later{\stream{X}}}) \rangle && \text{(\kl[stream dinaturality]{Dinaturality})}\\
    & = \langle \id{\now{X}} \mid \streamfun(\id{\later{\stream{X}}}) \rangle && \text{(Counitality of \(\cp\))}\\
    & = \langle \id{\now{X}} \mid \id{\later{\stream{X}}} \rangle &&    \text{(Coinduction)}
  \end{align*}
  These show that \(\streamfun\) is a functor.
  It is also monoidal because it is the identity on objects.
\end{proof}

\begin{lem}%
  \label{lemma:procfun-morphism-monoidal-action}
  The mapping \(\procfun\) from morphisms of \(\Stream(\cat{V},\totals{\cat{P}},\cat{P})\) to morphisms of \(\cCausal(\cat{V},\cat{P})\) preserves the operation \((\latercomp)\).
  For an effectful stream \(s \colon N \latercomp \stream{X} \to \stream{Y}\) and a morphism \(u \colon M \to N\) in \(\cat{P}\),
  \[u \platercomp \procfun(s) \cequiv \procfun(u \latercomp s).\]
\end{lem}
\begin{proof}
  Proceed by coinduction.
  \begin{align*}
    & \now{(u \platercomp \procfun(s))} &&\\
    & = (u \tensor \id{}) \dcomp \now{\procfun(s)} && \text{(\Cref{def:action-conditional-seq})}\\
    & = (u \tensor \id{}) \dcomp \now{s} \dcomp (\discard \tensor \id{}) && \text{(Definition of \(\procfun\))}\\
    & = \now{(u \latercomp s)} \dcomp (\discard \tensor \id{}) && \text{(Definition of \((\latercomp)\))}\\
    & = \now{\procfun(u \latercomp s)} && \text{(Definition of \(\procfun\))}\\
  \end{align*}
  For the tail of the action, consider a conditional \(m \colon M \tensor \now{\stream{X}} \tensor \now{\stream{Y}} \to N \tensor \now{\stream{X}}\) of \((u \tensor \id{}) \dcomp \cp \dcomp (\id{} \tensor \now{\procfun(s)})\), and a conditional \(c \colon N \tensor \now{\stream{X}} \tensor \now{\stream{Y}} \to \memory{s}\) of \(\now{s}\).
  Then, \((\id{} \tensor \cp) \dcomp (m \tensor \id{}) \dcomp c\) is a conditional of \((u \tensor \id{}) \dcomp \now{s}\).
  Let \(r \dcomp i\) be a range of \((u \tensor \id{}) \dcomp \now{s} \dcomp (\discard \tensor \id{})\).
  \begin{align*}
    & \later{(u \platercomp \procfun(s))} \\
    &= ((\id{} \tensor \cp) \dcomp (m \tensor \id{})) \platercomp \later{\procfun(s)} && \text{(\Cref{def:action-conditional-seq})}\\
    &= ((\id{} \tensor \cp) \dcomp (m \tensor \id{})) \platercomp \procfun(c \latercomp \later{s}) && \text{(Definition of \(\procfun\))}\\
    & \cequiv ((\id{} \tensor \cp) \dcomp (m \tensor \id{})) \platercomp (c \platercomp \procfun(\later{s})) && \text{(Coinduction)}\\
    &\cequiv ((\id{} \tensor \cp) \dcomp (m \tensor \id{}) \dcomp c) \platercomp \procfun(\later{s}) && \text{(\Cref{lemma:action-conditionalseq-compositions})}\\
    &=  ((\id{} \tensor \cp) \dcomp (m \tensor \id{}) \dcomp c) \platercomp \procfun(\later{(u \latercomp s)}) && \text{(Definition of \((\latercomp)\))}\\
    &\cequiv \procfun(((\id{} \tensor \cp) \dcomp (m \tensor \id{}) \dcomp c) \platercomp \later{(u \latercomp s)}) && \text{(Coinduction)}\\
    &= \later{\procfun(u \latercomp s)} && \text{(\kl{Conditionals} and definition of \(\procfun\))}
  \end{align*}
  By \Cref{lemma:action-conditionalseq-well-def}, we obtain that \(i \platercomp \later{(u \platercomp \procfun(s))} \cequiv i \platercomp \later{\procfun(u \latercomp s)}\), which gives that \(u \platercomp \procfun(s) \cequiv \procfun(u \latercomp s)\).
\end{proof}

\begin{prop}%
  \label{prop:processes-to-streams-faithful}
  Let \((\cat{V},\cat{P})\) be a \kl{copy-discard category} category with \kl{conditionals} and \kl{ranges}.
  The functor \(\streamfun\) is faithful, with right inverse \(\procfun\) (\Cref{def:iso-streams-conditional-sequences}).
\end{prop}
\begin{proof}
  The candidate right inverse functor \(\procfun \colon \Stream(\cat{V},\totals{\cat{P}},\cat{P}) \to \cCausal(\cat{V},\cat{P})\) is the identity on objects and, for an effectful stream \(s \colon \stream{X} \to \stream{Y}\), is defined coinductively.
  \begin{align*}
    \now{(\procfun(s))} & \defn \now{s} \dcomp (\discard \tensor \id{}) &
    \later{(\procfun(s))} & \defn \procfun(m \latercomp \later{s})
  \end{align*}
  We check that the mapping \(\procfun\) is well-defined.
  Since we haven't shown that \(\procfun\) does not depend on the choice of conditional \(m\), we will prove that it's well-defined for any choice of such conditional.
  Suppose that there are two streams \(s \dinat s'\) that are equivalent in one step, i.e.~there is a total morphism \(u \colon \memory{s'} \to \memory{s}\) in \(\cat{P}\) such that \(\now{s'} \dcomp (u \tensor \id{}) = \now{s}\) and \(u \latercomp \later{s} \dinat \later{s'}\).
  By \kl{totality} of \(u\), we show that \(\now{\procfun(s)} = \now{\procfun(s')}\).
  \begin{align*}
    \now{\procfun(s)}
    = \now{s} \dcomp (\discard \tensor \id{})
    = \now{s'} \dcomp ((u \dcomp \discard) \tensor \id{})
    = \now{s'} \dcomp (\discard \tensor \id{})
    = \now{\procfun(s')}
  \end{align*}
  We now show that the tails are equivalent.
  By definition of \(\procfun\),
  \[\later{\procfun(s')} = \procfun(n \latercomp \later{s'}) \quad\text{and}\quad \later{\procfun(s)} = \procfun(m \latercomp \later{s}) , \]
  for some conditional \(n\) of \(\now{s'}\) and some conditional \(m\) of \(\now{s}\).
  We have that \(n \dcomp u\) is also a conditional of \(\now{s}\) because
  \begin{equation}\label{eq:proc-conditionals-proof}
    \now{s} = \now{s'} \dcomp (u \tensor \id{}) = (\now{s} \dcomp (\discard \tensor \id{})) \latercomp (n \dcomp u) .
  \end{equation}
  Let \(r \dcomp i\) be a range of \(\now{s} \dcomp (\discard \tensor \id{})\).
  Then, \(i \dcomp m = i \dcomp n \dcomp u\) by the properties of ranges.
  We check that \(i \platercomp \later{\procfun(s)} \cequiv i \platercomp \later{\procfun(s')}\) for a range \(r \dcomp i\) of \(\now{s} \dcomp (\discard \tensor \id{})\).
  \begin{align*}
    & i \platercomp \later{\procfun(s)} && \\
    & = i \platercomp \procfun(m \latercomp \later{s}) && \text{(Definition of \(\procfun\))}\\
    & \cequiv i \platercomp (m \platercomp \procfun(\later{s})) && \text{(\Cref{lemma:procfun-morphism-monoidal-action})}\\
    & \cequiv (i \dcomp m) \platercomp \procfun(\later{s}) && \text{(\Cref{lemma:action-conditionalseq-compositions})}\\
    & \cequiv (i \dcomp n \dcomp u) \platercomp \procfun(\later{s}) && \text{(\kl{Ranges})}\\
    & \cequiv (i \dcomp n) \platercomp (u \platercomp \procfun(\later{s})) && \text{(\Cref{lemma:action-conditionalseq-compositions})}\\
    & \cequiv (i \dcomp n) \platercomp \procfun(u \platercomp \later{s}) && \text{(\Cref{lemma:procfun-morphism-monoidal-action})}\\
    & \cequiv i \platercomp (m \platercomp \procfun(\later{s'})) && \text{(Coinduction)}\\
    & \cequiv i \platercomp \procfun(m \latercomp \later{s'}) && \text{(\Cref{lemma:procfun-morphism-monoidal-action})}\\
    & = i \platercomp \later{\procfun(s')} &&
  \end{align*}
  We have shown that the mapping \(\procfun\) is well-defined.

  Now, we show that the definition of \(\procfun\) is independent of the choice of conditional.
  Suppose that there are two conditionals, \(m\) and \(n\), of \(\now{s}\) and consider a range \(r \dcomp i\) of \(\now{s} \dcomp (\discard \tensor \id{})\).
  Then, \(i \dcomp m = i \dcomp n\) by the properties of ranges.
  \begin{align*}
    & i \platercomp \procfun(m \latercomp \later{s}) &&\\
    & \cequiv \procfun(i \latercomp (m \latercomp \later{s})) && \text{(\Cref{lemma:procfun-morphism-monoidal-action})}\\
    & \cequiv \procfun((i \dcomp m) \latercomp \later{s}) && \text{(\Cref{lemma:monoidal-action-streams} and well-defined)}\\
    & = \procfun((i \dcomp n) \latercomp \later{s}) && \text{(\kl{Ranges})}\\
    & \cequiv \procfun(i \latercomp (n \latercomp \later{s})) && \text{(\Cref{lemma:monoidal-action-streams} and well-defined)}\\
    & \cequiv i \platercomp \procfun(n \latercomp \later{s}) && \text{(\Cref{lemma:procfun-morphism-monoidal-action})}
  \end{align*}
  This shows that \((\now{s} \dcomp (\discard \tensor \id{}) \mid \procfun(m \latercomp \later{s})) \cequiv (\now{s} \dcomp (\discard \tensor \id{}) \mid \procfun(n \latercomp \later{s}))\) and that the definition of \(\procfun\) does not depend on the chosen conditional \(m\).

  We now show that it is the right inverse of \(\streamfun\).
  Let \(p \colon \stream{X} \to \stream{Y}\) be a conditional sequence
  and let us compute \(\procfun(\streamfun(p))\) using coinduction.
  \begin{gather*}
    \now{\procfun(\streamfun(p))} =
    \now{\streamfun(p)} \dcomp (\discard \tensor \id{}) =
    (\now{p} \condcomp \id{\now{\stream{X}} \tensor \now{\stream{Y}}}) \dcomp (\discard \tensor \id{}) =
    \now{p}; \\
    \later{\procfun(\streamfun(p))} =
    \procfun(\id{\now{\stream{X}} \tensor \now{\stream{Y}}} \latercomp \later{\streamfun(p)}) =
    \procfun(\streamfun(\later{p})) =
    \later{p}.
  \end{gather*}
  Because both head and tail coincide, we conclude that
  \(\procfun(\streamfun(p)) = p\).
\end{proof}

We have shown that \kl{conditional sequences} faithfully correspond to \kl{effectful streams}.
We now show that this correspondence is also full.
Fullness gives a canonical representative to every \kl{effectful stream} \(f \colon \stream{X} \to \stream{Y}\);
the memory of the canonical representative is \(\now{\stream{X}} \tensor \now{\stream{Y}}\) and its first action is \((\now{f} \dcomp (\discard \tensor \id{})) \condcomp \id{}\); intuitively, the canonical representative stores in the memory objects all the previous inputs and outputs, \(M_{i} = X_{0} \tensor Y_{0} \tensor \cdots \tensor X_{i} \tensor Y_{i}\) for all \(i \in \naturals\).

\begin{prop}%
  \label{prop:normal-form-streams}%
  Let \((\cat{V},\cat{P})\) be a \kl{copy-discard category} category with \kl{conditionals} and \kl{ranges}.
  The functor \(\streamfun\) is full, with left inverse \(\procfun\) (\Cref{def:iso-streams-conditional-sequences}).
\end{prop}
\begin{proof}
  For a stream \(f \colon \stream{X} \to \stream{Y}\), the memory of \(\streamfun(\procfun(f))\) is \(\now{\stream{X}} \tensor \now{\stream{Y}}\) and its first action is
  \((\now{f} \dcomp (\discard \tensor \id{})) \condcomp \id{}\). We show that
  \(f \dinat \streamfun(\procfun(f))\).
  \begin{align*}
    & \streamfun(\procfun(f)) \\
    & = \langle (\now{f} \dcomp (\discard \tensor \id{})) \condcomp \id{} \mid \streamfun(\procfun(m \latercomp \later{f}))\rangle && \text{(\Cref{def:iso-streams-conditional-sequences})} \\
    &= \langle (\now{f} \dcomp (\discard \tensor \id{})) \condcomp \id{} \mid m \latercomp \streamfun(\procfun(\later{f}))\rangle && \text{(\Cref{lemma:procfun-morphism-monoidal-action,lemma:streamfun-morphism-monoidal-action})} \\
    & = \langle (\now{f} \dcomp (\discard \tensor \id{})) \condcomp m \mid \streamfun(\procfun(\later{f}))\rangle && \text{(\Cref{cor:sliding-qt-conditionals})}\\
    & = \langle \now{f} \mid \streamfun(\procfun(\later{f}))\rangle && \text{(Definition of \(m\))}\\
    & = \langle \now{f} \mid \later{f} \rangle && \text{(Coinduction)}\\
    & = f &&
  \end{align*}
  This concludes the proof.
\end{proof}

\begin{thm}%
  \label{th:streams-are-processes}%
  In a \kl{copy-discard category}, $(\cat{V},\cat{P})$, with \kl{conditionals} and \kl{ranges}, \kl{effectful streams} are monoidally isomorphic to \kl{causal processes},
  \[\Stream(\cat{V},\totals{\cat{P}},\cat{P}) \iso \Causal(\cat{V},\cat{P}).\]
\end{thm}
\begin{proof}
  By \Cref{prop:functor-processes-to-stream}, there is a monoidal functor $\streamfun ፡ \cCausal(\cat{V},\cat{P}) → \Stream(\cat{V},\totals{\cat{P}},\cat{P})$, which is faithful (\Cref{prop:processes-to-streams-faithful}) and full (\Cref{prop:normal-form-streams}).
  Then, streams are isomorphic to \kl{coinductive conditional sequences}, \(\cCausal(\cat{V},\cat{P}) \iso \Stream(\cat{V},\totals{\cat{P}},\cat{P})\).
  By \Cref{prop:coinductive-conditional-sequences}, \kl{causal processes} and \kl{conditional sequences} are also isomorphic, \(\cCausal(\cat{V},\cat{P}) \iso \Causal(\cat{V},\cat{P})\), and we obtain the thesis, \(\Stream(\cat{V},\totals{\cat{P}},\cat{P}) \iso \Causal(\cat{V},\cat{P})\).
\end{proof}

\begin{cor}\label{cor:iso}
  The following monoidal isomorphisms hold.
  \begin{align*}
    \Stream{(\Set,\Set,\Set)} &\iso \Proc{\Set, \Set}. \\
    \Stream{(\Set,\Set, \Par)} &\iso \Proc{\Set,\Par}. \\
    \Stream{(\Set,\Reltot,\Rel)} &\iso \Proc{\Set,\Rel}. \\
    \Stream{(\Set,\Stoch,\Stoch)} &\iso \Proc{\Set,\Stoch}. \\
    \Stream{(\Set,\Stoch,\subStoch)} &\iso \Proc{\Set,\subStoch}.
  \end{align*}
\end{cor}
\section{Conclusions}

We have introduced \kl{effectful Mealy machines}, a generalization of monoidal
Mealy machines to arbitrary \kl{effectful triples} as theories of processes.
Along with the definition, we generalized the main notions of equality: we
recovered \kl{bisimulation} via \kl{machine homomorphisms}; we recovered
\kl{trace equivalence} via \kl{effectful streams}.

Mealy machines with effects are relatively absent from the literature. We
attribute this to two barriers. The first is dinaturality: we have shown how
\kl{effectful streams} coincide with \kl{causal processes} in the commutative
case; it is thus tempting to stop at the commutative case, which enjoys an
easier characterization, and avoid the non-commutative, fully-effectful one. The
second is \kl{effectful triples}: when discussing effects, it is tempting to
limit ourselves to $\Set$-monads or monads with a cartesian base; however, even the Kleisli category of a $\Set$-comonad or comonad with a cartesian base generally falls outside this framework. We have seen how, in the monad case, we still recover the usual notions of \kl{bisimilarity} and \kl{trace equivalence}.

Further work on characterizing effectful \kl{causal processes} is warranted:
they connect dataflow programming with much existing literature on
\kl{copy-discard categories}, and allowed us to reuse concepts like that of
\kl{conditionals}. Further work on their dependent counterparts, following the
topos of trees~\cite{birkedal2012topos} or considering graded monoidal categories for graded semantics~\cite{milius2015generic,dorsch2019graded} is also likely to be fruitful. Finally,
we have mostly ignored the distributive structure of these categories, but it
will become important in programming applications.
 
\newpage

\section*{Acknowledgements}
We thank Giovanni de Felice for much encouragement and discussion. We thank
Pedro Azevedo de Amorim, Gabriele Tedeschi, Paweł Sobociński, and the anonymous
reviewers at NWPT and LiCS for suggestions and discussion.

\bibliographystyle{alphaurl}
\bibliography{bibliography}

\end{document}